%% file: jinping.tex
\RequirePackage{lineno}
\documentclass[dvipdfm,11pt,a4paper,doubleside]{article}
\usepackage{CJK}
\usepackage{graphicx}  
\usepackage{dcolumn}   
\usepackage{multirow}
\usepackage{bm}        
\usepackage{amssymb}   
\usepackage{amsmath}
\usepackage{epstopdf}  
\usepackage{color}     
\usepackage{slashbox}
\usepackage[affil-it]{authblk}  

\newcommand{\D}{\displaystyle}
\newcommand{\DF}[2]{\frac{\D#1}{\D#2}}

\begin{document}
\pagenumbering{Roman}




\title{
\vspace*{1cm}
\LARGE\bf Letter of Intent: \\
\vspace*{1cm}
\LARGE\bf Jinping Neutrino Experiment \\
\vspace*{1cm}
}

\newcommand{\OHO} {1}
\newcommand{\THE} {2}
\newcommand{\NCE} {3 }
\newcommand{\SDU} {4 }
\newcommand{\NKU} {5 }
\newcommand{\ECU} {6 }
\newcommand{\TCU} {7 }
\newcommand{\UMA} {8 }
\newcommand{\THC} {9 }
\newcommand{\ZSU} {10}
\newcommand{\AGS} {11}
\newcommand{\FUT} {12}
\newcommand{\GYY} {13}
\newcommand{\BNL} {14}
\newcommand{\GKD} {15}
\newcommand{\WUU} {16}
\newcommand{\TUD} {17}

\author[\OHO]{John~F.~Beacom}
\author[\THE]{Shaomin~Chen\thanks{Corresponding author, email: chenshaomin@tsinghua.edu.cn}}
\author[\THE]{Jianping~Cheng}
\author[\THE]{Sayed~N.~Doustimotlagh}
\author[\THE]{Yuanning~Gao}
\author[\THE]{Guanghua~Gong}
\author[\THE]{Hui~Gong}
\author[\THE]{Lei~Guo}
\author[\NCE]{Ran~Han}
\author[\THE]{Hong-Jian~He}
\author[\SDU]{Xingtao~Huang}
\author[\THE]{Jianmin~Li}
\author[\THE]{Jin~Li}
\author[\THE]{Mohan~Li}
\author[\NKU]{Xueqian~Li}
\author[\ECU]{Wei~Liao}
\author[\TCU]{Guey-Lin~Lin}
\author[\THE]{Zuowei~Liu}
\author[\UMA]{William~McDonough}
\author[\ZSU]{Jian~Tang}
\author[\THE]{Linyan~Wan}
\author[\THC]{Yuanqing~Wang}
\author[\THE]{Zhe~Wang\thanks{Corresponding author, email: wangzhe-hep@tsinghua.edu.cn}}
\author[\THC]{Zongyi~Wang}
\author[\THE]{Hanyu~Wei}
\author[\AGS]{Yufei~Xi}
\author[\FUT]{Ye~Xu}
\author[\THE]{Xun-Jie~Xu}
\author[\THE]{Zhenwei~Yang}
\author[\GYY]{Chunfa~Yao}
\author[\BNL]{Minfang~Yeh}
\author[\THE]{Qian~Yue}
\author[\THE]{Liming~Zhang}
\author[\THE]{Yang~Zhang}
\author[\THC]{Zhihong~Zhao}
\author[\GKD]{Yangheng~Zheng}
\author[\WUU]{Xiang~Zhou}
\author[\THE]{Xianglei~Zhu}
\author[\TUD]{Kai~Zuber}

\affil[\OHO]{\small Dept. of Physics, Dept. of Astronomy, and CCAPP, The Ohio State University, Columbus, OH 43210}
\affil[\THE]{Department of Engineering Physics, Tsinghua University, Beijing 100084}
\affil[\NCE]{Science and Technology on Reliability and Environmental Engineering Laboratory, Beijing Institute of Spacecraft
Environment Engineering, Beijing 100094}
\affil[\SDU]{School of Physics, Shandong University, Jinan 250100}
\affil[\NKU]{School of Physics, Nankai University, Tianjin 300371}
\affil[\ECU]{School of Science, East China University of Science and Technology, Shanghai 200237}
\affil[\TCU]{Institute of Physics, National Chiao-Tung University, Hsinchu}
\affil[\UMA]{University of Maryland, College Park, Maryland 20742}
\affil[\THC]{Department of Civil Engineering, Tsinghua University, Beijing 100084}
\affil[\ZSU]{School of Physics, Sun Yat-Sen University, Guangzhou 510275}
\affil[\AGS]{Institute of hydrogeology and environmental geology, Chinese Academy of Geological Sciences, Shijia Zhuang 050061}
\affil[\FUT]{Fujian University of Technology, Fujian 350118}
\affil[\GYY]{Department of Structural Steels, China Iron \& Steel Research Institute Group 100081}
\affil[\BNL]{Brookhaven National Laboratory, Upton, New York 11973}
\affil[\GKD]{School of Physical Sciences, University of Chinese Academy of Sciences, Beijing 100049}
\affil[\WUU]{School of Physics and Technology, Wuhan University, Wuhan 430072}
\affil[\TUD]{Institut f\"{u}r Kern- und Teilchenphysik, Technische Universit\"{a}t Dresden, Dresden 01069}

\date{January 1, 2016}

\maketitle


\clearpage
\newpage
\mbox{}
\newpage

\input{Executive}

\clearpage
\newpage


\numberwithin{equation}{section}
\numberwithin{figure}{section}
\numberwithin{table}{section}

\tableofcontents
\clearpage
\newpage

\pagenumbering{arabic}
\input{Site}

\clearpage
\newpage

\input{Detector}

\clearpage
\newpage

\input{Solar}

\clearpage
\newpage

\input{Geoneutrino}

\clearpage
\newpage

\input{SRN}

\clearpage
\newpage

\input{SNBurst}
\clearpage
\newpage

\input{DM}

\clearpage
\newpage

%
%

\section{Summary}
The Jinping underground lab with an extremely low cosmic-ray muon flux, 200 times lower than the Borexino site, 2 times lower than the SNO site and with a low reactor neutrino flux (1,200 km from the closest nuclear power plant) and with several other good features
is ideal to do low background neutrino experiments.
With 2,000-ton fiducial mass for the solar neutrino physics (equivalently 3,000-ton for the geo-neutrino and supernova relic neutrino physics), we found that Jinping will be able to discover the neutrinos from the CNO fusion cycles of the Sun, to precisely measure the transition phase for the solar neutrinos oscillation from the vacuum to the matter effect, to precisely determine the geo-neutrinos' flux, the U, Th ratio, and to help to determine the geo-neutrino models.
It will bring our knowledge for them to a brand new stage.
These physics goals can be fulfilled with the present mature techniques.
Efforts on increasing the target mass with multi-modular neutrino detectors and
developing the water-based scintillator technique will enable us to eventually enrich the Jinping discovery potentials in the study of solar neutrinos, geo-neutrinos, and supernova neutrinos, and dark matter.
With the deepest motivation and wish to explore the Sun, the Earth, the universe, and the primary particle - neutrino,
we like to propose a neutrino experiment at China JinPing underground Laboratory II.

\include{Acknowledgement}

\end{document}

%% file: Executive.tex
\section*{Executive Summary}

In this letter of intent, we propose the Jinping Neutrino Experiment (Jinping), a unique observatory for low-energy neutrino physics, astrophysics and geophysics.  The total detector mass would be about 4 kiloton of liquid scintillator or water-based liquid scintillator, with a fiducial mass of 2 kiloton for neutrino-electron scattering events and 3 kiloton for inverse-beta interaction events.  With low radioactivity backgrounds, based on existing technology, and low cosmogenic and reactor backgrounds, based on depth and location, Jinping would have unprecedented sensitivity.

\begin{figure}[!h]
  \vspace*{-2mm}
  \centering
  \includegraphics[width=11cm, clip]{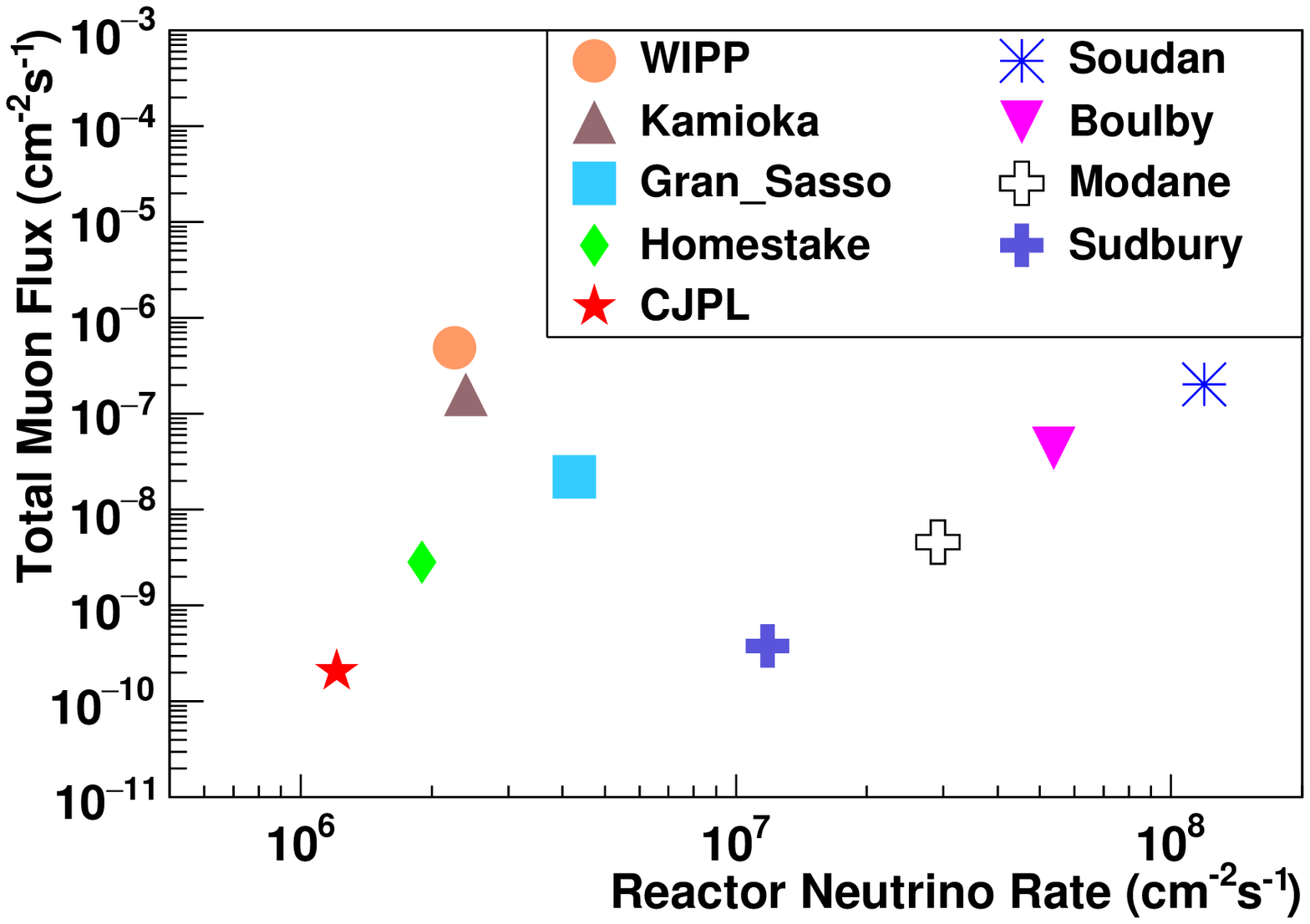}
  \caption{Jinping with the lowest muon flux and reactor neutrino background.}
  \label{fig:loc}
\end{figure}

China JinPing Laboratory (CJPL) has a number of unparalleled features (Fig.~\ref{fig:loc}): thickest overburden, lowest reactor neutrino background, dominant crustal geo-neutrino signal, lowest environmental radioactivity, longest solar neutrino path
through the Earth, etc. All of these attributes identify it as the world-best low-energy neutrino laboratory.
Moreover, CJPL has important practical advantages, including being within two hours of an airport, reachable on good roads, with drive-in access for trucks, and with excellent supporting infrastructure.  The first, small phase of the laboratory (CJPL I) is already in operation, hosting dark matter experiments.  The second, large phase (CJPL II) is already under construction, with $\sim$100,000 m$^3$ being excavated.


Neutrinos, though ghostly and mysterious, are a central character in interdisciplinary studies that connect fundamental issues in particle and nuclear physics, astrophysics and cosmology, and geo-science.  The weak interactions of neutrinos make them powerful probes of the deep interiors of sources like the Sun and Earth, including their power sources and composition.  And the quantum ability of neutrinos to shift identity through three-flavor oscillations in vacuum and matter make them exquisitely sensitive probes of new physics, including aspects of this mixing that have not yet been measured precisely.  The measurements made at Jinping would complement those at other detectors, together creating a richer view of neutrinos from natural sources as well as tests of neutrino properties.

We have conducted initial sensitivity studies for the Jinping detector based on assessments of the site and potential detector designs and give the expected discoveries and precision improvements for neutrino physics, astrophysics, and geo-science.

\begin{figure}[!h]
  \centering
  \includegraphics[angle=270, width=10cm, clip]{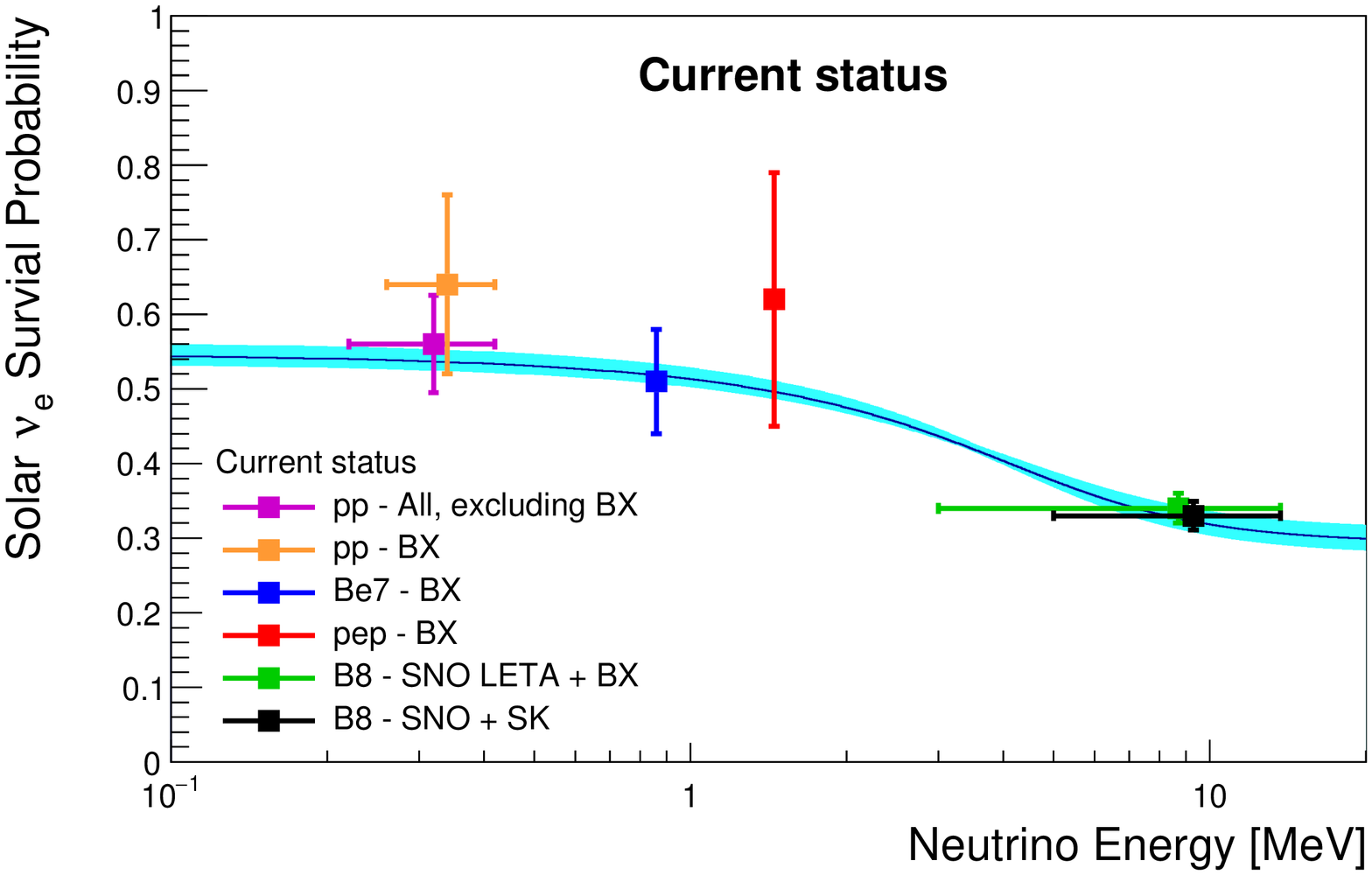}\\
  \includegraphics[angle=270, width=10cm, clip]{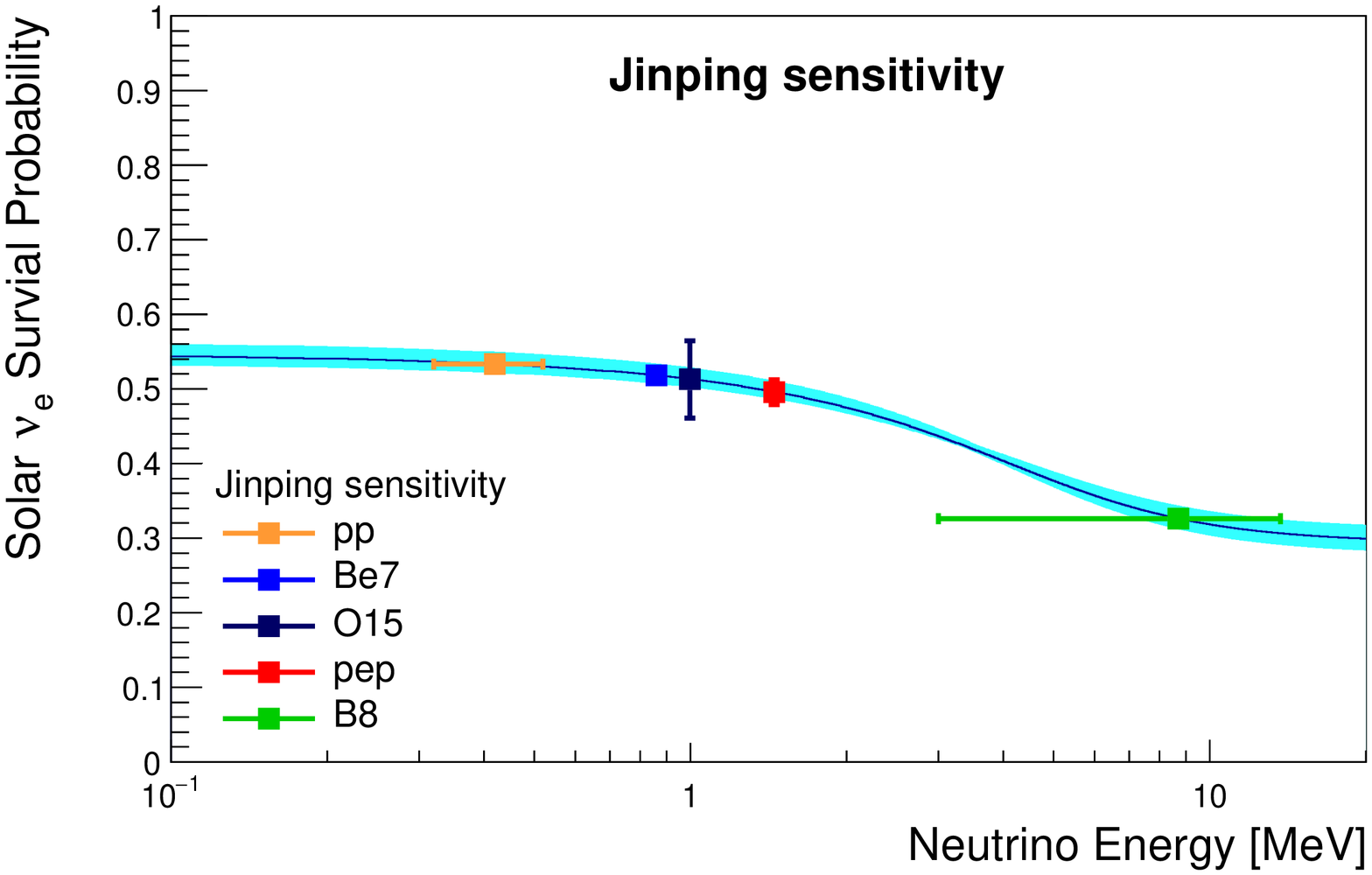}
  \caption{The transition of oscillation probability from the vacuum to matter effect as a function of the solar neutrino energy. The present experimental status and Jinping sensitivity are shown in the top and bottom plots, respectively.
  $^{15}$O horizontal error bar is not shown for clarity.}
  \label{fig:Solar}
\end{figure}

Jinping has a very strong potential to significantly
improve the measurements of neutrinos with a few MeV energy
from the interior solar fusion processes, including the components, fluxes
and spectra, see Fig.~\ref{fig:Solar}.
Jinping can precisely measure the transition phase for the solar neutrinos oscillation from the vacuum to the matter effect, providing a critical test for the Mikheyev-Smirnov-Wolfenstein (MSW) theory in the high density environment.
We predict a capacity to discover solar neutrinos from the carbon-nitrogen-oxygen (CNO) cycle with more than 5 sigma of statistical significance, discovering the energy source for massive stars and
shedding light on the metal abundance of the solar core and the homogeneous chemical assumption of the C, N, and O elements.
The proposed experiment can also resolve the high and low metallicity hypotheses by more than 5 sigma with known neutrino oscillation angles.
These measurements will give a much more comprehensive understanding of the Sun and of neutrino properties.

\begin{figure}[!ht]
\centering
\includegraphics[height=7cm, width=9cm, clip]{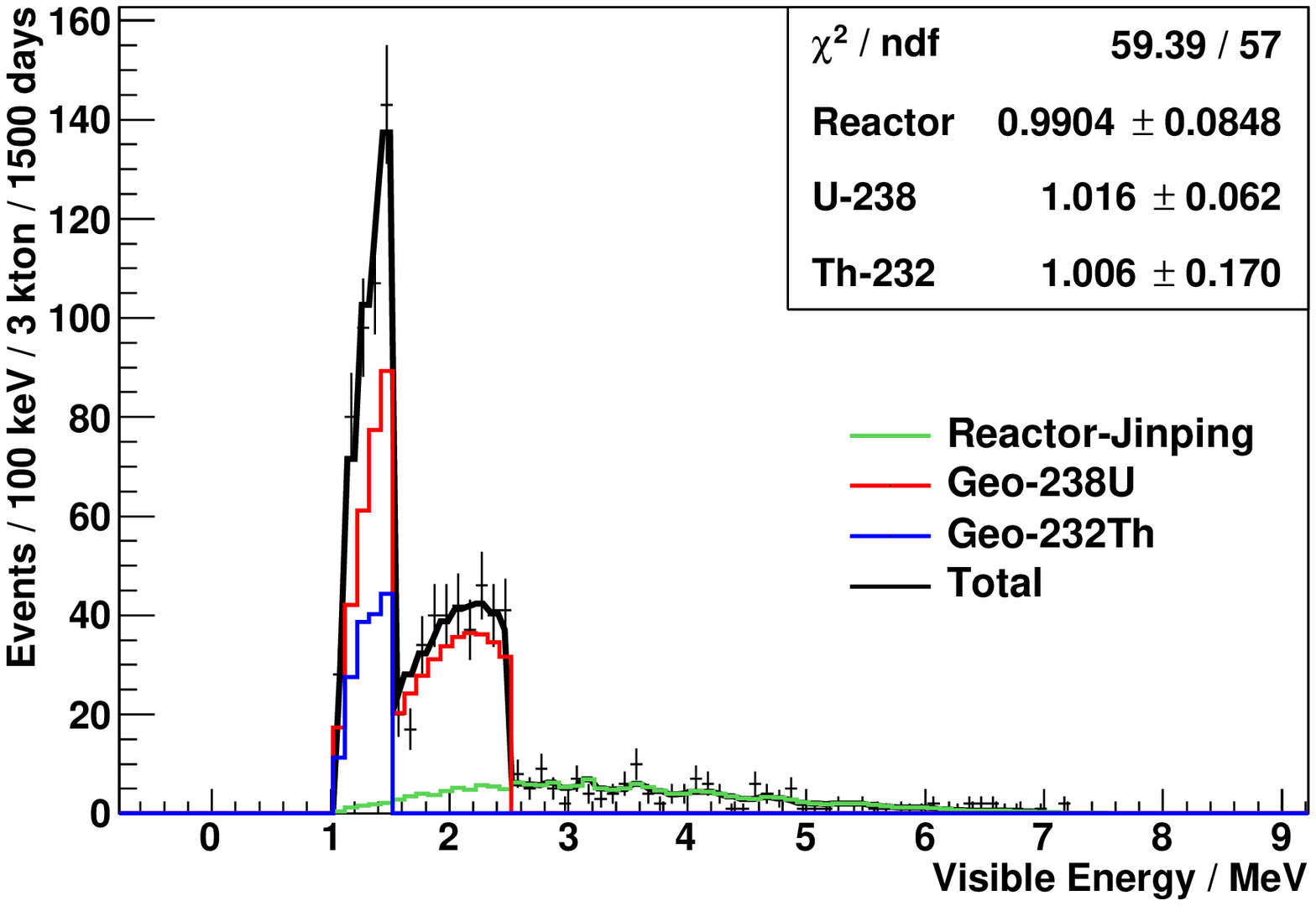}\\
\includegraphics[height=8cm, width=10cm, clip]{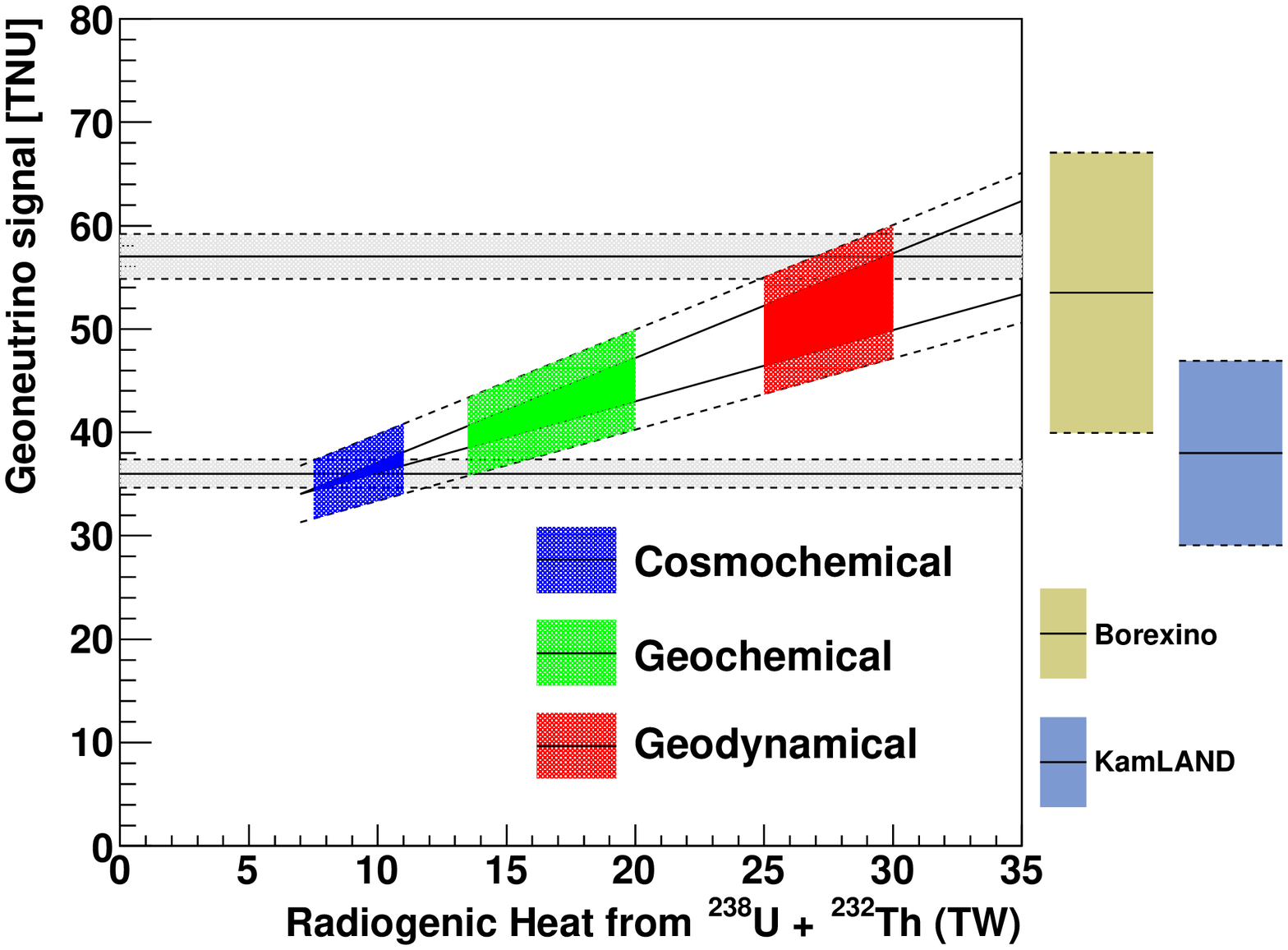}%
\caption{Jinping's sensitivity in measuring the U and Th components of geo-neutrinos separatively (top).
The geo-neutrino sensitivity vs geo-neutrino model predictions at Jinping (bottom).
The three filled regions in the plot delimit, from the left to the right, the cosmochemical, geochemical, and geodynamical models, respectively.
On the right, the Borexino and KamLAND measurements are shifted up according to the differences of their crust geoneutrino fractions with Jinping.
The two horizontal bars are plotted at two possible geo-neutrino flux assumptions with Jinping geo-neutrino measurement sensitivity.
}
\label{fig:Geo}
\end{figure}

Jinping can carry a precise measurement on
geo-neutrinos with an unambiguous separation on U and Th cascade decays from the dominant crustal anti-electron neutrinos, see Fig.~\ref{fig:Geo}.
The estimated event rates of 37 U and 9 Th geo-neutrino events/year/kton will be significantly above the expected $<$6
reactor neutrino events/year/kton. The ratio of U/Th can be determined to 10\%. We expect that the measurement from Jinping together with the Borexino and KamLAND results can give an extrapolation of the flux for the desired mantle neutrinos and reveal the mystery of the engine driving Earth's continental growth, mountain movement and distribution of heat producing elements.

These physics goals can be fulfilled using mature techniques and the unique opportunity of the CJPL.
Efforts to increase the target mass with multi-modular neutrino detectors and to develop water-based scintillator technique will enable us to eventually enrich the discovery potential of Jinping.
Promising sensitivity could be expected for neutrinos from a Milky Way supernova, the diffuse supernova neutrino background, and dark matter annihilation.

\clearpage
\newpage
\begin{figure}[h]
\centering
\includegraphics[width=165.10mm, clip]{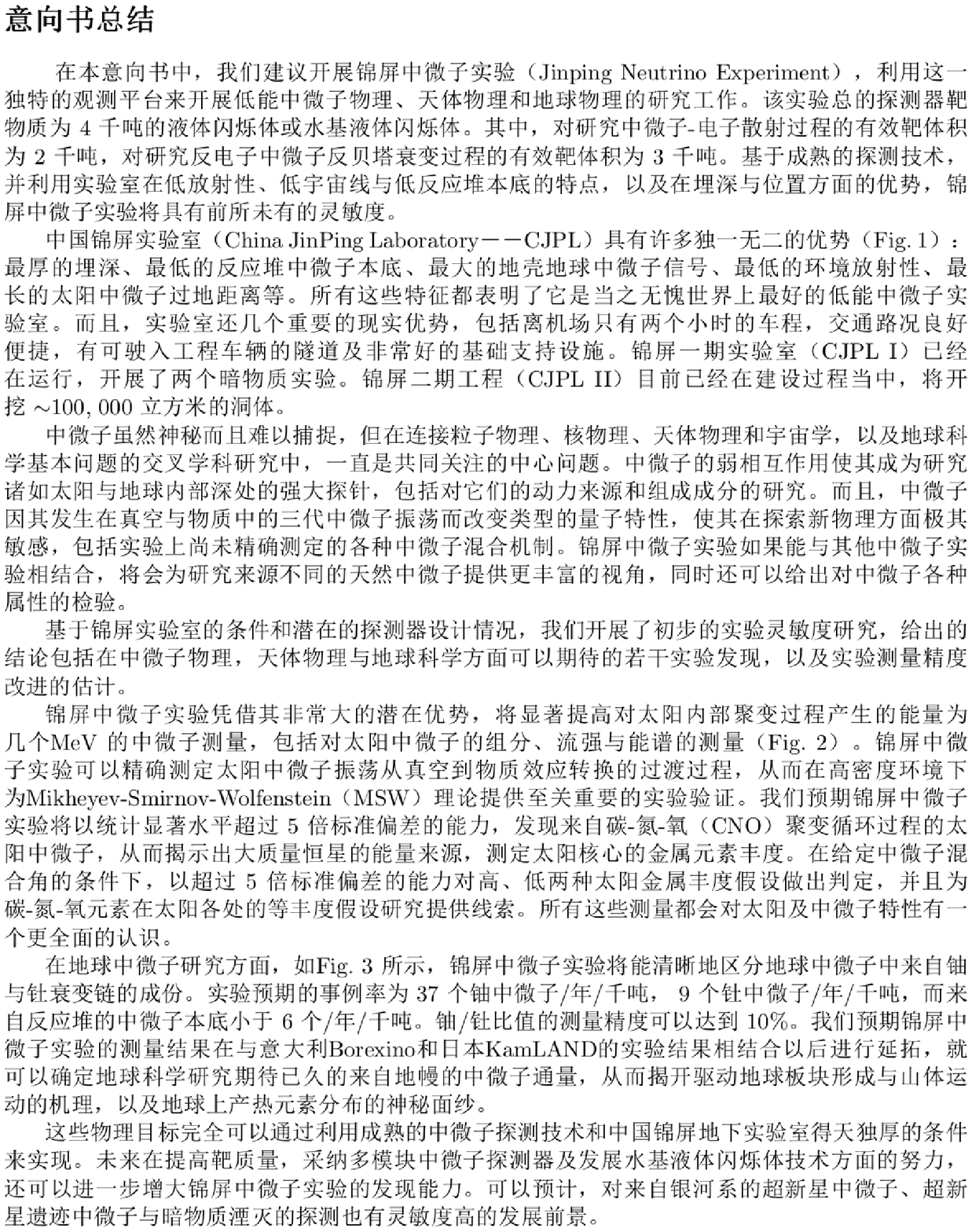}\\
\end{figure}

%% file: Site.tex
\section{Experimental site}

\subsection{Overview}
\begin{figure}[!h]
  \centering
  \includegraphics[width=7cm, height=7cm]{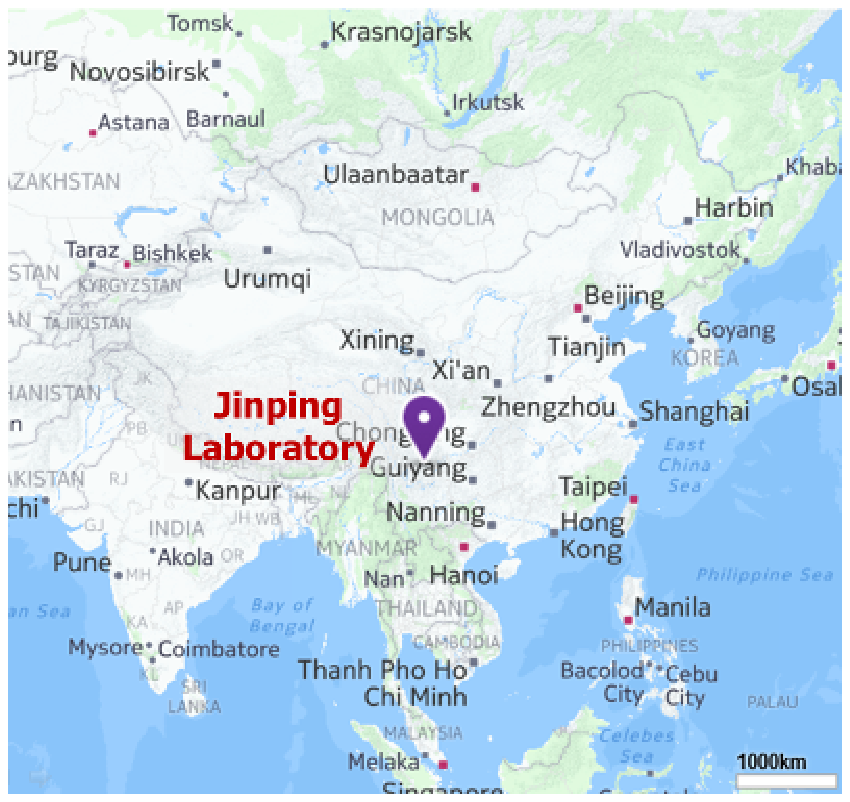}\hspace*{0.5cm}
  \includegraphics[width=7cm, height=7cm]{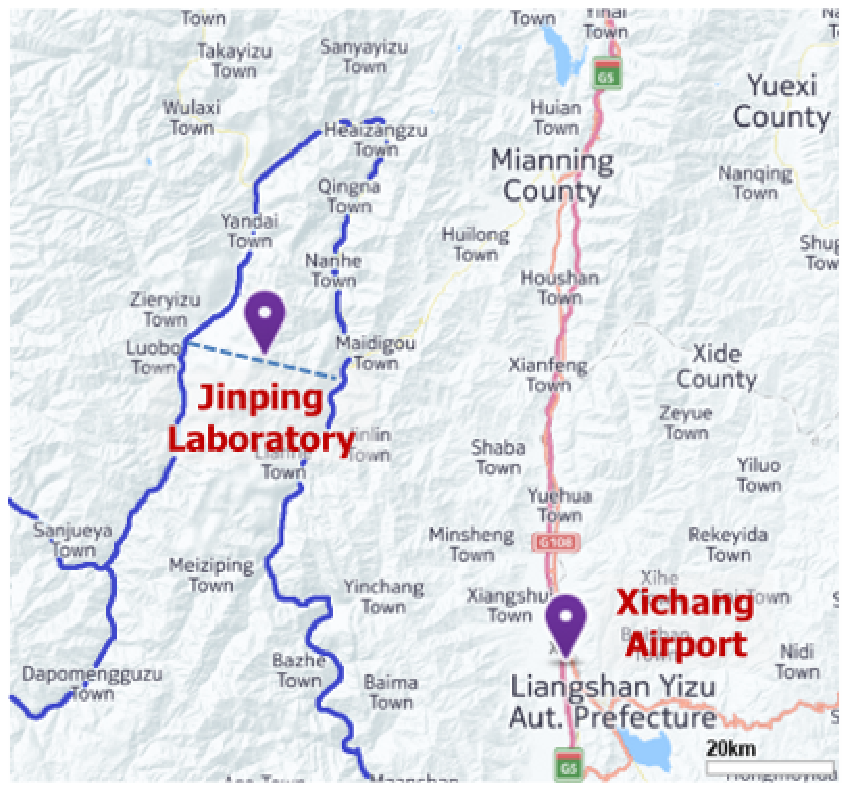}
  \caption{(Color online) China JinPing underground Laboratory (CJPL) is located in Jinping Mountain, Sichuan Province, China.
  Jinping Mountain is surrounded by Yalong River as the solid blue line in the right plot, and
  the position of Jinping tunnels is indicated by the dashed line. The distance to the closest airport is about 2 hours' drive. (Based on Yahoo Map) }
  \label{fig:JinpingSite}
\end{figure}

\begin{figure}[!tb]
  \centering
  \includegraphics[height=7cm]{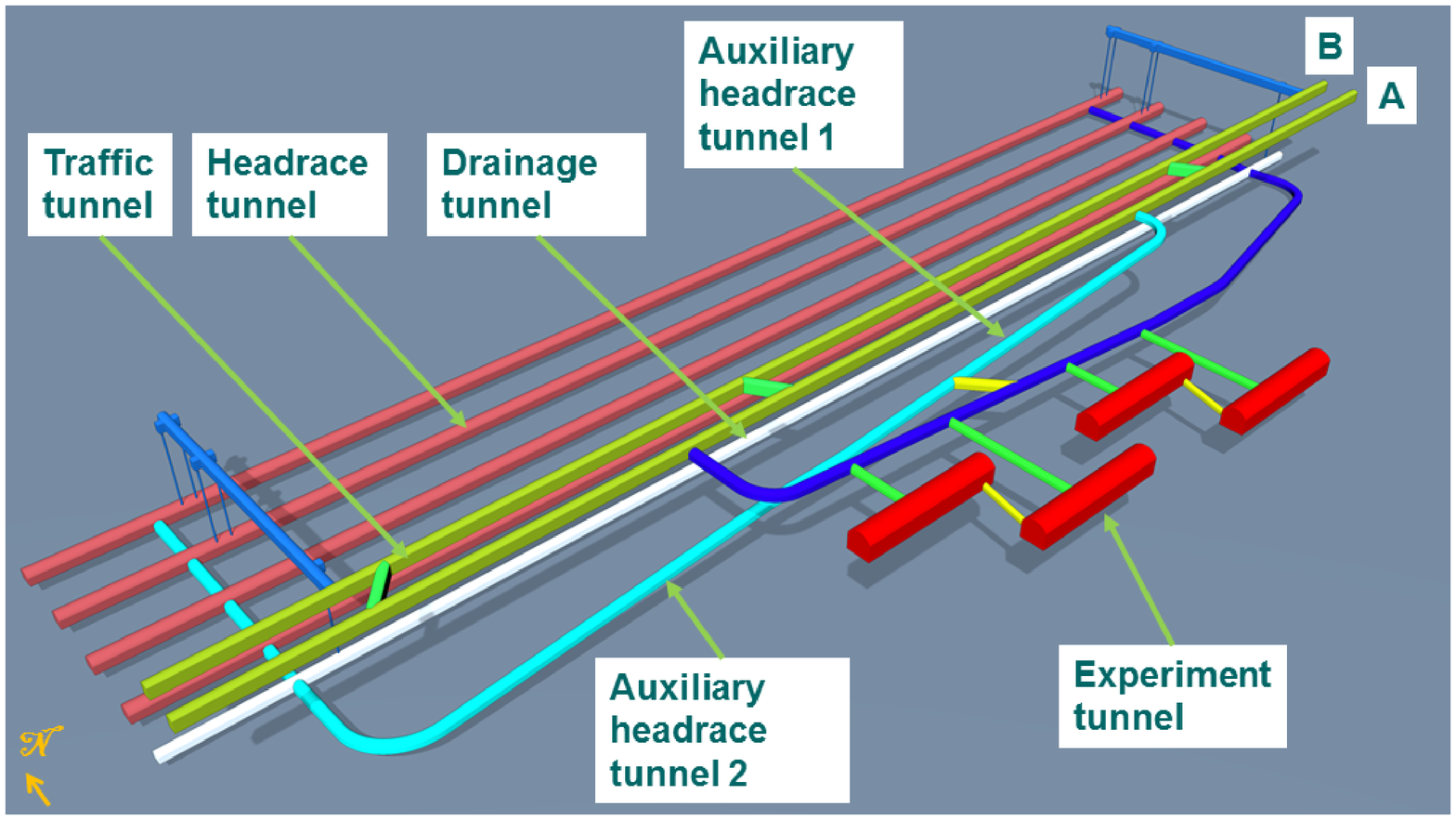}
  \caption{(Color online) Schematic of Jinping tunnels and Jinping phase II laboratories.}
  \label{fig:JPtunnels}
  \centering
  \includegraphics[width=12.6cm]{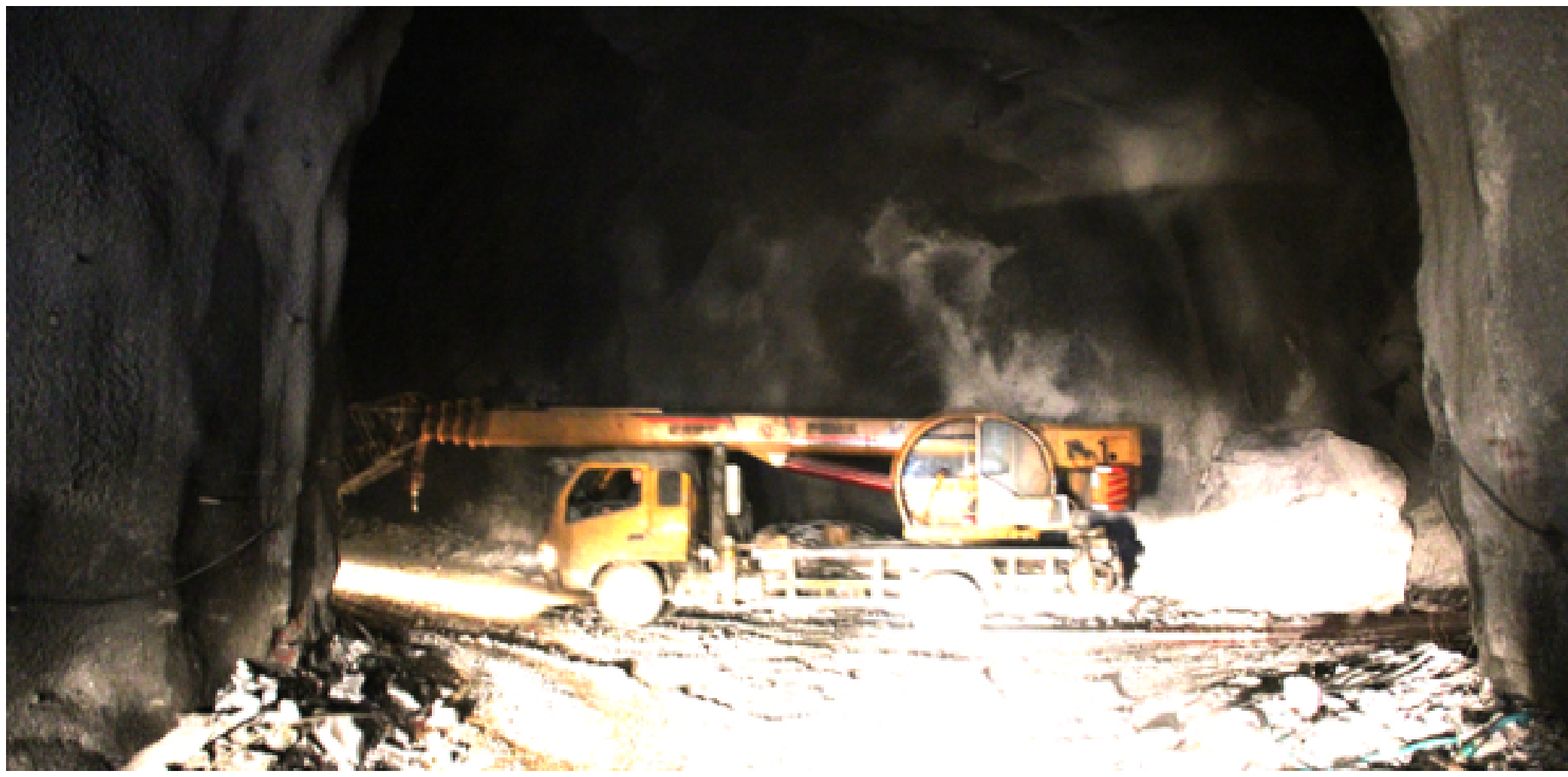}
  \caption{(Color online) A photograph of the construction site in one of the tunnels at CJPL II.}
  \label{fig:cons}
\end{figure}

China JinPing underground Laboratory (CJPL)~\cite{CJPL1} is one of the ideal sites to do low background experiments in the world.
The experimental site is located in Jinping Mountain, Sichuan Province, China (Fig.~\ref{fig:JinpingSite}).
Jinping Mountain measures
4,100-4,500 meters high and is surrounded by Yalong River. A 150-km river bend surrounds the mountain with a water
level difference of 312 m between both sides. China Yalong River Hydropower Development Company (Yalong Hydro, previously
known as Ertan Hydropower Development Company) has built the Jinping II Hydropower Station, including four headrace tunnels,
two traffic tunnels and one drainage tunnel across the Jinping Mountain (Fig.~\ref{fig:JPtunnels}).
The headrace tunnels are about 16.7 km in
length and 12.4-13.0 m in diameter, with a maximum overburden of 2,400 m (6,720 meter water equivalent assuming a constant rock density 2.8 g/cm$^3$).
More than 75\% of the tunnel depth is larger than 1,700 m. Two traffic tunnels are parallel to the headrace tunnels: \#A (5 m in width and 5.5 m in height) and \#B (6 m in width and 6.5 m in height). The drainage tunnel with a diameter of 7.2 m is located between traffic tunnel \#B and headrace tunnel \#4. All seven tunnels were finished in August 2008 and are all now maintained by Yalong Hydro.

The first phase of Jinping laboratory (CJPL I) was constructed in the middle of the traffic tunnels at the end of 2009.
The lab was used for two dark matter experiments: CDEX~\cite{CDEX1} and PandaX~\cite{PdX1}.
The second phase of Jinping laboratory (CJPL II) started in the end of 2014.
The construction plan is to build four 150 m long tunnels not far away from the traffic tunnels as shown in Fig.~\ref{fig:JPtunnels}.
Totally $\sim$100,000 m$^3$ is being evacuated~\cite{Jianmin}.
A photograph of the construction site in one of the tunnels at CJPL II can be seen in Fig.~\ref{fig:cons}.

We propose to use one of the new tunnels to build a neutrino experiment
with a fiducial target mass of two kilo-ton for solar neutrino physics, equivalently three kilo-ton for geo-neutrino and supernova relic neutrino physics.
The initial plan is to adopt the liquid-scintillator technique as the base line design, with a capacity of extension to a water-based scintillator detector.

\subsection{Geological conditions and geotechnical feasibility}

\begin{figure}[!tb]
  \centering
  \includegraphics[height=11cm]{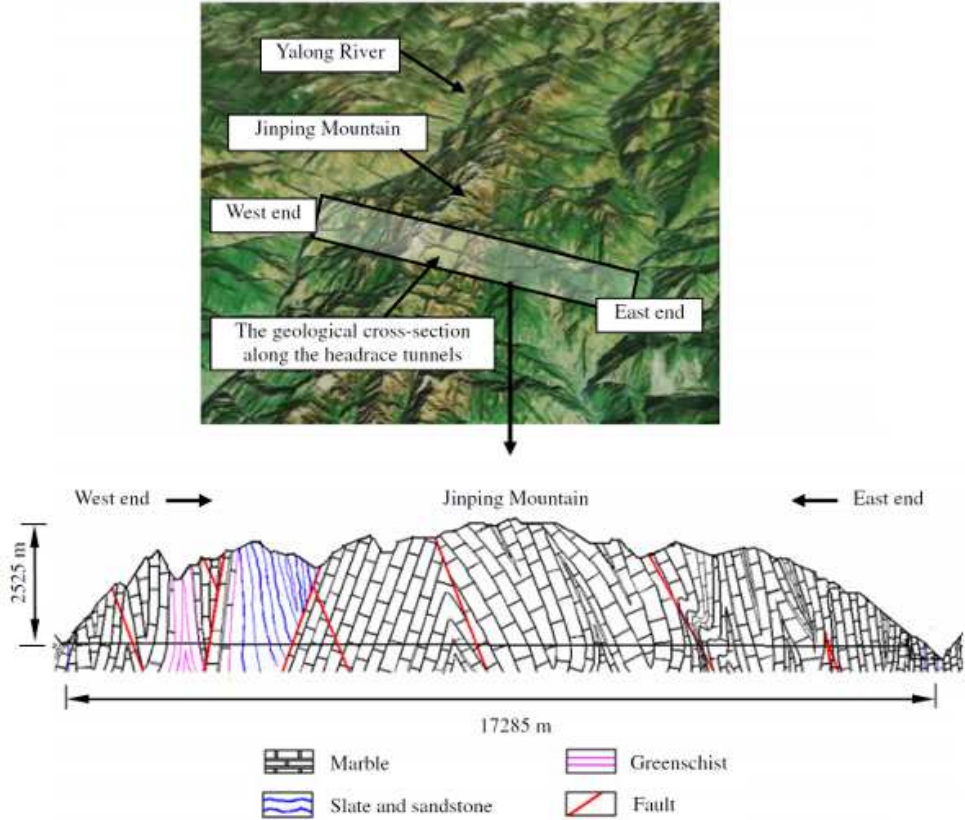}
  \caption{(Color online) Geological cross-section along the headrace tunnels of the Jinping II Hydropower Station~\cite{zh4}.}
  \label{fig:Crosssection}
\end{figure}

The Yalong River is located in the geomorphological level II ladder of the transition zone from the Tibetan Plateau to the Sichuan Basin. The altitude decreases from about 5,000 m in the northwest to approximately 2,000 m in the southeast. The Jinping Mountain extends along a nearly N-S direction. Many of the peaks are higher than 3,000 m in altitude, with a maximum relative elevation difference of 3,150 m. The main watershed lies along an N-S axis and is slightly oriented to the west. The regional distribution of mountain is basically consistent with the tectonic line. Generally the topography of the Jinping laboratory region appears as a undulating ground surface and has large differences in elevation (Fig.~\ref{fig:Crosssection}).

\subsubsection{In-situ stress}
\begin{figure}[!tb]
  \centering
  \includegraphics[width=13cm]{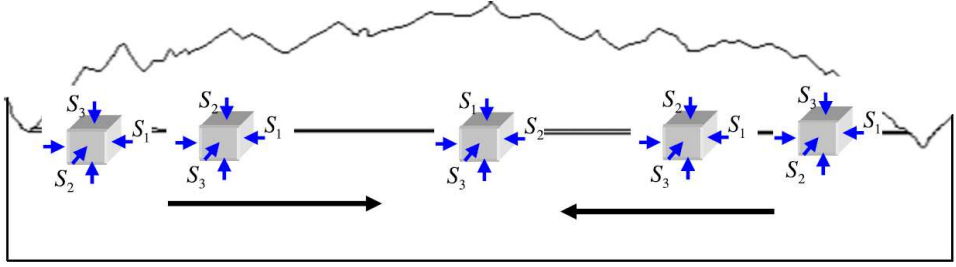}
  \caption{(Color online) Prediction of the macro-distribution of in-situ stress along the headrace tunnels of the Jinping II Hydropower Station~\cite{zh4}. The stress magnitudes are defined using the standard geologic/geophysical notation with S1$>$S2$>$S3, where S1 is the maximum principal stress, S2 is the intermediate principal stress, and S3 the minimum principal stress.}
  \label{fig:stress}
\end{figure}

The complicated mountainous topography, lithologies, geological structures (faulting and folding, and a complex geological history) directly lead to a sophisticated stress field in the Jinping Mountain (Fig.~\ref{fig:stress}). At the both ends of the tunnels near the river valley, horizontal stresses are larger than the vertical stress, and a TF-type (Thrust faulting) of stress regime is expected with the sequence of $\sigma_H>\sigma_h>\sigma_v$, where $\sigma_H$ is the maximum horizontal stress, $\sigma_h$ is the minimum horizontal stress, and $\sigma_v$ is the vertical stress. The Jinping laboratory site has an overburden of about 2,400 m, and the vertical stress is up to about 66 MPa, which is dominated by gravity stress. The maximum and minimum horizontal in-situ stresses are about 55 MPa and 44 MPa, respectively. The principal stress regime is NF (Normal faulting) with the sequence of $\sigma_v>\sigma_H>\sigma_h$. There might be a regime transition zone between the stress field near the river valley and that at the maximum buried depth, where the stress sequence would be $\sigma_H>\sigma_v>\sigma_h$ corresponding to SS (Strike-slip faulting) stress regime.

\subsubsection{Hydrological conditions}
The distribution of fractured and karst groundwater is complex due to local variations under hydrogeological conditions. At the eastern and western ends of the Jinping tunnels, karst geology structures were exposed by excavation, and karst water was predominant groundwater at both ends. However, in the middle of the tunnels and the experimental site under the large overburden, the groundwater distribution is inhomogeneous and probably concentrated on local tunnel regions, where rapid water influx may occur. The groundwater includes both fractured and karst water. Seven heavy water in-rush events were recorded during the excavation of the two traffic tunnels. Their water pressures ranged from 0.6 to 4.7 MPa and their influx flow rates ranged from 0.15 to 15.6 $m^3/s$~\cite{zh5}.

\subsubsection{Main rock mechanical parameters}
Along the Jinping tunnels, harder rocks (i.e.~marble and sandstone) and softer rocks (i.e.~chlorite schist and sand slate) compose the feature of rock mass, and marble is the main rock in the engineering region of Jinping Laboratory. The unconfined compressive strength (UCS) of Jinping marble is between 95 and 105 MPa, and its damage initiation stress is between 40 and 50 MPa.

\subsubsection{Excavation damage zone}
\begin{figure}[!h]
  \centering
  \includegraphics[width=9cm]{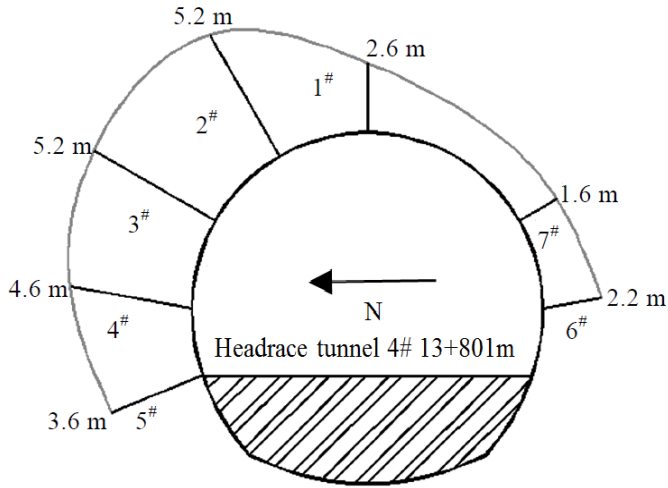}
  \caption{Excavation damage zone of Section 13+801m of Jinping headrace tunnel \#4~\cite{zh6}.}
  \label{fig:damage}
\end{figure}

Under high stress condition, the conflict between the strength of rock mass and the stress can lead to damage in the vicinity of the experimental hall. Acoustic testing and borehole television were used to monitor the size of the damage zone around Jinping headrace tunnel \#4, and the results showed that the damage zone appears an asymmetric shape, i.e. the depth of the north side of the headrace tunnel \#4 was much larger than that of the south side (Fig.~\ref{fig:damage}).

\subsubsection{Risks of rock bursts}
Rock-bursts were observed in the massive marble in the Jinping tunnels from approximately 1,700 m of overburden. More than 100 rock-burst events, of varying extents and severities, occurred during the excavation of the four headrace tunnels and the drainage tunnel with a depth greater than 1,900 m. Among them, a severe rock-burst event occurred during the TBM (Tunnel Boring Machine) tunneling in the drainage tunnel on 28 November 2009. Due to the instant impact of this rock-shock, all support systems were destroyed, the main beam of the TBM equipment was broken, and a space of about 30 m behind the cutter head was buried in debris. Seismic events, both before and during the rock-burst, were registered 2.0 on the Richter scale by the micro-seismic monitoring equipment.

\subsubsection{Future studies needed}
Overall, to host a large experimental hall for a kiloton scale neutrino detector, a couple of geotechnical challenges still exist, e.g., rock bursts, water influx, and large deformation. During the conceptual design phase, the conclusions to the following issues will be given:
(1) Experimental hall layout,
(2) Excavation and support method,
(3) Water Supply and Drainage,
(4) Electricity and Ventilation System, and
(5) Risks during the construction and operation.

\subsection{Rock radioactivity}
The radioactivity of the rock in Jinping tunnel is measured~\cite{JPRock} and the result is shown in Tab.~\ref{tab:rock} together with
the measurements in Sudbury~\cite{SNORock}, Gran sasso~\cite{BXRock}, and Kamioka~\cite{KMRock} underground laboratories.

\begin{table}[h]
\begin{center}
\begin{tabular}[c]{cccc} \hline\hline
Site               & $^{238}$U    & $^{232}$Th    & $^{40}$K \\\hline
 Jinping           & 1.8$\pm$0.2 ($^{226}$Ra)  & $<$0.27         & $<$1.1  \\
Sudbury            & 13.7$\pm$1.6 & 22.6$\pm$2.1  & 310$\pm$40  \\
Gran sasso hall A  & 116$\pm$12   & 12$\pm$0.4    & 307$\pm$8  \\
Gran sasso hall B  & 7.1$\pm$1.6  & 0.34$\pm$0.11 & 7$\pm$1.7  \\
Gran sasso hall C  & 11$\pm$2.3   & 0.37$\pm$0.13 & 4$\pm$1.9  \\
Kaminoka           & $\sim$12 & $\sim$10 & $\sim$520  \\\hline
\end{tabular}
\caption{Radioactivity contamination in Bq/kg for some underground laboratories.}
\label{tab:rock}
\end{center}
\end{table}

\subsection{Cosmic-ray muon flux}
\begin{figure}[!h]
  \centering
  \includegraphics[height=7cm]{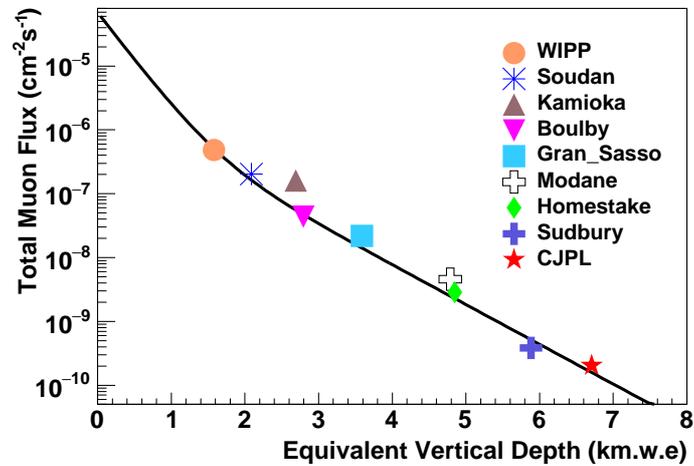}
  \caption{(Color online) Cosmic-ray muon flux of CJPL and a comparison with other laboratories.}
  \label{fig:MuFlux}
\end{figure}

Cosmic-ray muon itself can be easily detected and vetoed, but muon induced spallation backgrounds, especially fast neutrons and long lifetime isotopes are extremely dangerous for low background counting experiments. The rejection method usually includes a large buffer region to tag original muons and a long veto time window, which consumes a lot of space and detection efficiency.
However the cosmic-ray muon flux decreases sharply when the depth is getting larger.
According to the in-situ measurement~\cite{Mountain}, the muon flux is as low as $(2.0\pm0.4)\times10^{-10}/(\rm{cm}^2\cdot\rm{s})$. A comparison with other underground labs can be seen in Fig.~\ref{fig:MuFlux} and Fig.~\ref{fig:GeoFlux}~\cite{Mountain, HomestakeS}.

\subsection{Reactor neutrino background}
\begin{figure}[h]
  \centering
  \includegraphics[height=7cm]{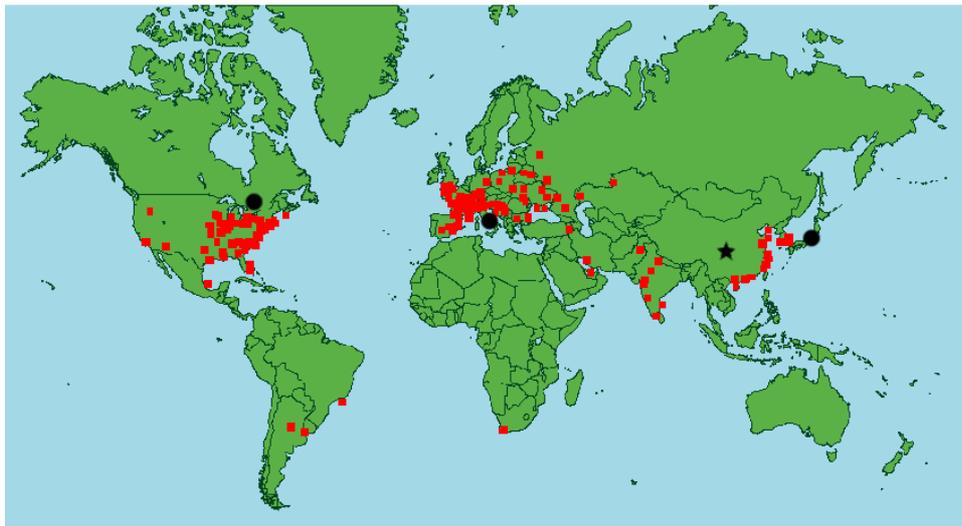}
  \caption{(Color online) World map with all the nuclear power plants in operation and under construction.
  SNO, Gran Sasso, Kamioka and Jinping laboratory locations are also marked.}
  \label{fig:WorldNPP}
\end{figure}

\begin{figure}[h]
   \vspace*{-3mm}
  \centering
  \includegraphics[height=8cm]{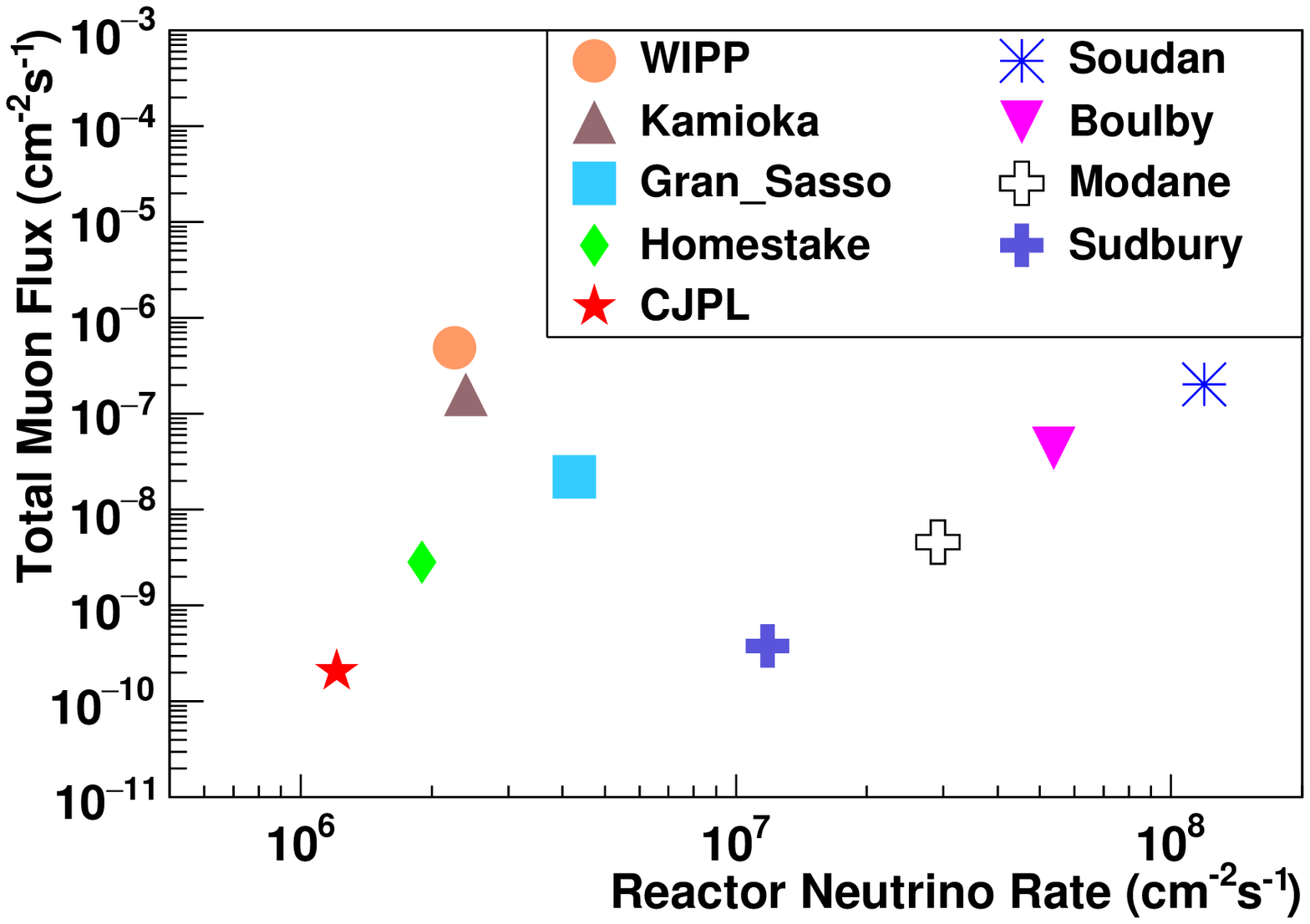}
  \caption{(Color online) Muon flux vs reactor neutrino background rate for various underground labs in the world.}
  \label{fig:GeoFlux}
\end{figure}

Jinping is also far away from all the nuclear power plants~\cite{IAEA} in operation and under construction.
A world map with all nuclear power plants and SNO, Gran Sasso, Kamioka, and Jinping laboratories
is shown in Fig.~\ref{fig:WorldNPP}.
A reactor background flux comparison with these laboratories is shown in Fig.~\ref{fig:GeoFlux}
The reactor electron antineutrino background at Jinping is rather low and will be explained in detail in later sections.

%% file: Detector.tex
\section{Detector concepts}

With the primary physics goals for low-energy neutrinos, the Jinping detector design follows the structure adopted by the recent underground neutrino experiments, and it will also consider the unique feature of deep underground and the tunnel structure.

Target mass is a key factor of the proposed neutrino experiment.
A few constraints must be considered to reach a certain level of target mass requirement.
\begin{itemize}
\item[1)] Both the tunnel size and shape limit the total volume of the experiment.
Especially the minimal dimension of the tunnel is important if a spherical detector is preferred.
This type of detector usually has the best uniformity for energy detection and
the most economical ratio of number of photon detection sensors (like photomultiplier tubes, PMTs) to volume.
\item[2)] The attenuation length is usually less than 20 meters for liquid scintillators,
and is much longer for water, and that of water-based liquid scintillator is still unknown.
A detector should not have a dimension significantly larger than the attenuation length of the chosen detection material.
\item[3)] A buffer zone for shielding external radiative gammas is extremely important to get a clean detector target region.
MeV-scale gammas have long attenuation lengths, which are about 20-40 cm in water.
Detector framework (steel, plastic, rubber), photon sensors, etc.~are all the sources of radioactive backgrounds for the signal detection.
The fiducial volume of the Borexino experiment is four meters away for all sides from its stainless steel sphere to shield external gammas.
\item[4)] Water shielding was constructed in previous experiments to detect cosmic-ray muons and shield fast muon-induced spallation neutrons generated in surrounding rock. The minimal thickness is around 1-2 meters.
Because of the low cosmogenic background and radioactive background, we will consider to combine the water shielding with the buffer zone.
\item[5)] Cost and risk for any large amount of civil construction and detector construction.
\item[6)] Other properties also put a secondary constrain to the size of the detector.
\end{itemize}

The waveform output from the PMT is rather important, since the pulse shape information can be used for particle identification in scintillator.
In the water-based liquid scintillator, a measurement of the waveform with a fine timing structure can be more useful,
for example, in separation between Cherenkov and scintillation radiations. This new feature consequently can be exploited to perform a particle identification among
gamma, electron, and proton, for instance.

In this section, we give a preliminary plan of the experimental hall and the neutrino detectors to express the concept.
A preliminary study from a simple test-stand is also shown to demonstrate the possible separation between Cherenkov and scintillation lights.
Considering the current level of technology and expected development, we also explain some thoughts
for the electronics used to precisely readout the PMT waveform.

\subsection{Experimental hall layout and neutrino detector}
\begin{figure}[!h]
  \centering
  \includegraphics[height=10cm]{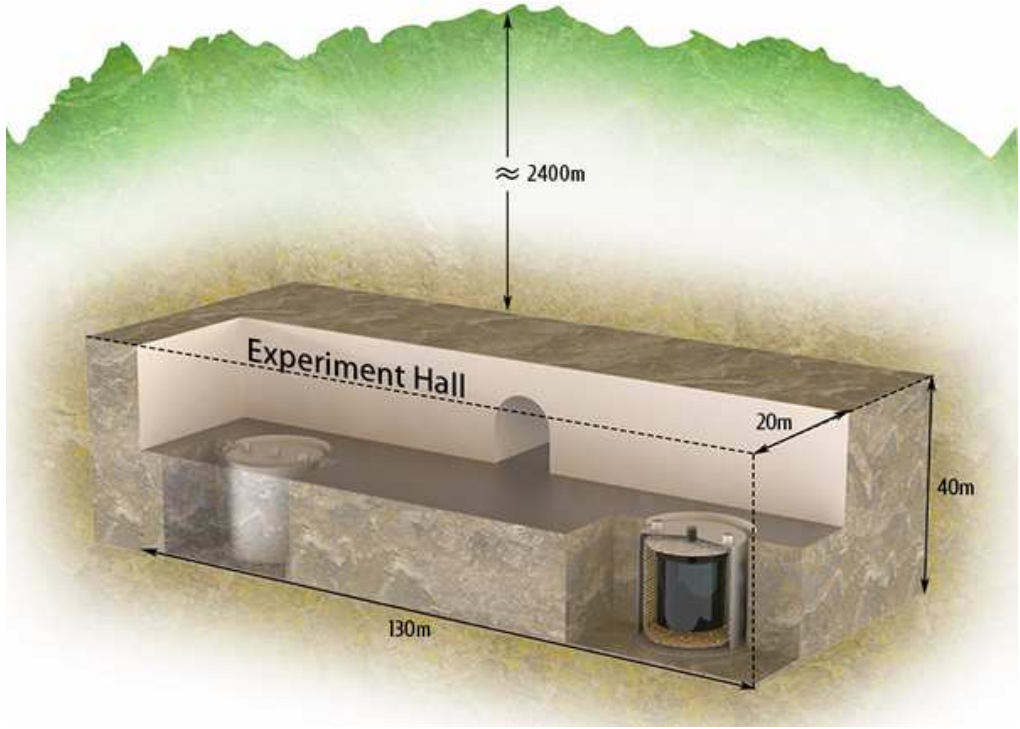}
  \caption{(Color online) Experimental hall layout.}
  \label{fig:hall}
\end{figure}

Within one experimental hall at CJPL II, two cylinder caverns can be evacuated along the tunnel as shown in Fig.~\ref{fig:hall}.
Two neutrino detectors can be deployed in the hall, and one detector in one cavern.
Each cavern is required to be around 20 m in both diameter and height.

\begin{figure}[!h]
  \centering
  \includegraphics[height=8cm]{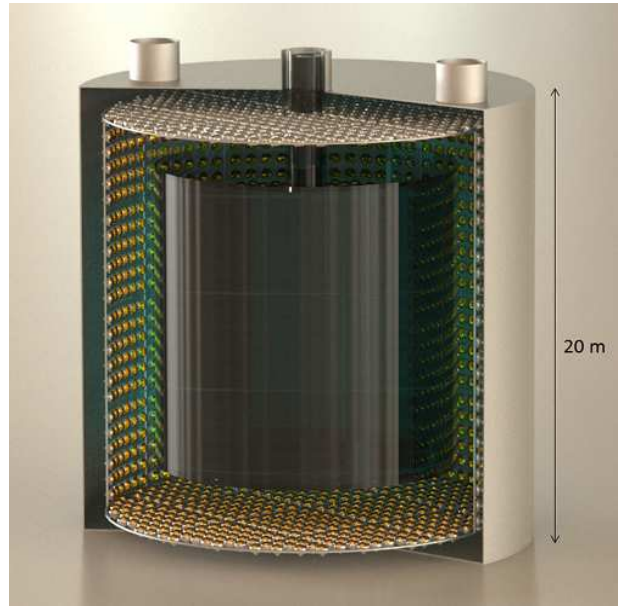}
  \caption{(Color online) The conceptual design for a cylindric neutrino detector at Jinping.}
  \label{fig:cylinder}
\end{figure}

\begin{figure}[!h]
  \centering
  \includegraphics[height=8cm]{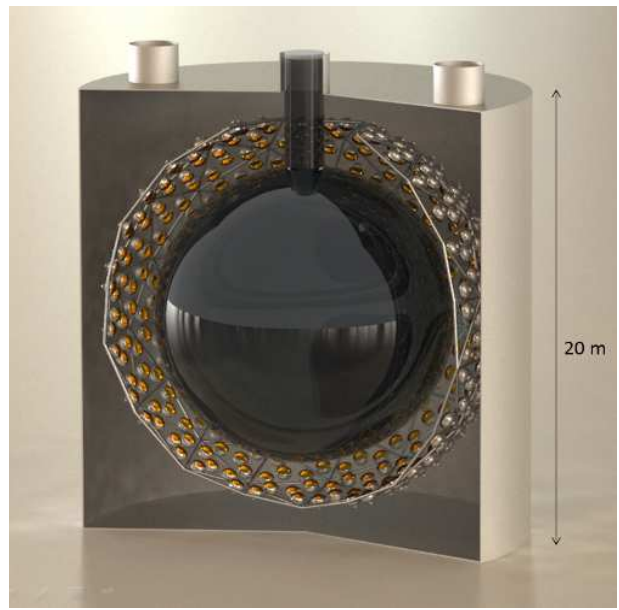}
  \caption{(Color online) The conceptual design for a spherical neutrino detector at Jinping.}
  \label{fig:sphere}
\end{figure}

A conceptual design of the neutrino detector can be seen in Fig.~\ref{fig:cylinder} or Fig.~\ref{fig:sphere}
for a cylinder or sphere inner vessel, respectively.
The central vessel is made of acrylic, and the height and diameter of the cylinder and
the diameter of the sphere are both 14 meters.
The vessel is filled with the target material, which can be either liquid scintillator or water-based liquid scintillator.
For the sphere inner vessel, the fiducial volume is a sphere of 12.8 m diameter, so that each neutrino detector will give 1 kiloton fiducial mass assuming the target material density is 0.9 g/cm$^3$.
For the cylinder inner vessel, the fiducial volume is a cylinder of 11.2 m diameter and 11.2 m height, and the fiducial mass is also 1 kiloton under the same density.

The central vessel will be sealed and surrounded with pure water.
Scintillation and Cherenkov lights originated from neutrino interaction with the target material in the central region will be collected by the PMT's.
These PMT's will be mounted on a supporting stainless steel structure,
and will be kept 2-3 m away from the central vessel to shield the gammas.
The outmost layer of the detector is a low radioactive stainless steel tank with 20 m in both diameter and height and
hosts the central vessel, PMTs, supporting structure, and pure water.

With two neutrino detectors, the total fiducial volume will be about 2 kiloton,
which is required for the solar neutrino study, in which the detection process is neutrino-electron scattering.
For the geo-neutrino and supernova relic neutrino physics,
the equivalent fiducial mass is 3 kiloton, because
the neutrino interaction is the inverse-beta-decay process,
which has a higher ratio of signal to background when requiring a prompt-delayed coincidence.

With such a design, we currently think it is the most economic one when balancing the need of 2 kiloton fiducial mass and the safe dimension of the CJPL II tunnel.
For the detector geometry, the spherical design has the best symmetry, which is a key issue to understand
the detector response for energy deposit.
In addition, this design has the largest surface to volume ratio, which minimizes the number
of PMT's and the number of channels of electronics.
On the other hand, the cylinder design can maximize the fiducial volume, which is also an important factor.
More studies or considerations are needed.

\subsection{Target material}
To study low-energy solar neutrinos and geoneutrinos, liquid scintillator with sufficient light yield is
a good option, since the technology is mature. Successful applications can be found in many experiments,
especially a very low radioactivity in the liquid scintillator has been achieved by the Borexino experiment.

However, the new water-based liquid scintillator is a very attracting option. Redundant measurements of a particle can be possible.
Cherenkov light can be used for the directional reconstruction of charged particles as in the Super Kamiokande Experiment,
while scintillation light can be used for the energy reconstruction of particles as in the Borexino Experiment.
Furthermore, Cherenkov light yield and scintillation light yield
have different dependencies on particle momentum. In principle, this feature can be exploited to identified  gamma, electron, muon, and proton.

\subsubsection{Liquid scintillator}
The light yield of liquid scintillator can be as high as $(1-2)\times10^{4}$~photons/MeV, which is sufficient
for the energy resolution required in later sections.

\subsubsection{Cherenkov and scintillation separation with LAB}
Water-based liquid scintillators are under development.
Linear alkyl benzene (LAB) is one important ingredient for the water-based liquid scintillator (WbLS)~\cite{Minfang}.
With a 20 liter container in a small test-stand~\cite{20L}, we
measured the time profile of scintillation lights in the LAB and
tested the separation between Cherenkov and scintillation lights.

\begin{figure}[!h]
  \centering
  \includegraphics[width=8cm]{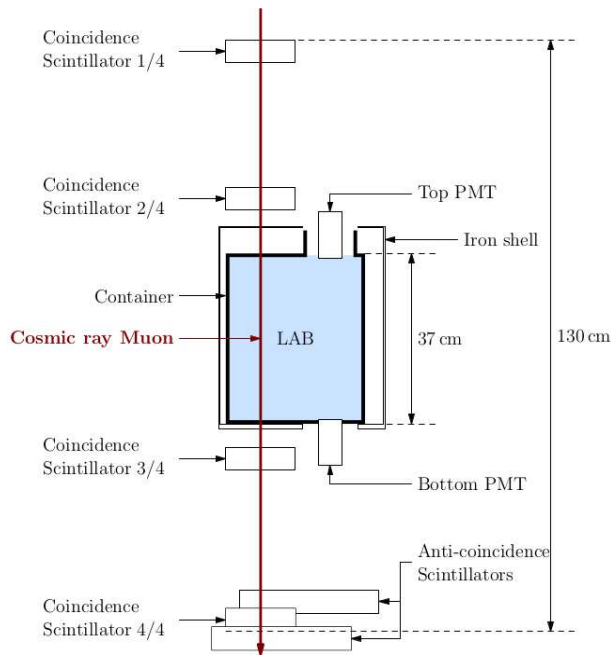}
  \caption{Schematic front view of the LAB test-stand.}
  \label{fig:teststand}
\end{figure}

A schematic drawing of the test-stand is shown in Fig.~\ref{fig:teststand}, and the total height is only 1.3 m.
The selected cosmic-ray muons can pass the 20 liter container, which holds LAB.
One PMT (bottom PMT) is placed facing the forward direction of Cherenkov lights generated by the
downward-going muons, the other one (top PMT) is placed facing the opposite direction and should not observe the direct Cherenkov light. The scintillation lights are generated uniformly
in all directions, and should produce symmetric signals on the two PMT's.

\begin{figure}[!h]
\begin{center}
\includegraphics[width=10cm]{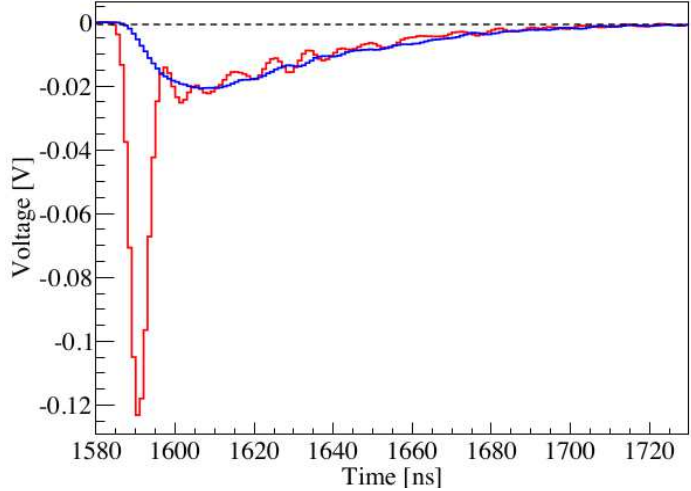}
\end{center}
\caption{Average waveforms of the top (blue) and bottom (red) PMTs.
The top PMT can only observe the scintillation lights, while the bottom PMT can observe both the scintillation and Cherenkov lights.
The extra oscillation structure, especially in the bottom channel, is consistent with the PMT's baseline ringing,
whose amplitude is less than 3 mV in the experiment.}
\label{fig:AverageAccumulateWave}
\end{figure}

The average waveforms of the top and bottom PMT's are shown in Fig.~\ref{fig:AverageAccumulateWave}.
A clear separation between the Cherenkov and scintillation lights is visible. The prompt ($<$10 ns) Cherenkov light
was clearly seen by the bottom PMT, and while the long and symmetric scintillation light was seen by both top and bottom PMT's.
The yield of scintillation light was estimated to be $1\times10^3$~photons/MeV.

Our study shows that the LAB's light yield is too low to reach a high energy resolution.
More efforts for water-based liquid scintillators are needed to fully explore the interesting physics property.

\subsection{Electronics}
Recording a precise PMT waveform is critical to distinguish the scintillation light and Cherenkov light, thus a waveform sampling with 1 GHz FLASH ADC (FADC) will be applied as the baseline technology for the Jinping neutrino experiment. Multi channels with different gains will be used to cover the dynamic range from 1 photo-electron (PE) to 100 PE.

For each PMT, a separate electronics module which includes a High Voltage generator, a base divider, an FADC sampler and a processing circuit, will be installed at the end of each PMT in a water tight housing. The signal from PMT does not need to pass through any coaxial cable which may degrade the signal quality and timing resolution. The reliability of the circuit and the housing structure is a major design challenge.

Each PMT works in self-trigger mode: once the signal level goes above a certain threshold, like 0.3 p.e, the sample data in a certain readout window around the over-threshold points will be stored and transferred. The window size is adjustable up to a few micro-seconds. All the electronics contain synchronized time-tick counters for aligning sample fragments among PMTs.

Out of the water, the back-end electronics will provide the data acquisition, clock synchronization and control service.
The PMT electronics and back-end electronics will be connected via multi-pair twist cables which will carry the low voltage power supply, dedicated clock/time signal, upstream and downstream data links. An encoding algorithm with variable lengths will be applied to the data stream to reduce the band width requirement.

\subsection{Simulation studies}
\begin{figure}[!h]
\begin{center}
\includegraphics[width=12cm]{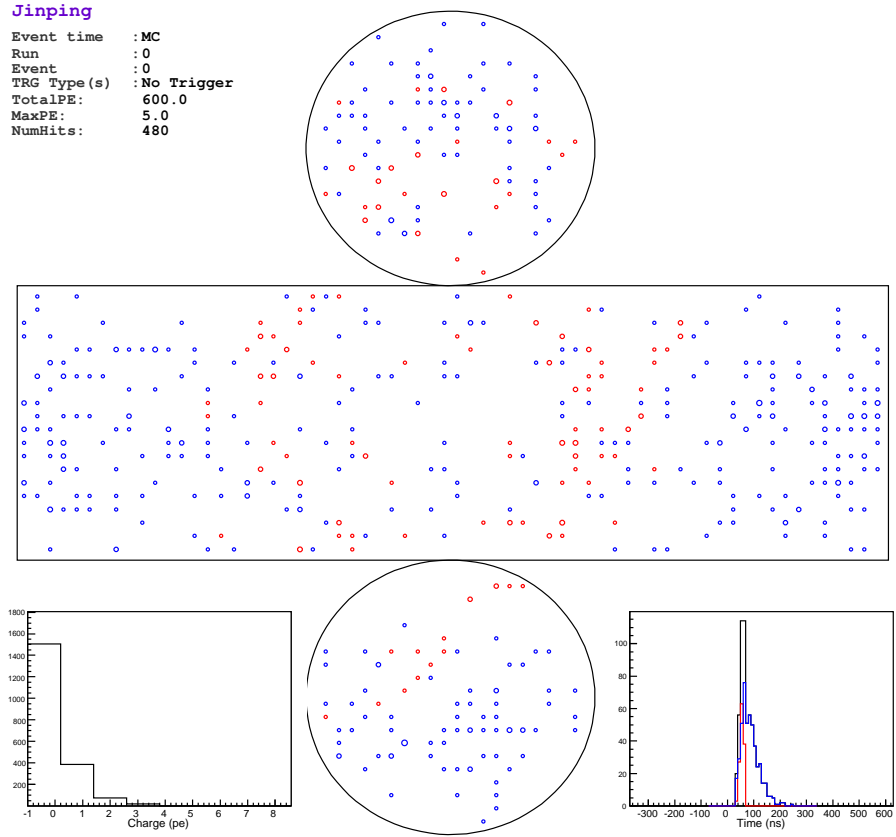}
\end{center}
\caption{An event display of a 7 MeV electron.
This simulation is for a cylindric neutrino detector with LAB as the target material.
The coverage of PMT photo cathode is 30\%.
Each circle indicates a PMT with at least one photo-electron (PE) detected.
The red circles are for those originated with the prompt Cherenkov radiations based on the truth information,
and the blue ones are for the scintillation light.
The bottom-left panel is for the distribution of the number of PE's for all the PMT's and
the bottom-right panel is for the time distribution.
A Cherenkov ring is visible in this plot.
}
\label{fig:EvtDsp}
\end{figure}

We have conducted simulation studies to optimize the detector design to take the great advantage of
the Jinping underground laboratory and to achieve the physics goals.
The following work was done with the Geant4 simulation package together with the customized geometry, light emission model, PMT response, etc.
Figure~\ref{fig:EvtDsp} shows an event display for a 7 MeV electron, which can be produced via the neutrino-electron scattering in LAB according to our measurement.
For a demonstration purpose, we use different colors for the Cherenkov and Scintillation lights, respectively.

%% file: Solar.tex
\newpage
\section{Solar neutrino}
\subsection{Introduction}
\label{sub:sol:Intro}
Particles from sources at cosmic distances are of great interest.
Neutrinos, as a stellar probe, are featured by their extremely low interaction cross sections.
Unlike gammas, optical photons and protons, neutrinos can easily reach our detectors
without being interrupted by matter on their path.
The original status, i.e. energy and direction, can therefore be maximally maintained,
except that the flavors will oscillate among the three families of neutrinos,
and consequently more information about the initial interactions of neutrino productions can be probed.
The attributes of neutrinos make them powerful probes of the deep interiors of
sources like the Sun, providing ways to test the models of solar evolution and solar neutrino oscillation.



\subsubsection{Solar models}
Solar models, neutrino theories, and solar neutrino experiments
have a rapid development over the past half a century,
and the remarkable history has been documented in ~\cite{jnb1, HaxtonNt, McDonald, thirty}
and references therein.
Nowadays the Sun is described by the Standard Solar Model (SSM)~\cite{JBHomepage},
which relies on about 20 parameters.
\begin{itemize}
\item The first group of them are the current solar age, luminosity, mass, and radius.
\item The primordial abundances of key elements, He, C, N, O, Ne, Mg, Si, S, Ar, and Fe.
\item The cross-section of nuclear reactions, including $p$($p$, $e^+\nu_e$)$d$, $d$($p$, $\gamma$)$^3$He, $^3$He($^3$He, 2$p$)$^4$He, $^3$He($^4$He, $\gamma$)$^7$Be, $^3$He(p, $e^+\nu_e$)$^4$He, $^7$Be($e^-$, $\nu_e$)$^7$Li,
    $p$($e^-p$, $\nu_e$)$d$, $^7$Be($p$, $\gamma$)$^8$B, $^{14}$N($p$, $\gamma$)$^{15}$O.
\end{itemize}

The evolution starts with a cluster of homogeneous gas of H, He, C, N, etc.
Nuclear fusion reactions burn H, He to heavier elements and emits gammas, electrons, positions, etc.
The Sun fuels itself by both the $pp$ and CNO fusion processes, in which the $pp$-cylce contributes 99\% of the total for energy production.
The whole processes are constrained by the boundary conditions of the current solar status.
The transport of energy in the central region is primarily through the inverse bremsstrahlung process of photons, and the calculated radiative opacity depends upon the chemical composition and the modeling of complex atomic processes.
In the outer region, the energy is brought to the surface by the convective motions.
The interior of the Sun is assumed to be spherical symmetric and to be at the balance of the gravity, radiation, and particle pressure.

The present composition of the solar surface is presumed to reflect the initial abundances of all of the elements that are as heavy as carbon~\cite{UniMetal}. These metal elements are assumed to be chemically homogeneous throughout the Sun, except for a minor correction due to diffusion. Suggestions have been proposed to argue the assumption of composition. CNO neutrinos measurement could be a direct test of the solar-core metallicity~\cite{HaxtonAb, CompS}.

The nuclear reaction cross-sections are from theoretical calculations and/or terrestrial measurements~\cite{fusionI, fusionII}.
For example, the cross-section for the initial $p$($p$, $e^+\nu_e$)$d$ process is too low to be measured in a laboratory, and has to be calculated with the theory of weak interaction. The $^3$He($^4$He, $\gamma$)$^7$Be cross-section is measured in the laboratory, and the result must be extrapolated to the solar Gamow peak with correct theoretical consideration.
More experimental efforts on this regard can be found in LUNA~\cite{LUNA}, JUNA~\cite{JUNA}, and~\cite{WPLiu} etc.

Electron neutrinos can oscillate to muon and tau neutrinos in the three-flavor framework in vacuum or low electron-density material~\cite{pontecorvo, mns}. However, this quantum ability of neutrinos is changed in the high electron-density environment in the Sun's interior, also known as the matter, Mikheyev-Smirnov-Wolfenstein (MSW) effect~\cite{MSW-W, MSW-MS}.

The SSM describes the whole life of the Sun from the pre main-sequence time to the current day, even to the future.
The study of the solar neutrinos directly tests the theory of stellar evolution and nuclear energy generation and neutrino oscillation. The knowledge of the Sun is critical to further understand the stars at distant space.

\subsubsection{Solar neutrino experiments}
The first triumph of solar neutrino flux measurement is
for the $\nu_e$ component detected using a $^{37}$Cl detector at Homestake~\cite{Homestake},
but it is a big surprise that the measurement is only 30\% of the prediction.
The following steady experimental efforts by SAGE ($^{71}$Ga detector)~\cite{SAGE},
GALLEX ($^{71}$Ga detector)~\cite{GALLEX}, GNO ($^{71}$Ga detector)~\cite{GNO}, Kamiokande (water Cherenkov detector)~\cite{Kamiokande}, and Super
Kamiokande (water Cherenkov detector)~\cite{SuperK} all confirmed the Homestake measurement.
Later the SNO~\cite{SNO} experiment used a heavy water detector to make a measurement sensitive to all the flavors, whose result agrees with the SSM prediction.
Today we understand that
electron neutrinos, $\nu_e$, generated through the fusion processes occurring inside the Sun may oscillate to other flavors, $\nu_\mu$ or $\nu_\tau$, and this process is affected by MSW effect in the surrounding dense materials
in the Sun.

Recently the Borexino~\cite{BxPhaseI} experiment successfully identified the low energy $^7$Be,  $pep$  and $pp$ solar neutrinos, and measured the fluxes in agreement with the SSM prediction when taking the oscillation into account. A new era of precision measurement of the solar neutrino has begun.

\subsubsection{Helioseismology}
Solar oscillation is first found in~\cite{Helio} by studying the velocity shifts in absorption lines formed in the solar surface. The surface of the Sun is divided in to patches which oscillate with velocity amplitudes of order 0.5 km~s$^{-1}$ and periods of order of 5 minutes. The phenomenon can be used to deduce a precise sound speed profile in the Sun and correspondingly the density and pressure profile.
Helioseismology is the other method can be used to study the interior of the Sun.

\subsubsection{Issues}
The remaining issues about the property of neutrinos and the solar model~\cite{HaxtonProg, OsciForm, jnb2} are summarized below.
\begin{itemize}
\item{Discovery of the missing solar neutrino components from a more precise measurement on the
    fluxes~\cite{HaxtonNt, HaxtonAb, Roadmap}.
    The CNO neutrinos are believed to dominate the fueling processes inside the high temperature
    massive stars, which however have not yet been observed by any neutrino experiment.
    A good approach is to search these neutrinos from the Sun even though the relative small flux.
    The $hep$ neutrinos are also still missing from any experimental measurements.}

\item{Precise measurements on all the solar neutrino components will provide not only a tighter constrain on the solar model,
    but also a high statistics observation of $pp$ and others in real time,
    all of which might give completely new insights into the energy production and fluctuation of the stars.
    In addition, a precise measurements on the solar neutrino flux can play a key role in the study of following problems.}

\item{Solution of the metallicity problem. As discussed in reference~\cite{MetalProb, MetalProb2}, an improved solar model prediction is available with the input of the most up-to-date photospheric abundance of metals, which is 30\% lower than earlier results. The new calculation based on the low metallicity assumption predicts lower fluxes for several neutrino components than those based on the previous high metallicity assumption.
    The next generation of solar neutrino experiments are expected to resolve the conflict.

    A precise measurement of the CNO neutrino is especially important for the metallicity problem ~\cite{HaxtonAb, CompS}.
    The relation of the solar neutrino fluxes and helioseismology can be seen in Fig.~\ref{fig:helio}.
    The CNO neutrino flux predicted by the SSM has a closer dependence on the abundance of the metal elements than other components.
    The variance of metallicity will change the density and pressure profile and affect the fluxes of others indirectly.

\begin{figure}[!htb]
  \vspace{0.2cm}
  \centering
  \includegraphics[width=12cm]{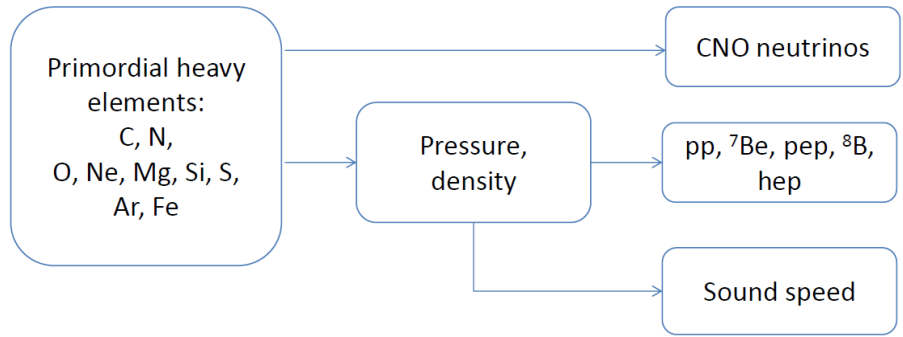}
  \caption{(Color online) The relation of solar neutrino fluxes and sound speed measurement of helioseismology.}
  \label{fig:helio}
\end{figure}
    }

\item{A full picture of MSW effect in the solar electron neutrino oscillation.
    The oscillation of low energy $\nu_e$, $<1$ MeV, likely occurs in vacuum.
    As the neutrino energy increases,
    the MSW effect on solar neutrino oscillation emerges and becomes dominant due to the high electron density environment of the Sun's interior,    and the transition of $\nu_e$ to the other flavors
    will eventually reach maximum.
    However this transition region from vacuum to matter is still poorly constrained by experiments~\cite{NoStandard, Bonventre, Friedland, newRev}.}

\item{Precise measurements of $\theta_{12}$ and $\Delta m^2_{21}$.
    There is currently a 2$\sigma$ tension in the $\Delta m^2_{21}$ measurement between solar and reactor measurements.
    The Sun emits electron-neutrinos, while reactors emits electron-antineutrinos.
    Both neutrinos can be used to improve the measurement of PMNS matrix and provide a test of CP invariance as well.
    A more precise value of $\theta_{12}$ will help to define a better lower edge of the inverted neutrino mass hierarchy, and is thus important for neutrinoless double beta decay experiments in the future.
}

\item{Observation of the $\nu_e$ regeneration inside the Earth.
    The Earth should in principle have a terrestrial matter effect on solar neutrinos.
    A regeneration of these neutrinos will give rise to a flux asymmetry for the electron flavor solar neutrinos during the daytime and the nighttime~\cite{Bright, SurfaceDen}.
    An indication of the day-night asymmetry  has already been observed with a  2.7$\sigma$ significance at Super Kamiokande~\cite{SK}.}

\item{Probing the sub-leading effects in addition to neutrino oscillation. The Sun serves as an ideal neutrino source to probe new physics especially for those through a secondary effect of the standard scheme~\cite{NoStandard, newRev}.
    In addition, the expected upturn behavior has not been observed yet for the solar neutrino oscillation from the matter to the vacuum effect.
    This might point to a non-standard neutrino interaction.
    Other interesting topics can also be studied with solar neutrinos, such as  new neutrino states, sterile neutrinos, effects of violation of fundamental symmetries, new dynamics of neutrino propagation, probes of space and time.
    }

\end{itemize}

Below are arranged as follows.
Section~\ref{sub:sol:introduction} introduces the simulation setup for the sensitivity study,
including neutrino oscillation probability, solar neutrino model, detector configuration, etc.
Section~\ref{sub:sol:systematic} presents the consideration on the systematics
which can affect the studies of the solar model and MSW effects.
Section~\ref{sub:sol:precision} gives the sensitivity for identifying each solar neutrino component.
Section~\ref{sub:sol:tran} ~\ref{sub:sol:daynight} and \ref{sub:sol:metal} discuss the
potentials for studying the transition of vacuum-matter oscillation, day-night asymmetry, and the test of
high and low metallicity models, respectively.




\subsection{Simulation study}
\label{sub:sol:introduction}
A simulation study was done with some default settings at Jinping, including the expected signal and background levels, energy resolution, target mass, and live time in order to give an estimate of sensitivity for each physics topic.

\subsubsection{Solar neutrino model}
\begin{figure}[!htb]
  \vspace{0.2cm}
  \centering
  \includegraphics[angle=270, width=9cm]{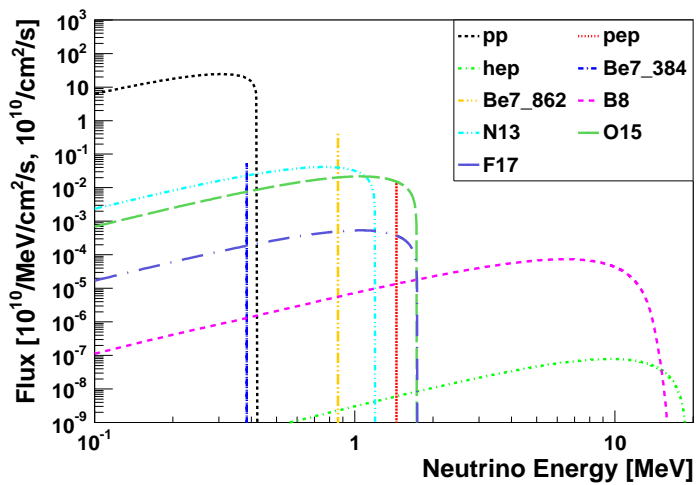}
  \caption{(Color online) Solar neutrino energy spectra and fluxes with the high metallicity hypotheses, where the unit for continuous spectra is $10^{10}/$MeV/cm$^{2}/$s, and for discrete lines is $10^{10}/$cm$^{2}$/s.}
  \label{fig:solarSpec}
\end{figure}

The neutrino energy spectra for all the solar neutrino components were taken from Ref.~\cite{JBHomepage}. The average neutrino flux predictions on the Earth without the oscillation effect were from Ref.~\cite{MetalProb2} and~\cite{LowMetal} for the high and low metallicity hypotheses, respectively. The correlations between the neutrino components estimated in reference~\cite{10000model} were used in the study.
The spectra with high metallicity flux prediction are shown in Fig.~\ref{fig:solarSpec} and all the numerical values  are listed in Table~\ref{tab:solarFlux}.

\begin{table}[!htb]
\centering
\small
\begin{tabular}{l lll}
\hline\hline
          & E$_{Max}$ or E$_{Line}$& Flux (GS98) high metallicity    & Flux (AGS09) low metallicity     \\
          &   [MeV]      & [$\times10^{10}$s$^{-1}$cm$^{-2}$]  & [$\times10^{10}$s$^{-1}$cm$^{-2}$]   \\ \hline
$pp$      & 0.42 MeV     & $5.98(1\pm0.006)$               &  $6.03(1\pm0.006)$               \\
$^7$Be    & 0.38 MeV     & $0.053(1\pm0.07)$               &  $0.048(1\pm0.07)$               \\
          & 0.86 MeV     & $0.447(1\pm0.07)$               &  $0.408(1\pm0.07)$               \\
$pep$     & 1.45 MeV     & $0.0144(1\pm0.012)$             &  $0.0147(1\pm0.012)$           \\
$^{13}$N  & 1.19 MeV     & $0.0296(1\pm0.14)$              &  $0.0217(1\pm0.14)$              \\
$^{15}$O  & 1.73 MeV     & $0.0223(1\pm0.15)$              &  $0.0156(1\pm0.15)$              \\
$^{17}$F  & 1.74 MeV     & $5.52\times10^{-4}(1\pm0.17)$   &  $3.40\times10^{-4}(1\pm0.17)$   \\
$^8$B     & 15.8 MeV     & $5.58\times10^{-4}(1\pm0.14)$   &  $4.59\times10^{-4}(1\pm0.14)$   \\
$hep$     & 18.5 MeV     & $8.04\times10^{-7}(1\pm0.30)$   &  $8.31\times10^{-7}(1\pm0.30)$   \\
\hline
\end{tabular}
\caption{Solar neutrino flux predictions without oscillation based on the high and low metallicity hypotheses~\cite{MetalProb2, LowMetal}. The production branching ratios for the 0.38 and 0.86 MeV $^7$Be lines are 0.1052 and 0.8948, respectively.}
\label{tab:solarFlux}
\end{table}

\subsubsection{Oscillation probability}
The propagation path of solar neutrinos from the Sun to the Earth can be divided into three parts: 1) from the inner core to the surface of the Sun; 2) from the surface of the Sun to the surface of the Earth; 3) the path through the Earth during the nighttime.

\begin{figure}[!htb]
  \centering
  \includegraphics[angle=270, width=9cm]{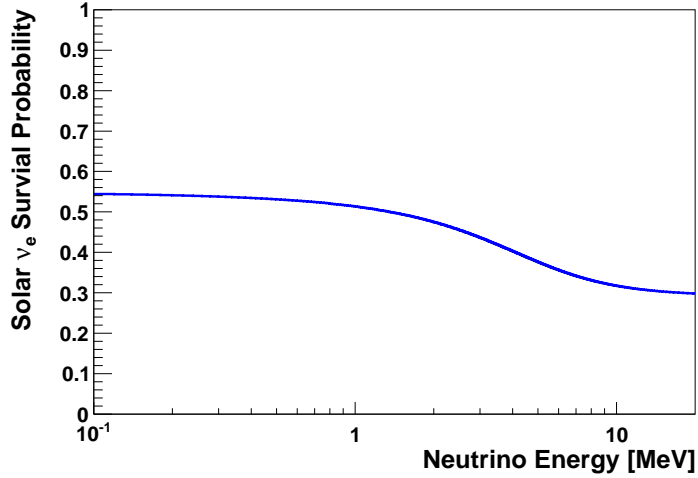}
  \caption{(Color online) Solar electron neutrino survival probability.}
  \label{fig:OsciProb}
\end{figure}

The survival probability of solar electron neutrinos with energy $E_{\nu}$ from the inner core to the surface of the Sun must include the matter effect~\cite{MSW-W, MSW-MS} and can be approximated by the following formula~\cite{OsciForm},
\begin{equation}
P^{\odot}_{ee}=\cos^4\theta_{13}(\frac{1}{2}+\frac{1}{2}\cos2\theta^M_{12}\cos2\theta_{12}),
\label{eq:Pee}
\end{equation}
where the mixing angle in matter is
\begin{equation}
\cos2\theta^M_{12}=\frac{\cos2\theta_{12}-\beta}{\sqrt{(\cos2\theta_{12}-\beta)^2+\sin^22\theta_{12}}},
\end{equation}
with
\begin{equation}
\beta=\frac{2\sqrt{2}G_F\cos^2\theta_{13}n_eE_{\nu}}{\Delta m^2_{12}},
\end{equation}
where $G_F$ is the Fermi coupling constant and $n_e$ is the density of electrons in the neutrino production place of the Sun. The calculation is under a good assumption of adiabatic evolution~\cite{Liao},
so that the density of electrons varies slowly and does not causes any exchange among the mass eigenstates after being created.
For the solar case initially only $n_e$ are produced by the fusion processes.
With $\sin^2\theta_{12}$=0.307, $\sin^2\theta_{13}$=0.0241, $\Delta m^2_{12}=7.54\times10^{-5}\ \rm{eV}^2$, and $n_e=6\times10^{25}/\rm{cm}^3$~\cite{n_e} in the inner core of the Sun,
the survival probability of $n_e$ as a function of neutrino energy can be obtained as  shown in Fig.~\ref{fig:OsciProb}.
No consideration was taken for the generation of radius distribution of each component.
Correspondingly the appearance probability of $\nu_{\mu}$ and $\nu_{\tau}$ is
\begin{equation}
P^{\odot}_{e\mu(\tau)}=1-P^{\odot}_{ee}.
\end{equation}

The second part is from the surface of the Sun to the surface of the Earth. The mass eigenstates of neutrinos emerging from the surface of the Sun
are treated to be decoherent~\cite{decoherent} due to the sizable width of the energy spectrum of each neutrino component, even for the $^7$Be neutrinos~\cite{dc-Be7}.
The amplitudes for all the mass eigenstates keep unchanged and decoherent all the way to the Earth.
The fluxes per unit area only decrease by a factor of the Earth-Sun distance squared with
a percent-level annual modulation effect due to the eccentric orbit of the Earth.
The above oscillation probability $P^{\odot}_{ee}$ is sufficient for most of the studies.

To study the day-night asymmetry of solar neutrino flux, a full numerical
calculation was done under the framework of the three-generation oscillation~\cite{solar-3f}.
Assuming that both the adiabatic condition and the decoherent property are still valid,
we can obtain the probability of detecting a solar electron neutrino
\begin{equation}
P_{ee} = \sum_{i=1,2,3}P^{\odot}_{ei}P^{\oplus}_{ie},
\end{equation}
where $P^{\odot}_{ei}$ is the probability of solar electron neutrinos surviving as their mass eigenstates $\nu_{i}$ ($i=1,2,3$) on the surface of the Sun,
and $P^{\oplus}_{ie}$ is the appearance probability of solar electron neutrinos for the i-th mass eigenstate $\nu_{i}$ after passing through the
Earth. $P^{\odot}_{ei}$ is determined by the local $n_e$ in the production place of $\nu_e$ in the Sun. $P^{\oplus}_{ie}$ should be calculated numerically through a multi-shell model of the Earth
as explained below. The result for the day time is the same as Eq.~\ref{eq:Pee}.

\subsubsection{Earth shell model}
\begin{figure}[h]
  \centering
  \vspace{0.2cm}
  \includegraphics[angle=270, width=13cm]{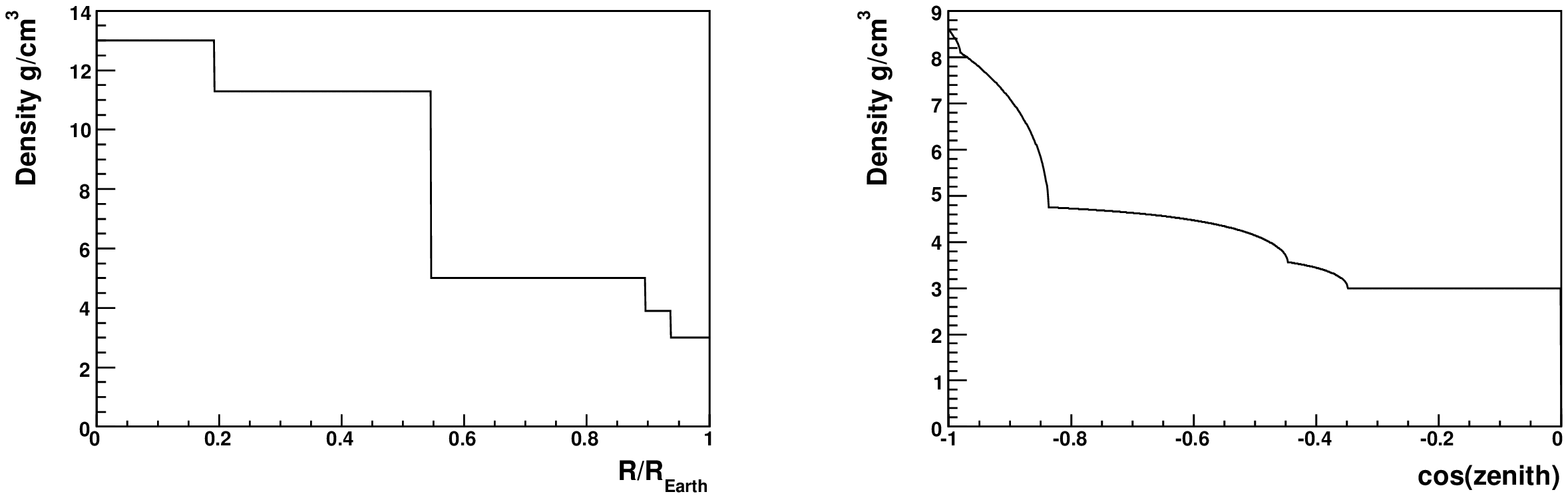}
  \caption{Depth dependence of the density for continental experiments. Left is the density vs.~radius, and the right is the average density vs.~cosine of zenith angle.}
  \label{fig:EarthJinping}
  \vspace{0.2cm}
  \includegraphics[angle=270, width=13cm]{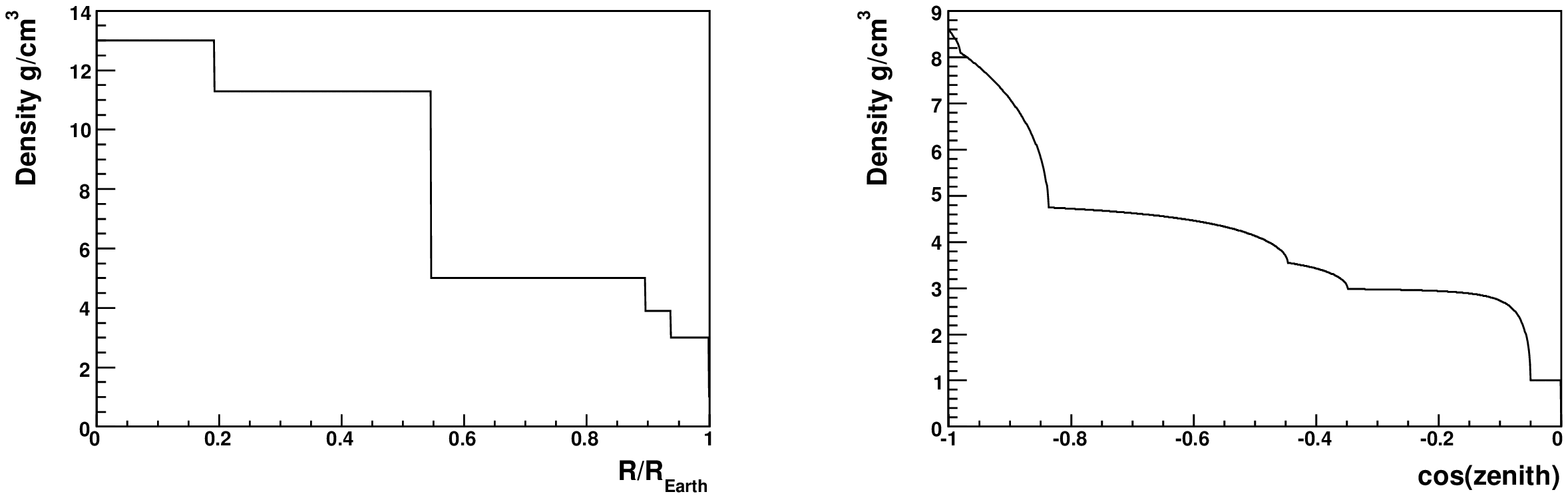}
  \caption{Depth dependence of the density for experiments next to ocean. Left is the density vs.~radius, and the right is the average density vs.~cosine of zenith angle.}
  \label{fig:EarthJapan}
\end{figure}

Two different 6-shell models of the Earth~\cite{Earth1, Earth2} were used. One density profile is for continental region, i.e.\ Jinping, which is all surrounded by rock,
whose density is 3 g/cm$^3$, and the other one is for ocean, i.e.\ Super-Kaminokande, which is surrounded by a 8 km deep ocean, whose density
is 1 g/cm$^3$~\cite{SKSolarI}. The two different shell assumptions are shown in Fig.~\ref{fig:EarthJinping} and Fig.~\ref{fig:EarthJapan}, respectively. The ratio of $n_e$ to the density is set to 0.47 mol/cm$^3$ for the inner shells
and 0.50 mol/cm$^3$ for the outer shells, respectively.  As will be seen later, the $\nu_e$ regeneration in the Earth is very sensitive to the surface density.

\subsubsection{Elastic scattering cross section}
\begin{figure}[!htb]
  \centering
  \includegraphics[angle=270, width=16cm]{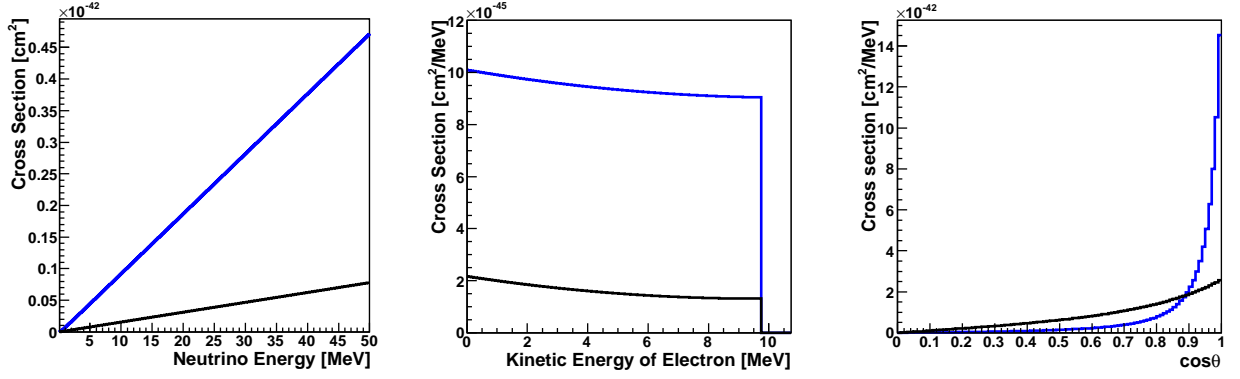}
  \caption{(Color online)
  Left: the total electron scattering cross section as a function of neutrino energy for $\nu_e$ (blue) and $\nu_{\mu,\tau}$ (black);
  middle: the differential scattering cross section as a function of the kinetic energy of recoiling electron for a 10 MeV $\nu_e$ (blue) and $\nu_{\mu,\tau}$ (black);
  right: the distribution of the cosine angle between the recoiling electron and initial neutrino direction for a 10 MeV $\nu_e$ (blue) and a 1 MeV $\nu_e$ (black).}
  \label{fig:NueeXSec}
\end{figure}

The neutrino electron elastic scattering process will be used to detect solar neutrinos.
The scattered electron's energy and direction can be measured and used to derive the incoming neutrino energy and direction.
The differential scattering cross-sections as a function of the kinetic energy of the recoil electron, $T_e$, and neutrino energy, $E_{\nu}$,
in the electron rest frame can be written, for example, in Ref.~\cite{NueXSec} as:
\begin{equation}
\frac{d\sigma(E_{\nu},T_e)}{dT_e}=\frac{\sigma_0}{m_e}\left[g_1^2+g_2^2(1-\frac{T_e}{E_{\nu}})^2-g_1g_2\frac{m_eT_e}{E_{\nu}^2}\right],
\end{equation}
with
\begin{equation}
\sigma_0=\frac{2G_F^2m_e^2}{\pi}\simeq88.06\times10^{-46}cm^2,
\end{equation}
where $m_e$ is the electron mass. Depending on the flavor of the neutrino, $g_1$ and $g_2$ are:
\begin{equation}
\begin{split}
g_1^{(\nu_e)}&=g_2^{(\bar\nu_e)}=\frac{1}{2}+\sin^2\theta_W\simeq0.73,\\
g_2^{(\nu_e)}&=g_1^{(\bar\nu_e)}=\sin^2\theta_W\simeq0.23,
\end{split}
\end{equation}
where $\theta_W$ is the Weinberg angle, then for $\nu_{\mu,\tau}$ they are
\begin{equation}
\begin{split}
g_1^{(\nu_{\mu,\tau})}&=g_2^{(\bar\nu_{\mu,\tau})}=-\frac{1}{2}+\sin^2\theta_W\simeq-0.27,\\
g_2^{(\nu_{\mu,\tau})}&=g_1^{(\bar\nu_{\mu,\tau})}=\sin^2\theta_W\simeq0.23.
\end{split}
\end{equation}
The total $\nu_e$ electron scattering cross section, differential cross section as a function of  $T_e$ and the cosine angle between the recoiling electron and initial neutrino direction are shown in Fig.~\ref{fig:NueeXSec}.

\subsubsection{Detectable electron spectrum}
\begin{figure}[!htb]
  \centering
  \vspace{0.5cm}
  \includegraphics[angle=270, width=9cm]{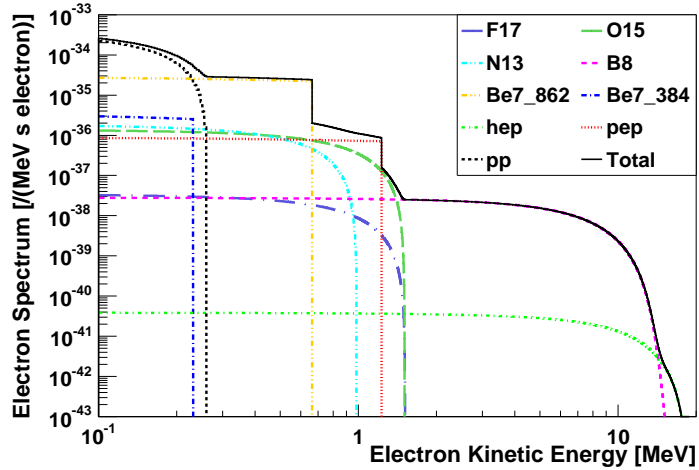}
  \caption{(Color online) Kinetic energy distribution of recoiling electrons for each solar neutrino component, in which the MSW oscillation and the high metallicity hypotheses are both considered.}
  \label{fig:ElecSpec}
\end{figure}

Since the observed spectrum of electron kinetic energy contains all the contributions from electron-, muon- and tau-neutrinos,
the electron kinetic energy spectrum becomes
\begin{equation}
R_{\nu}=N_e \Phi_{\nu} \int dE_{\nu} \frac{d\lambda}{dE_{\nu}} \int \left\{ \frac{d\sigma_e(E_{\nu},T_e)}{dT_e}P_{ee}(E_{\nu}) + \frac{d\sigma_{\mu,\tau}(E_{\nu},T_e)}{dT_e}\left[1-P_{ee}(E_{\nu})\right]  \right\} dT_e,
\end{equation}
where $N_e$ is the number of electrons in the target, $\Phi_{\nu}$ is the neutrino flux of the Sun, ${d\lambda}/{dE_{\nu}}$ is the differential energy spectrum of the solar neutrinos,
$\frac{d\sigma_{e}}{dT_e}$ ($\frac{d\sigma_{\mu,\tau}}{dT_e}$) is the differential scattering cross section as a function of electron kinetic energy
for $\nu_{e}$ ($\nu_{\mu,\tau}$), and $P_{ee}$ is the $\nu_e$ survival probability.
The recoiling electron spectra of the solar neutrinos for all the fusion processes can be seen in Fig.~\ref{fig:ElecSpec}.
The number of electron candidates for the high and low metallicity hypotheses and the effective number of electron candidates
with a 200 keV energy threshold are shown in Tab.~\ref{tab:EvtRate}, where the number of electrons per 100 tons was assumed to be $3.307\times10^{31}$~\cite{BxPhaseI}.

\begin{figure}[!htb]
\vspace{0.3cm}
  \centering
  \includegraphics[angle=270, width=14cm]{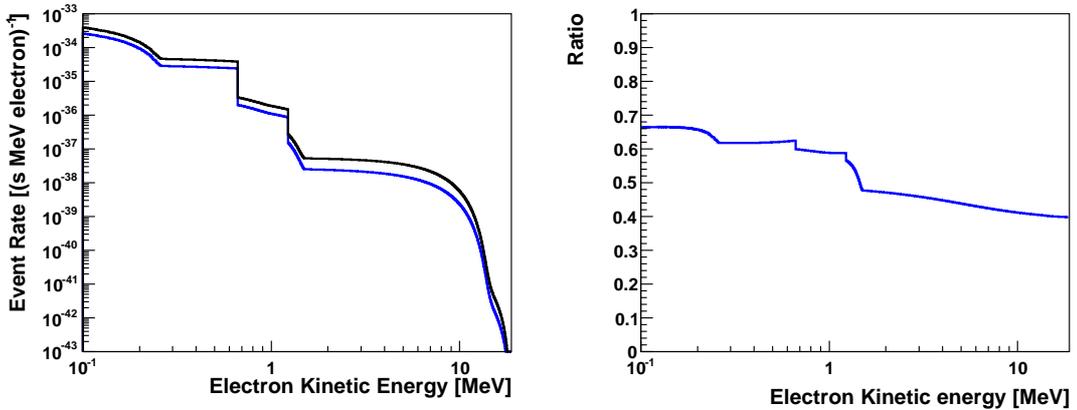}
  \caption{(Color online) Left: Event rate as a function of the kinetic energy of the recoiling electrons with
  (blue) and without (black) oscillation. Right: the relative ratio between them.}
  \label{fig:ElecSpecNoOs}
\end{figure}

Without a precise electron direction measurement and reconstruction, it is very hard to estimate the original neutrino energy (see Fig.~\ref{fig:NueeXSec}). The smooth MSW
oscillation transition in term of neutrino energy differs in electron kinetic energy.  Figure.~\ref{fig:ElecSpecNoOs} shows the observed electron kinetic energy spectra
with and without the oscillation effect together with the relative ratio between them.

\begin{table}[!htb]
\centering
\small
\begin{tabular}{l llll }
\hline\hline
Electron Event       & $>$0 keV (GS98)& $>$0 keV (AGS09)& $>$200 keV (GS98) & $>$200 keV (AGS09) \\
Rate [/day 100 ton]  & high metallicity & low metallicity & high metallicity & low metallicity    \\\hline
$pp$                &$132.59\pm0.80$     & $133.70\pm0.80$   & $4.557\pm0.027$     & $4.595\pm0.028$   \\
$^7$Be (0.38 MeV) &$1.93\pm0.13$       & $1.76\pm0.12$     & $0.228\pm0.016$    & $0.208\pm0.015$    \\
$^7$Be (0.86 MeV) &$46.9\pm3.3$        & $42.8\pm3.0$      & $31.6\pm2.2$       & $28.8\pm2.0$   \\
 $pep$                &$2.735\pm0.033$     & $2.792\pm0.034$   & $2.244\pm0.027$    & $2.291\pm0.028$   \\
$^{13}$N          &$2.45\pm0.34$       & $1.80\pm0.25$     & $1.48\pm0.21$      & $1.09\pm0.15$   \\
$^{15}$O          &$2.78\pm0.42$       & $1.95\pm0.29$     & $2.03\pm0.31$      & $1.42\pm0.21$   \\
$^{17}$F          &$0.069\pm0.012$     & $0.0426\pm0.0072$ & $0.0506\pm0.0086$  & $0.0312\pm0.0053$   \\
$^8$B             &$0.443\pm0.062$     & $0.364\pm0.051$   & $0.427\pm0.060$    & $0.351\pm0.049$   \\
hep               &$0.0009\pm0.0003$   & $0.0009\pm0.0003$ & $0.0009\pm0.0003$  & $0.0009\pm0.0003$   \\
\hline
\end{tabular}
\caption{Expected electron event rates for different thresholds and metallicity hypotheses.
The uncertainties are all from the solar model prediction only.}
\label{tab:EvtRate}
\end{table}

\subsubsection{Fiducial target mass}
In our sensitivity study the fiducial target masses are set to be 1,000, 2,000, and 4,000 tons, respectively.

\subsubsection{Detector response model}
Three types of target materials were studied for the detection of the recoiling electrons through the neutrino electron elastic scattering process.

Liquid scintillator with features in its high light yield, low detecting threshold, has been successfully applied in the Borexino experiment~\cite{BxPhaseI},
and is also an option in the SNO+ experiment.
The liquid scintillator detector response can be approximated by a simple characteristic resolution function. The non-uniform and non-linear detector energy responses
can both be corrected, so they are not necessary to be included in this study.
The SNO+ experiment inherited the almost doubled photocathode coverage from the SNO experiment~\cite{SnoCoverage} compared to  Borexino~\cite{BxCoverage},
then a doubled light yield was considered possible in this study.
Liquid scintillator was used as a reference material for this study.

Water is the second option under our consideration.
The technique developed by the Super Kaminokande experiment~\cite{SuperK2} is very mature,
however limited by a threshold about 3 MeV.

Water-based scintillators~\cite{Yeh} is the third option for the attracting feature to separate scintillating and Cherenkov lights and to provide an additional information
for energy reconstruction and background suppression.

Three typical energy resolutions were tested in this study and their values in terms of photo-electron/MeV (PE/MeV) and corresponding resolution functions are summarized in Table~\ref{tab:DetModel}.
\begin{table}[!h]
\centering
\small
\begin{tabular}{ l l l }
\hline\hline
Light yield    & Resolution function ($\sigma_E/E$)             & Experiment      \\ \hline
200 PE/MeV     & $1/\sqrt{200E/{\rm MeV}}$                & Super Kamiokande, water-like                \\
500 PE/MeV     & $1/\sqrt{500E/{\rm MeV}}$                & Borexino, liquid scintillator   \\
1,000 PE/MeV    & $1/\sqrt{1000E/{\rm MeV}}$               & SNO+, high light yield liquid scintillator   \\
\hline
\end{tabular}
\caption{Three types of light yields and resolution functions for the detector response.}
\label{tab:DetModel}
\end{table}

\subsubsection{Background assumption}
\label{sub:sol:bkg}

\begin{figure}[!tb]
  \centering
  \vspace*{5mm}
  \includegraphics[angle=270, width=16cm]{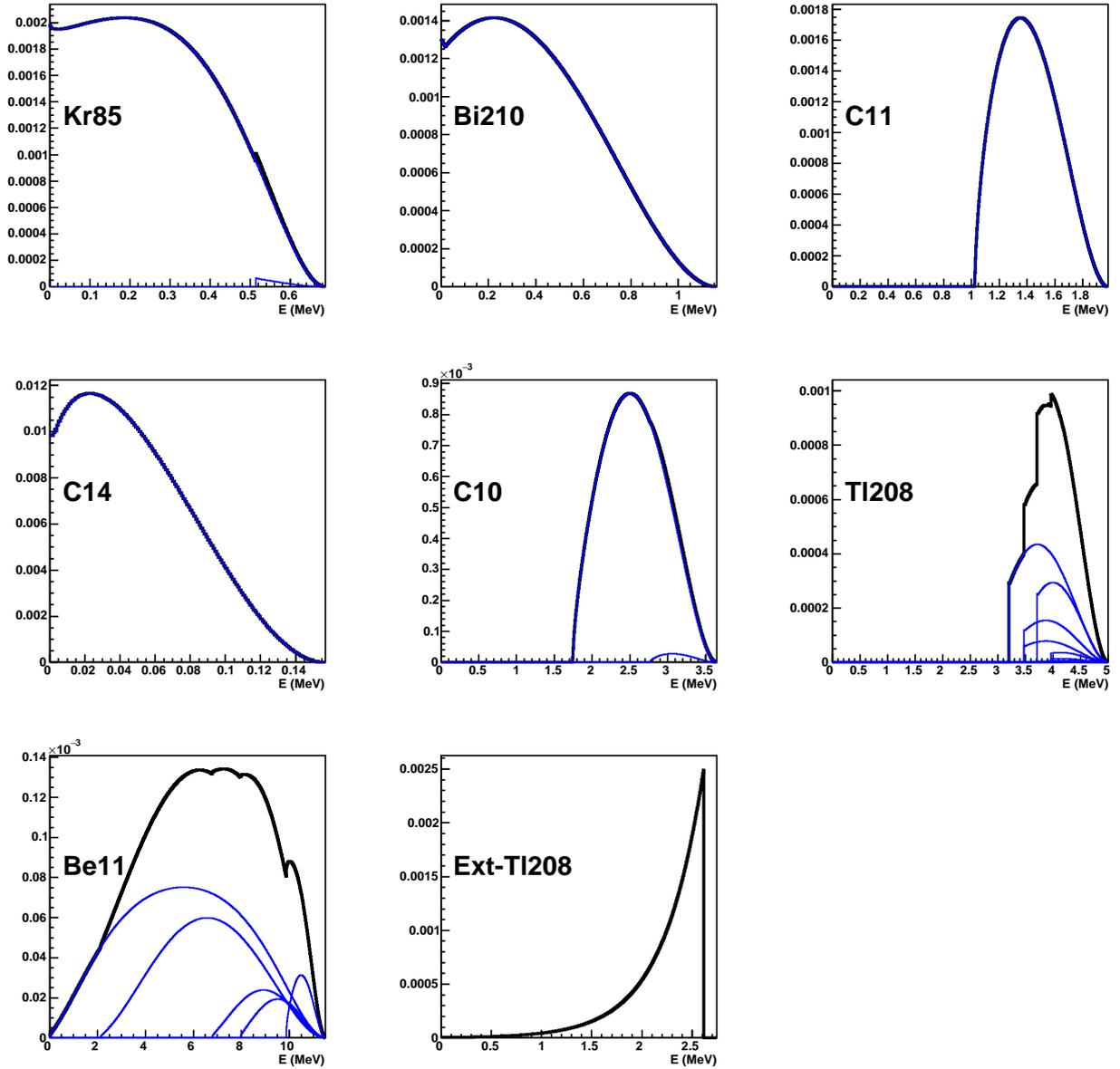}
  \caption{(Color online) Visible energy spectra for cosmic-ray muon induced, and residual radioactive backgrounds (first 7 plots). The black curves are  for the total visible energy spectra and the blue ones are for the corresponding sub branches.
  Also shown is the visible energy spectrum for the external 2.6 MeV gamma background ($^{208}$Tl) (bottom middle plot).}
  \label{fig:BkgSpec}
\end{figure}

There are mainly three categories of backgrounds. 1) Cosmic-ray muon induced spallation backgrounds. With the overburden of Jinping, these backgrounds will be
a factor of 200 lower than those in Borexino or a factor of 2 lower than SNO. 2) Internal radioactive beta or gamma backgrounds. They are the residual background remaining in detecting material regardless of the depth. We assume that these background can be reduced by purification down to  the same level as Borexino. 3) Environment radioactive background. They present as external gammas for a central detector volume.
Borexino background rates were applied in our study with a scale according to the surface area.

For simplicity,
no quenching effect was considered in the following study, so that for sequential beta and gamma decays,
all the gamma energies and beta kinetic energies were added linearly without considering the decay structure of excited states.
For the positrons from the beta+ decays, twice of electron mass was added the detected energy for the positron annihilation.

The visible energy spectra for category 1) and 2) are shown in Fig.~\ref{fig:BkgSpec}.
The external gamma background was modeled by an exponential distribution, motivated by Ref.~\cite{BxPhaseI}. An example of the energy distribution for the 2.6 MeV gamma background from the external $^{208}$Tl is also shown in Fig.~\ref{fig:BkgSpec}, where
the decay constant was assumed to be 0.4 MeV, which is related to the gamma ray attenuation length and fiducial volume buffer dimension.

A summary of the event rates for all the backgrounds can be found in Table~\ref{tab:solarBkg}. Details are given in below.

The Borexino-I $^7$Be refers to the analysis result of the phase-I $^7$Be measurement at Borexino~\cite{BxPhaseI, BxBe7I, BxBe7II}, from which the fiducial volume mass, live time, and background rates of $^{14}$C, $^{85}$Kr, $^{210}$Bi, and $^{11}$C were extracted. The values for $^{14}$C and $^{11}$C were used for the Jinping study. Other background values, $^{10}$C, $^{208}$Tl, $^{11}$Be, and Ext-$^{208}$Tl were extracted from the other references discussed below.

The Borexino-I  $pep$  refers the analysis result of the phase-I  $pep$  measurement at Borexino~\cite{BxPhaseI, BxPep}, where the data of 598.3 live days was scaled down by 48.5\% for
the final selection efficiency. With the technique of three-fold coincidence (TFC), the background rate of $^{10}$C was suppressed.
The $^{10}$C background rate without the TFC technique was taken as a standard value of the Borexino experiment, and then
scaled to Jinping. The most significant and presentative external gamma background, Ext-$^{208}$Tl, as a major background was extracted from this analysis and used for the Jinping study.

Borexino-I $^8$B is for the phase-I $^8$B analysis at Borexino~\cite{BxB8}, where the energy beyond 3 MeV was discussed. The reported rates of the high
energy backgrounds $^{208}$Tl and $^{11}$Be were taken to be the standard values for the Borexino experiment, and used for the Jinping study.

The second phase of the Borexino experiment has a much lower $^{85}$Kr and $^{210}$Bi background rates~\cite{BxPp}. A sample with double live days of data was assumed, in order to compare the phase one analysis. The measured background rates of $^{85}$Kr and $^{210}$Bi of the second phase were used for the Jinping study.

To test the SNO+ proposal, the same background assumptions were taken.


\begin{table}[!tb]
\centering
\small
\begin{tabular*}{0.79\paperwidth}{ @{\extracolsep{\fill}} l l l l l l l l l l l l}
\hline\hline
                    & Mass & Time & Resolution & $^{14}$C         &$^{85}$Kr& $^{210}$Bi & $^{11}$C & $^{10}$C & $^{208}$Tl & $^{11}$Be & Ext-$^{208}$Tl \\
                    & [100 ton] & [day] & [PE/MeV] & \multicolumn{8}{c}{[Counts/day/100 ton]} \\ \hline
Borexino-I $^7$Be   & 0.7547 & 740.7 & 500 & $3.46\times10^6$ & 31.2 & 41.0 & 28.5 & 0.62 & 0.084 & 0.032 & 2.52 \\
Borexino-I $pep$    & 0.7130 & 290.2 & 500 & $3.46\times10^6$ & 31.2 & 41.0 & 2.48 & 0.18 & 0.084 & 0.032 & 2.52 \\
Borexino-I $^8$B    & 1      & 345.3 & 500 & $3.46\times10^6$ & 31.2 & 41.0 & 28.5 & 0.62 & 0.084 & 0.032 & 2.52 \\
Borexino-II $^7$Be  & 0.7547 & 1480  & 500 & $3.46\times10^6$ &  1   & 25.0 & 28.5 & 0.62 & 0.084 & 0.032 & 2.52 \\
Borexino-II $pep$   & 0.7130 & 580   & 500 & $3.46\times10^6$ &  1   & 25.0 & 2.48 & 0.18 & 0.084 & 0.032 & 2.52 \\
Borexino-II $^8$B   & 1      & 690   & 500 & $3.46\times10^6$ &  1   & 25.0 & 28.5 & 0.62 & 0.084 & 0.032 & 2.52 \\
SNO+                & 5      & 1500  & 1000& $3.46\times10^6$ &  1   & 25.0 & 0.29 & 0.0062& 0.084& 0.00032&1.47 \\
Jinping             & scan   & 1500  & scan& $3.46\times10^6$ &  1   & 25.0 & 0.15 & 0.0031& 0.084& 0.00016&1.17 \\
\hline
\end{tabular*}
\caption{A summary of the fiducial mass, live time, and backgrounds for all the known running or planed solar neutrino experiments. See the text in Sec.~\ref{sub:sol:bkg} for the references and calculation methods for each experiment or analysis. Jinping's fiducial mass and resolution will be scanned in the study.}
\label{tab:solarBkg}
\end{table}

\subsubsection{Total spectrum}
\begin{figure}[h]
  \vspace*{0.5cm}
  \centering
  \includegraphics[angle=270, width=9cm]{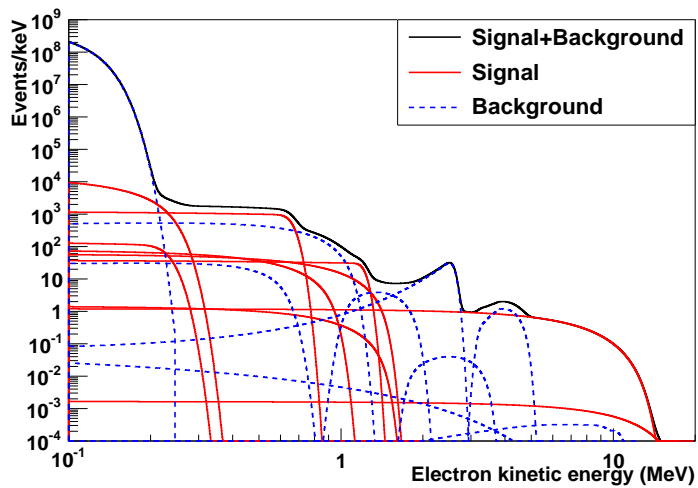}
  \caption{(Color online) An example plot of the total expected spectrum including all the neutrino and background components at Jinping with 1 kton liquid scintillator scheme, 1,500 live days of data, and 500 PE/MeV of energy resolution.}
  \label{fig:JinpingScin500}
\end{figure}

An example plot of the total expected spectrum including all the neutrino and background components at Jinping with 1 kton liquid scintillator scheme, 1,500 live days of data, and
500 PE/MeV of the detector response is shown in Fig.~\ref{fig:JinpingScin500}. Simulated samples were fitted and analyzed for each physics topic below and the corresponding discovery sensitivities will be reported.

\subsection{Systematics on the flux measurement}
\label{sub:sol:systematic}

Two systematic uncertainties were considered for the measurement on the solar neutrino flux.
One is the fiducial volume definition, which is only related to the bias of vertex reconstruction rather than the resolution.
A 1\% systematic uncertainty was assumed for the fiducial volume cut.
The other one is from the energy response  of detector. With the experience of the
Borexino experiment and the recent Daya Bay experiment~\cite{nonlinear},
we believe that
the uncertainty from the non-linearity and non-uniformity effect in the energy reconstruction can be controlled down to the level of 1\%.
With a large data sample expected at Jinping,
we assumed there was no fitting procedure error as introduced by Borexino analysis.
In total, 1.5\% systematic uncertainty is assigned to all the flux measurements.

\subsection{Precision for each component measurement}
\label{sub:sol:precision}
\begin{figure}[!hp]
  \centering
  \includegraphics[angle=270, width=9cm]{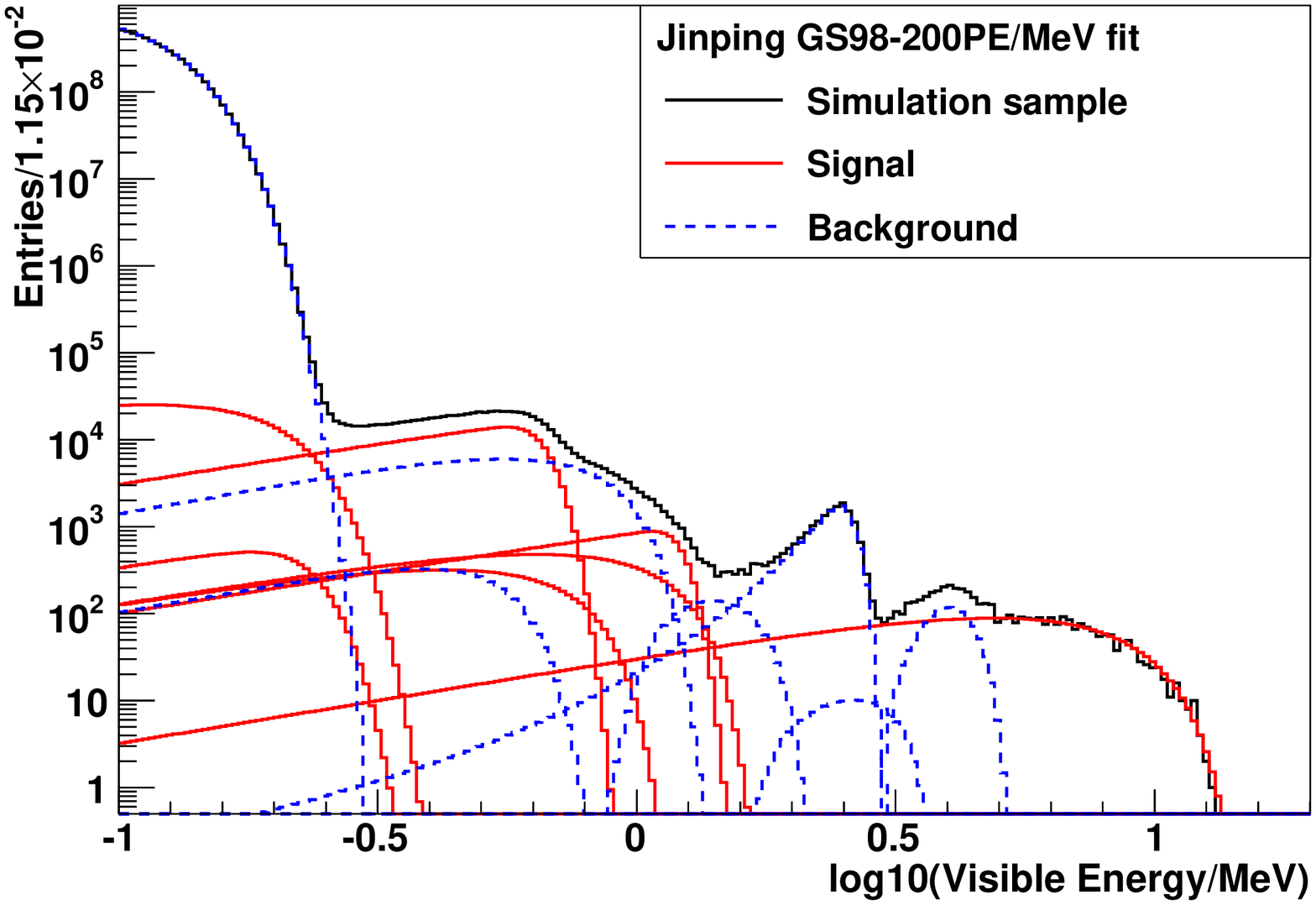}

  \vspace*{0.8cm}
  \includegraphics[angle=270, width=9cm]{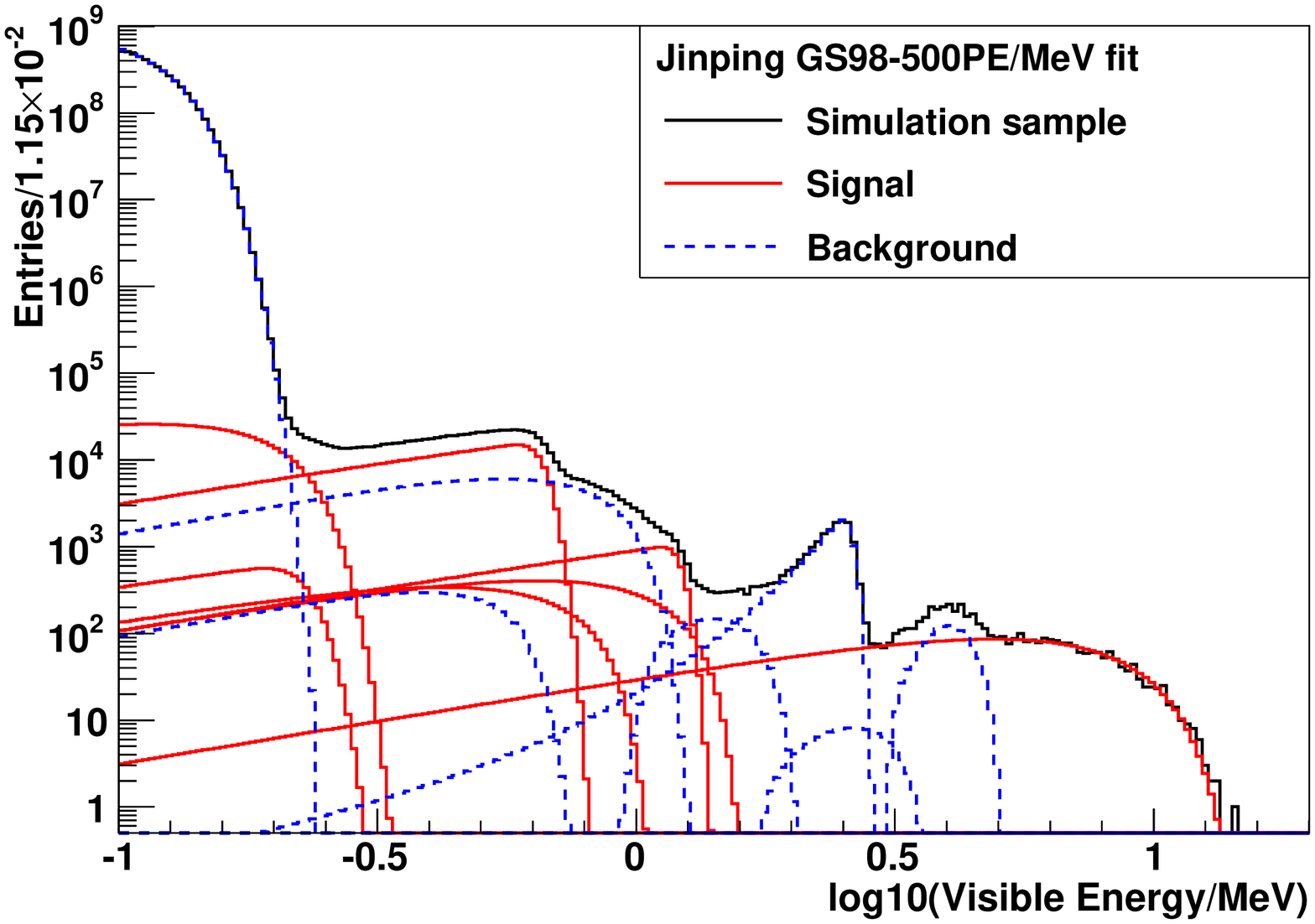}

  \vspace*{0.8cm}
  \includegraphics[angle=270, width=9cm]{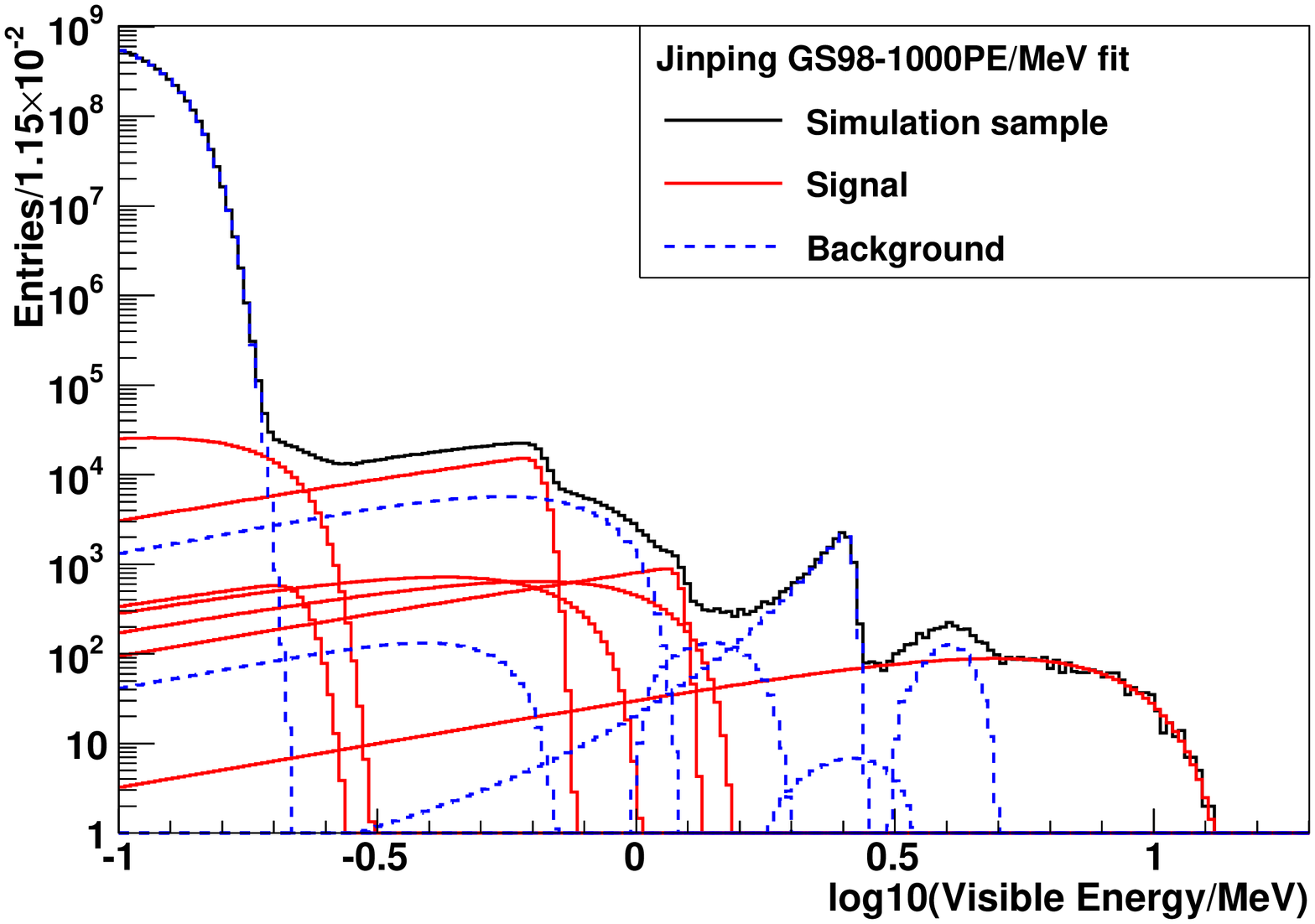}
  \caption{(Color online) Fit results for the simulation samples with a fixed 1,000-ton target mass and three different energy resolutions.
  Black curves are for the simulated samples, and red and blue curves are for the fit results.
  $^7$Be and  $pep$  neutrinos have a sharp edge structure in the energy distributions, while $^8$B neutrinos give a very broad one, so that they are not so sensitive to the detector resolution. However $pp$ and CNO neutrinos rely on a precise determination of all the other neutrino and background components, and are therefore subject to the detector resolution.}
  \label{fig:JinpingFit}
\end{figure}

Simulations with the inputs from Table~\ref{tab:EvtRate}, various target masses and energy resolutions were done to evaluate the expected precisions. Fitting examples are shown in Fig.~\ref{fig:JinpingFit}.
The $^7$Be 0.38 MeV line to 0.86 MeV ratio was fixed according to Table~\ref{tab:solarFlux}.
Since the characteristic line shapes for $^{15}$O and $^{17}$F were not distinguishable, only $^{15}$O component was considered in the fitter.
The $hep$ neutrino contribution was not significant in the fit, and was ignored.
Table~\ref{tab:500 $pep$ recision} lists the
statistical and systematic precisions for all the solar neutrino components with the high and low metallicity models.

\begin{table}[h]
\centering
\small
\begin{tabular}{cclll} \hline\hline
               & Neutrino & \multicolumn{3}{c}{Energy resolution}         \\
               & component& 200 PE/MeV & 500 PE/MeV & 1000 PE/MeV  \\ \hline
               & $pp$       & 0.02       & 0.007      & 0.005        \\
               & $^7$Be   & 0.007      & 0.006      & 0.005        \\
Fiducial mass  &  $pep$       & 0.07       & 0.05       & 0.04         \\
1,000 ton       & $^{13}$N & NA         & 0.5 (NA)   & 0.3 (0.4)    \\
               & $^{15}$O & 0.3        & 0.2 (0.4)  & 0.1 (0.2)    \\
               & $^8$B    & 0.02       & 0.02       & 0.02         \\\hline

               & $pp$       & 0.01       & 0.005      & 0.004        \\
               & $^7$Be   & 0.005      & 0.004      & 0.004        \\
Fiducial mass  &  $pep$       & 0.06       & 0.03       & 0.03         \\
2,000 ton       & $^{13}$N & 0.4        & 0.3        & 0.2 (0.3)    \\
               & $^{15}$O & 0.2        & 0.1        & 0.08 (0.1)   \\
               & $^8$B    & 0.02       & 0.02       & 0.02         \\\hline

               & $pp$       & 0.01       & 0.004      & 0.003        \\
               & $^7$Be   & 0.004      & 0.003      & 0.003        \\
Fiducial mass  &  $pep$       & 0.04       & 0.03       & 0.02         \\
4,000 ton       & $^{13}$N & 0.3        & 0.2 (0.3)  & 0.2 (0.3)    \\
               & $^{15}$O & 0.1 (0.2)  & 0.07 (0.1) & 0.06 (0.09)  \\
               & $^8$B    & 0.01       & 0.01       & 0.01         \\\hline
\end{tabular}
\caption{Relative statistical precision of solar neutrino fluxes
for three different target masses and energy resolutions.
The default results are for the high metallicity assumption and
the ones in the parentheses are for low metallicity if significantly different.
NA is marked when the relative uncertainty is greater than 50\%.
The $^7$Be 0.38 MeV line to 0.86 MeV ratio was fixed according to Table~\ref{tab:solarFlux}.
Since the energy line shapes for the $^{15}$O and $^{17}$F cannot be distinguished, we use the one for $^{15}$O only.
The \emph{hep} neutrino contribution was ignored in the fit due to the low statistics.}
\label{tab:500 $pep$ recision}
\end{table}

\subsubsection{Improvement on the known neutrino components}
\textbf{$pp$ neutrino:}
As shown in Fig.~\ref{fig:JinpingFit}, the electron energy from the $pp$ neutrino elastic scattering is slightly higher than that from the main
background $^{14}$C, and the best signal region for detecting the $pp$ neutrinos is around 0.2 - 0.3 MeV.
The statistical uncertainty on the $pp$ neutrino flux is very sensitive to the energy resolution of the detector, which
can reach 1\% with the 500 PE/MeV light yield.
The total uncertainty will be dominated by the systematic uncertainty.
We hope to control the dominant systematic uncertainty and reduce the total uncertainty down below 1\%,
and this will help to explore the expected difference between the neutrino luminosity and the optical luminosity.

\textbf{$^7$Be neutrino:}
$^7$Be and $^8$B neutrinos are critical to distinguish the high and low metallicity hypotheses.
The $^7$Be neutrino flux can be measured statistically better than 1\%, which is less dependent on the energy resolution due to the characteristic sharp turn.
The total flux uncertainty is dominated by the systematic uncertainty.

\textbf{$^8$B neutrino:}
$^8$B neutrinos suffer the largest matter effect, which is sensitive to the vaccum-matter transition phase and the day-night flux asymmetry.
The relatively high energy of $^8$B neutrinos makes them less contaminated by other backgrounds, and because of the broad energy spectrum, the study of $^8$B neutrinos does not rely on
the energy resolution much. The statistical precision on the flux of $^8$Be neutrinos is expected to be about 1\% - 2\% , which is limited by the target mass,
and is comparable to the systematic uncertainty.

\textbf{$pep$  neutrino:}
The distinguishable structure of the $pep$  neutrino spectrum, like the $^7$Be neutrinos, makes them be easily identified. With the three energy resolution options considered, the sensitivity can all reach beyond 7\% and even 3\% if the target mass can go to 2,000 tons. It can reach 3\%. The  $pep$  neutrinos are one of the key ingredients in the study of the solar model and the vacuum-matter oscillation transition.

\subsubsection{Discovery of the CNO neutrinos}
\textbf{CNO neutrino:}
The flux of CNO neutrinos strongly depends on the metallicity hypotheses and itself is an very interesting subject since the CNO neutrinos are from the
main fueling process of high temperature stars, while the $pp$ process is dominant in the Sun because of the relatively low temperature.
The major background for the $^{13}$N and $^{15}$O neutrinos detection are
the $^7$Be and  $pep$  neutrinos, and $^{85}$Kr and $^{210}$Bi decays.
An effective identification of the other neutrinos and backgrounds will help to resolve the CNO neutrinos, and this relies on the energy resolution.
With a resolution of 500 PE/MeV or better, a discovery of $^{15}$O neutrinos at Jinping will be possible (the relative error is less than 30\%) with the metallicity assumption. With a larger target mass, for example 2,000 tons, the discovery potential will be more significant.

\subsection{Matter-vacuum transition phase}
\label{sub:sol:tran}

\begin{figure}[!h]
  \centering
  \includegraphics[angle=270, width=11cm]{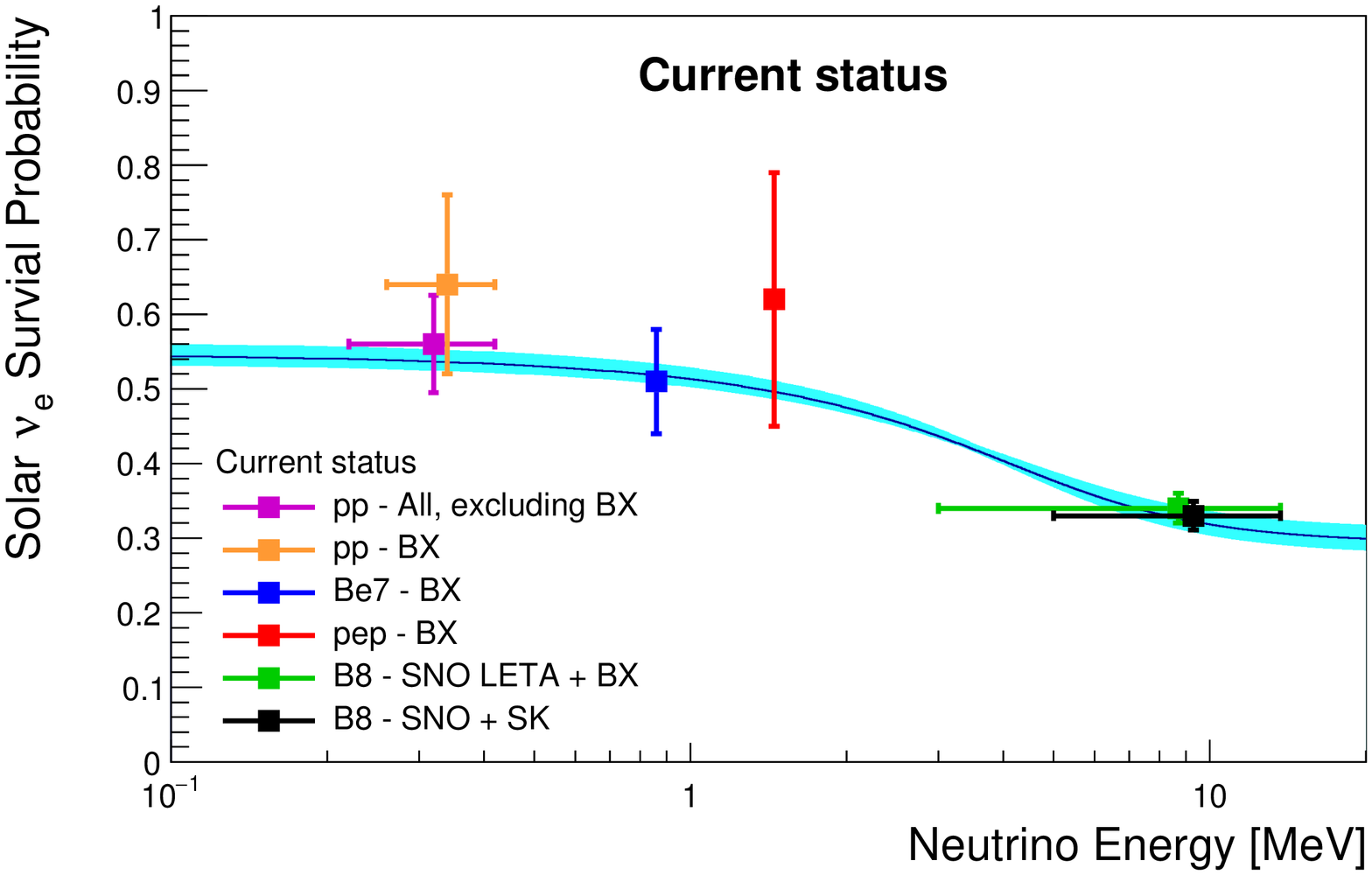}
  \caption{(Color online) The transition of oscillation probability from the vacuum to matter effect as a function of neutrino energy. Data points plotted are the present measurements~\cite{BxBe7II, BxPep, BxPp}.
  The solid line is for the theoretical prediction, while the shaded area is obtained by marginalizing $\theta_{12}$, $\theta_{13}$, and $\Delta m^2_{12}$ with the present experimental uncertainty.}
  \label{fig:EnTranCurr}

  \vspace*{0.8cm}
  \includegraphics[angle=270, width=11cm]{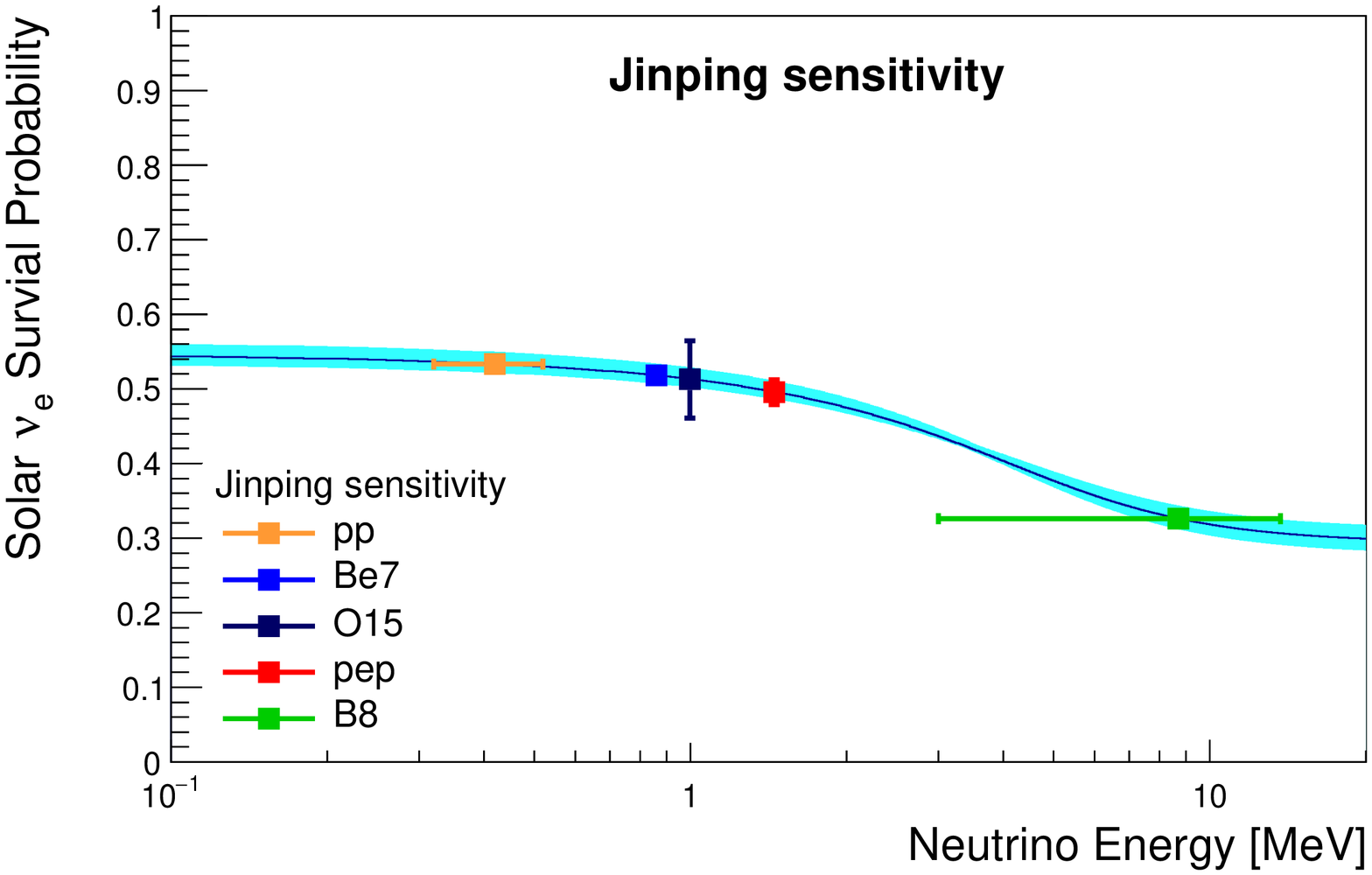}
  \caption{(Color online) The transition of oscillation probability from the vacuum to matter effect as a function of neutrino energy with  Jinping's sensitivities.
  Here we assume a 2,000-ton target mass, 1,500-day exposure, a resolution of 500 PE/MeV, and the low metallicity hypothesis, where the solid line is for the theoretical prediction, the shaded area is obtained by marginalizing $\theta_{12}$, $\theta_{13}$, and $\Delta m^2_{12}$
  with the present experimental uncertainty, the five points with error bars are the simulation results for $pp$, $^7$Be,  $pep$, $^{15}$O and $^8$B,
  in which the central values are set to the true ones, and the y-error bars include both statistical and systematic uncertainties and the x-error bars correspond to the range of measurement, while the $^{15}$O x-error is omitted for a clear view.}
  \label{fig:EnTrans}
\end{figure}

\begin{figure}[!h]
  \centering
  \vspace{0.3cm}
  \includegraphics[angle=270, width=9cm]{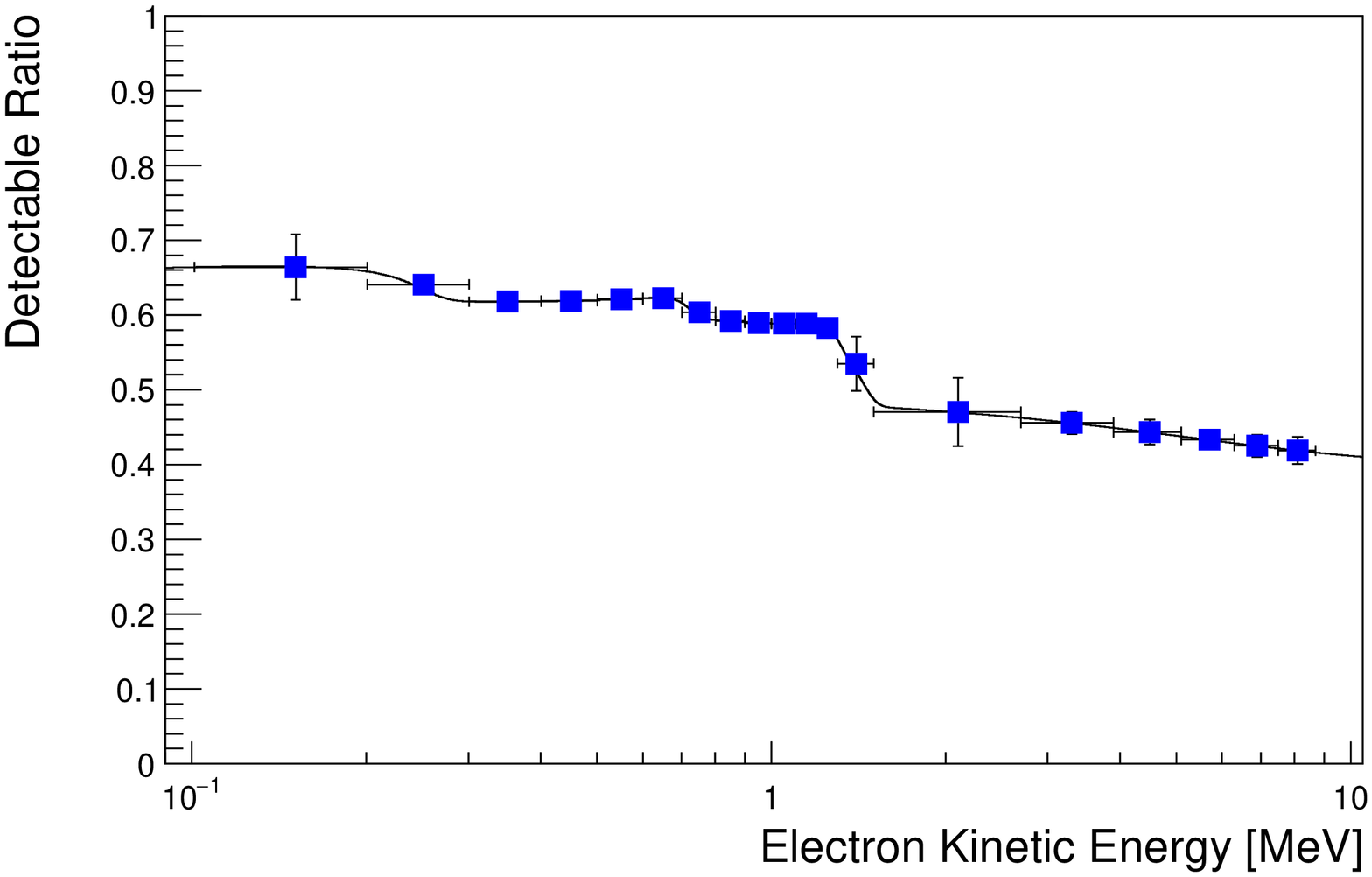}
  \caption{(Color online) The transition of oscillation probability from the vacuum to matter effect as a function of kinetic energy of recoiling electrons.
  Here we assumed a 2,000-ton target mass, 1,500-day exposure, a resolution of 500 PE/MeV, and the low metallicity hypothesis. The solid line is for the
  theoretical prediction and the dots with error bars are for the simulation.
  The statistical uncertainties are from both backgrounds and signals.
  Because of the large correlation among the points, the systematic uncertainties are not included.}
  \label{fig:KeTrans}
\end{figure}

The transition of oscillation probability from the matter-governed region to the pure vacuum-like region is a very interesting phenomenon of the MSW effect as
shown in Fig.~\ref{fig:OsciProb} and~\ref{fig:ElecSpecNoOs}. This effect has been studied by Borexino~\cite{BxPhaseI, BxB8},
Super Kaminokande~\cite{SuperK2}, SNO~\cite{SnoSumm}, and previous experiments, however, experimentally the oscillation in the transition region is still loosely constrained, the current status is shown in Fig.~\ref{fig:EnTranCurr}.
With the Jinping simulation, the expected flux measurements are compared with the predictions with neutrino energy and recoiling electron kinetic energy,
which are shown in Fig.~\ref{fig:EnTrans} and~\ref{fig:KeTrans}, respectively. For Fig.~\ref{fig:KeTrans},
the uncertainty of each bin is conservatively treated as the square root of the full statistics of each bin including all the backgrounds and signals, and,
for a better performance of demonstration, the bin ranges were adjusted according to the statistics of each individual signal region.

\subsection{Day-night asymmetry}
\label{sub:sol:daynight}
\begin{figure}[!h]
  \centering
  \vspace*{0.3cm}
  \includegraphics[angle=270, width=7.5cm]{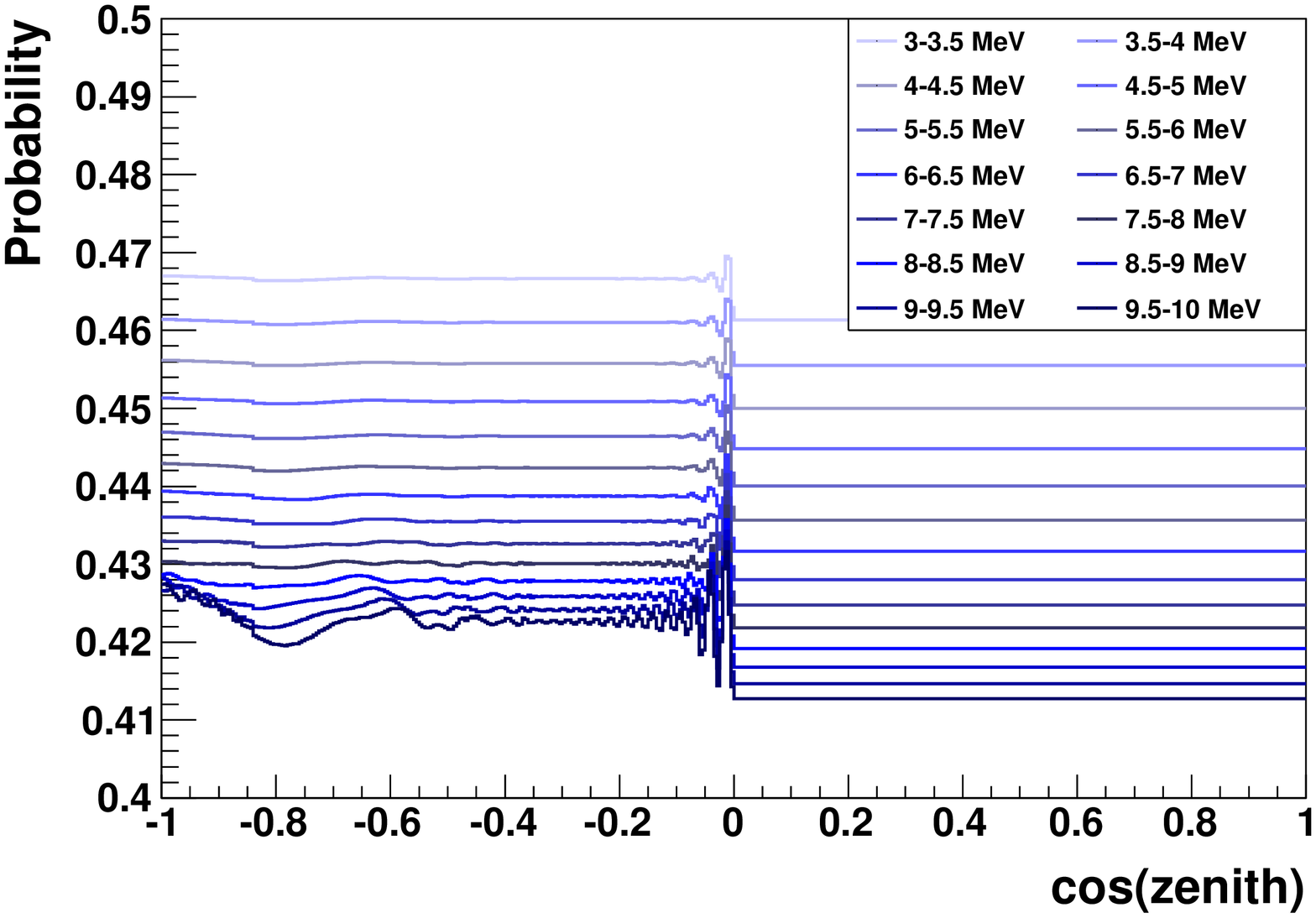}
  \vspace*{0.3cm}
  \includegraphics[angle=270, width=7.5cm]{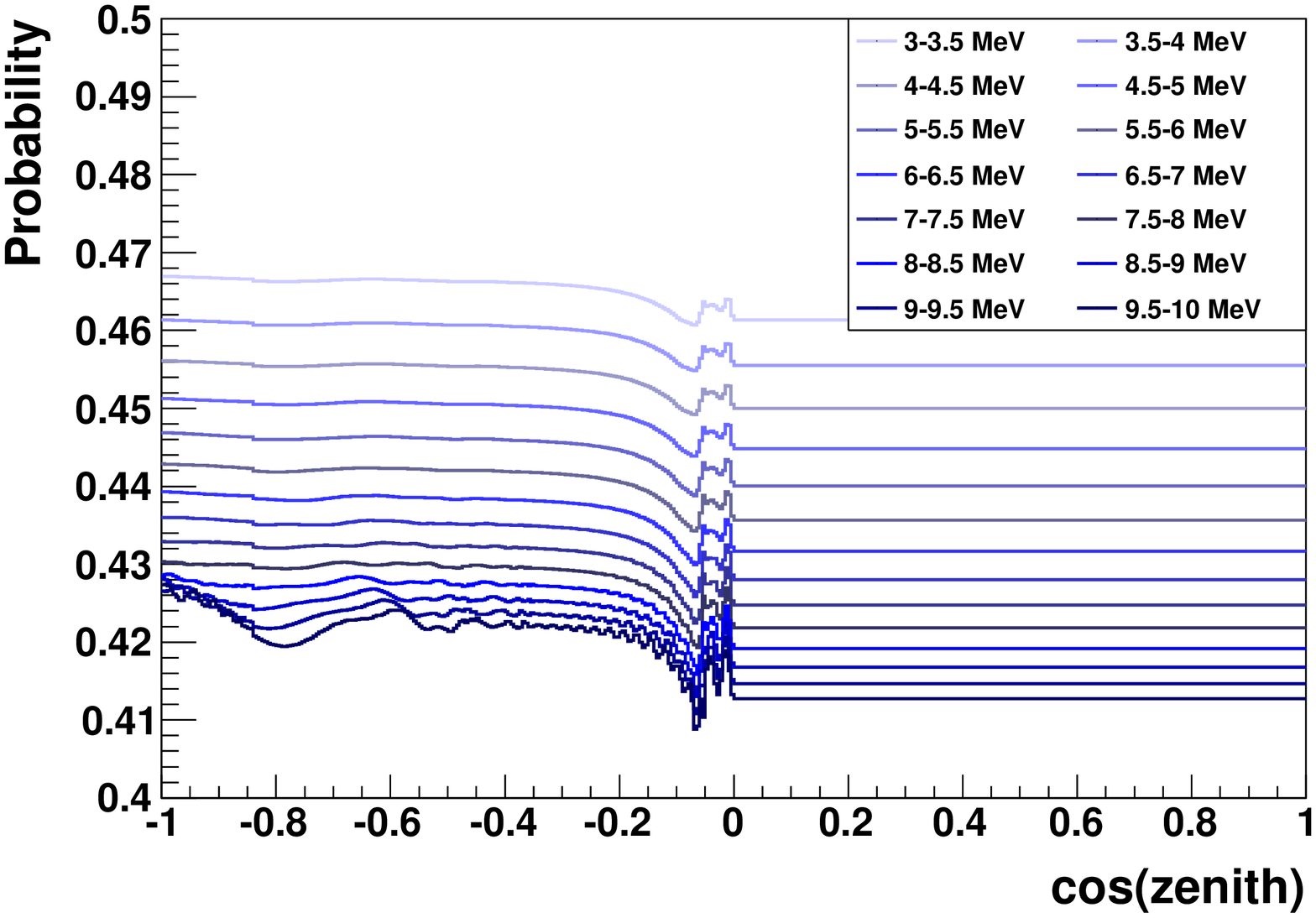}
  \caption{(Color online) Day-night asymmetry for a continental site (left) and an ocean site (right).}
  \label{fig:dn}
\end{figure}

After solar neutrinos passing through the Earth, electron neutrinos may be regenerated because of the MSW matter effect~\cite{Bright},
which leads to a slightly higher survival probability during the night than that during the day time.
The survival probability is very sensitive to $\Delta m^2_{21}$ and the density profile at the nearby surface of the Earth~\cite{SurfaceDen, SK},
which can cause a day-night asymmetry varying from 1-3\%
for the rates of solar neutrinos.
The asymmetry variation for a continental site and a ocean site is shown in Fig.~\ref{fig:dn}.
The sensitivity for the asymmetry above 3 MeV at Jinping can be directly estimated with the expected statistics of $^8$B events, since the background contamination above 3 MeV
is not significant. The separation of the fluxes between the daytime and nighttime is
\begin{equation}
s = 2(N-D)/(N+D),
\end{equation}
where $N$ and $D$ are the signal event counts during the night and day. The uncertainty of $s$ is
\begin{equation}
\sigma_s \approx 2/\sqrt{N+D} = 2/\sqrt{N_{\rm B}},
\end{equation}
where $N_{\rm B}$ is the total statistics of $^8$B neutrino events.
According to Table~\ref{tab:EvtRate}, with a 5-year data-taking, a 2,000-ton detector, the total statistics is around 10,000, which is weak in obtaining a conclusive measurement on the day-night asymmetry.

\subsection{Metallicity problem}
\label{sub:sol:metal}
With the expected improvements on the measurement of solar neutrino fluxes,
we carried out a simulation study on a hypothesis test for the solar models with the high and low metallicities.
The evaluation was done assuming 2,000-ton target mass and 1,500-day exposure.

The ability to distinguish two hypotheses, separation $S$, is defined as
\begin{equation}
S = \left[ \sum_{i,j} (F_{h,i}-F_{l,i}) (V^{-1})_{ij} (F_{h,j}-F_{l,j}) \right]^{1/2},
\label{eq:Sep}
\end{equation}
where $F_{h,i}$ and $F_{l,i}$ are the predicted neutrino fluxes for the i-th component with the high (h) and low (l) metallicities, respectively, and $V$ is the covariance matrix, which follows the usual definition as
\begin{equation}
V_{ij}=\sigma_i \sigma_j \rho_{ij},
\end{equation}
where $\sigma_i$ is the uncertainty of the i-th component and $\rho_{ij}$ gives the correlation between the
i-th and j-th components.

Firstly, an optimistic calculation of the separation, $S_{opt}$, between the two hypotheses was done assuming all
of the flux measurements were independent. By simplifying Eq.~\ref{eq:Sep},
\begin{equation}
S_{opt} = \left[ \sum_{i} S_{opt,i}^2 \right]^{1/2} = \left[ \sum_{i} (F_{h,i}-F_{l,i})^2 / \sigma_i^2 \right]^{1/2},
\label{eq:Sep_i}
\end{equation}
where $S_{opt,i}$ gives the separation achieved for the i-th component.
Table~\ref{tab:SepOpt} gives the inputs for calculating $S_{opt}$, including
the flux difference between the two hypotheses, the expected experimental uncertainties, $\sigma_{exp,i}$, $S_{opt,i}$, and the theoretical uncertainties $\sigma_{theory,i}$.
The final result is
\begin{equation}
S_{opt} = 9.6,
\end{equation}
which can be treated as a 9.6-sigma rejection to the high metallicity hypothesis and vice versa.
The most powerful separations are expected from the $^7$Be, $^{15}$O, and $^8$B neutrinos.

\begin{table}[!h]
\centering
\small
\begin{tabular}{lcccc}
\hline\hline
          & $F_{h,i}-F_{l,i}$ & $\sigma_{exp,i}$  & $|S_{opt,i}|$ &   $\sigma_{theory,i}$ \\\hline
pp        &		 0.05  & 0.10              &	0.53    &	 0.036             \\
$^7$Be    &		-0.44  & 0.078             &	5.7     &	 0.32              \\
$pep$     &		 0.03  & 0.048             &	0.62    &	 0.018             \\
$^{13}$N  &		-0.79  & 0.89              &	0.89    &	 0.30              \\
$^{15}$O  &		-0.67  & 0.23              &	3.0     &	 0.28              \\
$^{17}$F  &		-2.12  & -                 &	-       &	 0.76              \\
$^8$B     &		-0.99  & 0.14              &	7.1     &	 0.64              \\
$hep$     &		 0.27  & -                 &	-       &	 2.49              \\
\hline
\end{tabular}
\caption{Details of the high and low metallicity hypotheses test.
The second column shows the flux difference, i.e. GS98-AGS09.
The third column gives the expected absolute experimental error, $\sigma_{exp,i}$,
which is calculated according to the GS98 flux estimation and the expected statistical and systematic errors.
The fourth column is the separation for the i-th component, $|S_{opt,i}|$, as defined in Eq.~\ref{eq:Sep_i}.
In the last column, the absolute theoretical uncertainty for the i-th component, $\sigma_{theory,i}$, is presented for comparison.
The flux difference, $\sigma_{exp,i}$, and $\sigma_{theory,i}$ are in units
of $10^{10}$ ($pp$), $10^9$ (7Be), $10^8$ ( $pep$ , $^{13}$N, $^{15}$O), $10^6$ ($^{8}$B, $^{17}$F), and $10^3$ ($hep$) $\rm{cm}^{-2}\rm{s}^{-1}$.
}
\label{tab:SepOpt}
\end{table}

Secondly, a conservative estimation of model separation $S_{cons}$ was calculated taking into account the contribution from the $^7$Be, $^8$B, and $^{15}$O neutrinos, and the experimental correlations among them.
The correlations stem from the fitting procedure to separate the $^7$Be and $^8$B components, since the statistical precisions are relatively high as listed in Table~\ref{tab:500 $pep$ recision}.
The major sources of systematic errors, target mass and energy response, might be fully correlated among all the neutrino components for the worst case.
It should be noted that the relative systematic uncertainty of 1.5\% is dominant for the $^7$Be neutrinos and significant to the $^8$B neutrinos. As a result,
the correlation between them is as high as 50\%, degrading the power to distinguish the models.
The correlation matrix between them is given below
\begin{equation}
\rho=\begin{bmatrix}
         1 &     0.1581 &     0.5799  \\
    0.1581 &          1 &     0.1104  \\
    0.5799 &     0.1104 &          1
        \end{bmatrix},
\end{equation}
where the rows and columns are arranged as $^7$Be, $^{15}$O, and $^8$B,
and the calculation gives
\begin{equation}
S_{cons} = 7.6.
\end{equation}

Finally a calculation of $S_{theory}$ was done with all the experimental and theoretical uncertainties and correlations for the $^7$Be, $^{15}$O, and $^8$B neutrinos.
Because of the large theoretical uncertainties on the $^7$Be and $^8$B and the associated  correlation, their impact becomes quite weak.
The contribution from the $^{15}$O neutrinos is also not so significant because of the large theoretical uncertainty.
The correlation matrix is
\begin{equation}
\rho=\begin{bmatrix}
       1  &    0.2012  &    0.8804 \\
  0.2012  &         1  &    0.3229 \\
  0.8804  &    0.3229  &         1
        \end{bmatrix}.
\end{equation}
The result is
\begin{equation}
S_{theory} = 1.2.
\end{equation}

As a conclusion, with the proposed Jinping neutrino experiment, the experimental resolving capability for the high and low metallicity hypotheses is within 7$\sim$10 $\sigma$'s
with the current values of mixing angles.
The uncertainties of the mixing angles should also have impacts on resolving the problem.
We expect the uncertainties will be improved with terrestrial experiments.


\subsection{Conclusion and summary}
Based on the discussion in this section, with 2,000-ton fiducial mass and 500 PE/MeV resolution, the expected outcome for the proposed Jinping neutrino experiment running over 5 years
is very promising to make a discovery of the CNO neutrinos and to significantly improve the measurements of the $pp$, $^7$Be,  $pep$ neutrino fluxes.
The experiment can also provide a much stronger constrain on the vacuum-matter transition to the MSW effect, and
have the experimental capability to distinguish the high and low metallicity hypotheses.
Due to the limited target mass, physics relying on the statistics of $^8$B neutrinos cannot be precisely probed, for example, the day-night asymmetry.
In this study, we didn't discuss the subject of possible improvement on the measurement of neutrino mixing angles
and the possibility to rule out other new physics.

%% file: Geoneutrino.tex
\section{Geo-neutrinos}
\label{sec:geoneutrino}

\subsection{Introduction}
Identifying and understanding the Earth's energy budget is a fundamental question in geology as it defines the
power that drives plate tectonics, mantle convection, and the geodynamo~\cite{Geology}.
The motivation of understanding geo-neutrinos starts with the wish to understand our planet.

The Earth's total heat flow is currently estimated to be 46$\pm$3 TW~\cite{HeatFlow}.
The driving power comes presumably from two sources:
1) the Earth is a radioactive engine and the radioactive energy from the mantle is part of the energy source;
2) the planet is still consuming its initial inheritance of primordial energy that resulted from
the accretion of the planet and the gravitational differentiation of iron sinking to the center of the Earth.

There are basically three types of predictions for the radiogenic power of the Earth.
But the estimation of the Earth's radiogenic power still has a lot of uncertainties and covers a wide power range~\cite{GeoReview}.
The prediction has to take into account the composition of the Earth, chemical layering in the mantle, mantle convection, and the power source of the geodynamo, etc.
The heat predicted by these three types of models are:
1) the cosmochemical models have the lowest result ($\sim$10 TW);
2) the traditional geodynamical models have the highest prediction ($\sim$30 TW),
and later more developed models require less radiogenic power (15-17 TW);
and 3) the geochemical models predict the median amount ($\sim$20 TW).

One of the best ways of experimentally measuring the radiogenic power
is to measure the amount of neutrinos coming from the interior of the Earth.
Neutrinos as the daughter products from radioactive decay chains can provide
new insights into the Earth.
The study of geology with geo-neutrinos
only became practical recently by the advent of large underground neutrino detectors,
i.e. the KamLAND~\cite{KLGeo1, KLGeo2, KLGeo3} and Borexino~\cite{BXGeo1, BXGeo2, BxNew} experiments.

\begin{figure}[!h]
\centering
\includegraphics[width=15cm]{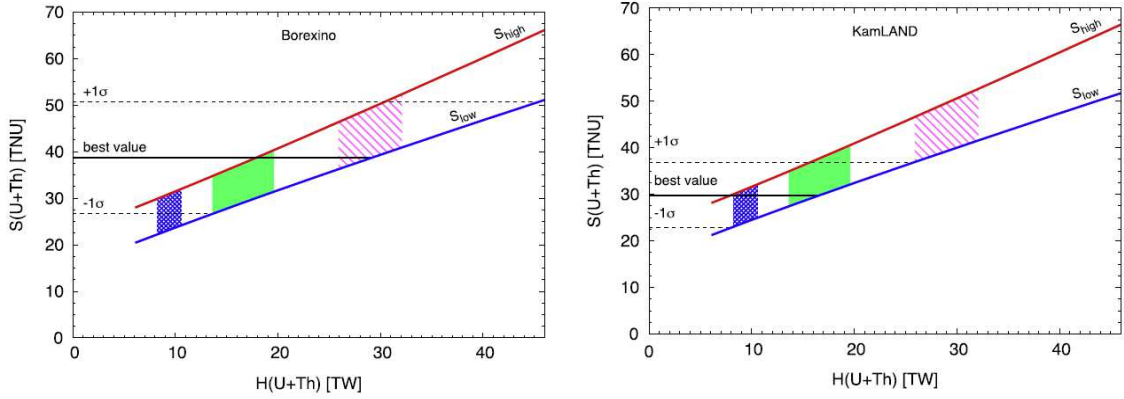}%
\caption{The expected geo-neutrino signal band in Borexino (left) and in KamLAND (right)
as a function of radiogenic heat released in radioactive decays of U and Th~\cite{GeoReview}. The Borexino and KamLAND results are indicated by the horizontal lines, respectively. The three filled regions delimit,
from the left to the right, the cosmochemical, geochemical, and geodynamical models, respectively.}
\label{fig:GeoStatus}
\end{figure}

Figure~\ref{fig:GeoStatus} shows the current status of geo-neutrino studies compiled by Ref.~\cite{GeoReview}.
The heat correlated with neutrinos is around 20 TW, which indicates that we are still
consuming the primordial gravitational power when the planet Earth formed,
but the uncertainty is quite significant.
Furthermore, current geo-neutrino measurements bear large uncertainties, experimentally and theoretically, to evaluate any Earth models.
Firstly, the geo-neutrino flux measurements have a large experimental uncertainty, due to the low statistics of signal and
high reactor neutrino background.
Secondly, the prediction of the total geo-neutrino flux consists of the contribution from the near- and far-field crust and from the deep mantle.
The mantle neutrinos are of the most theoretical interest, however, the corresponding
fraction is only from 15\% to 30\% for any given continental experiments~\cite{YuHuang},
and the distribution of heat producing elements is not precisely known.
Calculating the geo-neutrino contribution from the continental crust is done by
integrating data from geophysical~\cite{CrustErr, CrustErr2, CrustErr3} and geochemical~\cite{YuHuang}
surveys of continents, with detailed regional studies for the first 500 km surrounding the detector~\cite{McNew}
as this region typically contributes half of the total geo-neutrino signal.
Critical constraints on Earth models will come from precise measurements of the Earth's geoneutrino flux.

A natural nuclear fission reactor with a power output of 3-10 TW at the center of the earth
has been proposed as the energy source of the earth magnetic field, some introductions can be found, for example, in Ref.~\cite{GeoReactor}. Experimental result may will critically assess such kind of assumptions and set limits on its presumed power contribution.

Due to the exemplary location far away from nuclear power plants, Jinping is an ideal site to precisely measure the geo-neutrino flux and to probe geo-reactors.
Next we will introduce geo-neutrino signals in Section~\ref{sec:geosig}, the critical backgrounds from reactor neutrinos and other contributors in Section~\ref{sec:geobkg}, and the sensitivity to study geo-neutrinos at Jinping in Section~\ref{sec:geosen}. The sensitivity for geo-reactors will be briefly discussed in section~\ref{sec:georct}.

\subsection{Geo-neutrino signal}
\label{sec:geosig}
This section will discuss the spectrum and flux of the geo-neutrinos, together with the detection.

\subsubsection{Geo-neutrino spectrum and flux}
\begin{figure}[!h]
\centering
\includegraphics[angle=270, width=9cm]{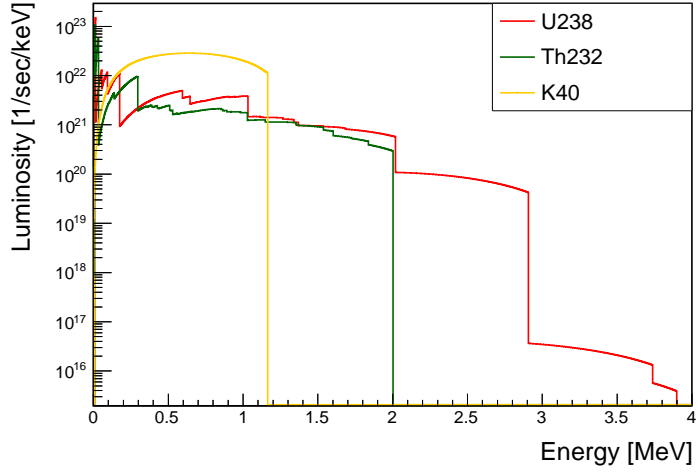}%
\caption{Geoneutrino spectrum.}
\label{fig:GeoNuSpec}
\end{figure}

Natural $\beta$-decays from $^{238}$U and $^{232}$Th families,
and $^{40}$K are believed to be the most heat producing isotopes. They heat the Earth's interior
through decays into stable nuclei as following
\begin{equation}
\begin{aligned}
^{238}_{92}\mbox{U} &\rightarrow^{206}_{82}\mbox{Pb} + 8\alpha
+ 6\beta^- + 6\bar{\nu}_e + 51.698~\mbox{MeV},\\
^{232}_{90}\mbox{Th} &\rightarrow^{208}_{82}\mbox{Pb} + 6\alpha
+ 4\beta^- + 4\bar{\nu}_e + 42.652~\rm{MeV},\\
^{40}_{19}\mbox{K} &\rightarrow^{40}_{20}\mbox{Ca} + \beta^-
+ \bar{\nu}_e + 1.311~\rm{MeV}~~~(\rm{BR} = 89.3\%),\\
^{40}_{19}\mbox{K}  + \beta^- &\rightarrow^{40}_{18}\mbox{Ar}
+ \nu_e + 1.505~\rm{MeV}~~~(\rm{BR} = 10.7\%).
\end{aligned}
\end{equation}
The predicted spectra of the antineutrinos~\cite{GeoSpec} are shown in Fig.~\ref{fig:GeoNuSpec}.
\vspace*{5mm}

It should be noted that the fluxes of geo-neutrinos depend on the geo-models and locations.
We use the prediction in Ref.~\cite{YuHuang} to test Jinping's sensitivity.
The mantle neutrino contribution assumes a modern mantle having 13 TW of power.
A chondritic mass Th/U ratio of 3.9 was fixed.
Since Jinping is near the Himalaya area,
the crust neutrino contribution at Jinping is assumed to be 1.5 times larger than that at  Kamioka.

\subsubsection{Geo-neutrino detection}
In principle, neutrinos can be detected via either the elastic scattering process or
the inverse beat decay (IBD) reaction. Owing to the low cross-section and the potential solar neutrino background, we do not expect the first process can be used at Jinping.
The electron antineutrinos will be detected using the IBD reaction chain~\cite{XSEC}
\begin{equation}
\bar{\nu}_e + p \rightarrow e^+ + n,\ \mbox{followed~by}~n + \mbox{H} \rightarrow d + \gamma~(2.2~\rm{MeV}).
\label{IBDProcess}
\end{equation}
Since the reaction has a threshold of 1.8 MeV, only the highest energy portion of the $^{238}$U and $^{232}$Th decay chains are measurable.
Depending on the detector type, the energy of electron antineutrino can be approximately calculated by either $T_{e^+}+1.8$ MeV in water or $T_{e^+}+0.78$ MeV in scintillator. Here, $T_{e^+}$ is
the detected energy of positron, and the tiny neutron recoil energy is neglected.
The 1.0 MeV difference in scintillator is due to the fact that
the detected energy of positron is actually the sum of the positron kinetic energy and
annihilation $\gamma$'s energy. Because of the soft geo-neutrino spectra and the high Cherenkov threshold,
the measurement on geo-neutrinos can only be performed in scintillator detectors or water-based scintillator detectors.

\subsection{Geo-neutrino backgrounds}
\label{sec:geobkg}
\subsubsection{Reactor antineutrino background}
Reactor electron antineutrinos can form an irreducible background to the detection of geo-neutrinos. The only way to reduce the reactor neutrino flux is to use the $1/r^2$ law and place the detector far away from nuclear power plants. Fortunately, the location of Jinping is at least 1,200 km away from nuclear power plants, and is therefore the best site for the geo-neutrino experiment among all the existing experiments.
Below are the details for the study of the reactor antineutrino background at Jinping.

\subsubsection{Differential neutrino flux of a single reactor}

\begin{figure}[h]
\centering
\includegraphics[width=9cm]{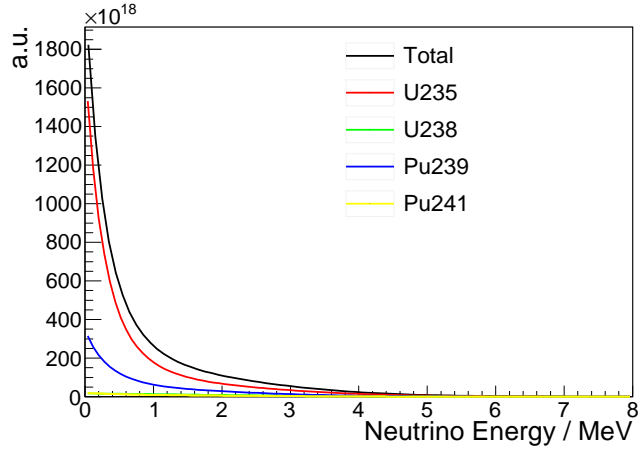}%
\caption{Reactor antineutrino spectrum for each isotope and their sum.}
\label{fig:component}
\end{figure}

Reactor antineutrinos are primarily from the beta decays of four main fissile nuclei $\mathrm{{}^{235}U}$, $\mathrm{{}^{238}U}$, $\mathrm{{}^{239}Pu}$, and $\mathrm{{}^{241}Pu}$. The differential $\bar{\nu}_e$ flux, $\phi(E_{\nu})$, for a reactor is estimated as
\begin{equation}
\phi(E_{\nu})=\frac{W_{th}}{\sum_{i}f_{i}e_{i}}\sum_{i}f_{i} S_{i}(E_{\nu}),
\label{equ:Neutrino_Flux}
\end{equation}
where $i$ sums over the four isotopes, $W_{th}$ is the thermal power of a reactor which can be found in the IAEA~\cite{industry, RctEst},
$f_i$ ($\sum_{i}f_i=1$) is the fission fraction of each isotope,
$e_i$ is the average energy released per fission of each isotope, and
$S_{i}(E_{\nu})$ is the antineutrino spectrum per fission of each isotope.
A set of typical fission fractions, $f_i$, and the average energy released
per fission, $e_i$, are listed in Table.~\ref{tab:EnergyReleased}.
The spectrum of each isotope, $S_{i}(E_{\nu})$, and their sum are shown in Fig.~\ref{fig:component}, respectively.

\begin{table}[h]
\begin{center}
\vspace*{0cm}
\begin{tabular}[c]{ccc} \hline\hline
Isotope               & $f_i$& $e_i$ [MeV/fission]   \\\hline
$\mathrm{{}^{235}U}$  & 0.58 & $202.36\pm0.26$     \\
$\mathrm{{}^{238}U}$  & 0.07 & $205.99\pm0.52$     \\
$\mathrm{{}^{239}Pu}$ & 0.30 & $211.12\pm0.34$     \\
$\mathrm{{}^{241}Pu}$ & 0.05 & $214.26\pm0.33$     \\\hline
\end{tabular}
\caption{Fission fraction and average released energy of each isotope.}
\label{tab:EnergyReleased}
\vspace*{-0.5cm}
\end{center}
\end{table}

\subsubsection{Total differential reactor neutrino flux}

\begin{figure}[h]
\vspace*{0.5cm}
\centering
\includegraphics[width=9cm]{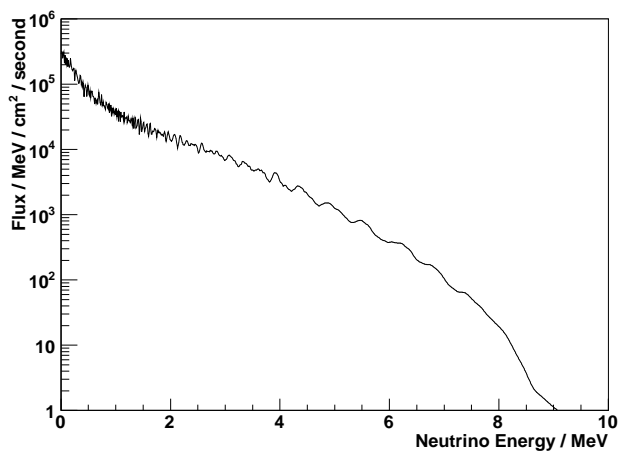}
\caption{Total differential reactor neutrino flux at Jinping.}
\label{fig:Reactor}
\end{figure}

To get the total reactor neutrino background spectrum at Jinping, $\phi_{Jinping}(E_{\nu})$, we used the thermal powers of all the currently running and under construction nuclear power plants  from the IAEA~\cite{industry}, and took into account the electron antineutrino survival probability. $\phi_{Jinping}(E_{\nu})$ is expressed as
\begin{equation}
\phi_{Jinping}(E_{\nu}) = \sum_i^{Reactors}\phi_i(E_{\nu}) P_{\bar\nu_e\to\bar\nu_e}(E_{\nu},L) \frac{1}{4\pi L^2},
\end{equation}
with
\begin{equation}
P_{\bar\nu_e\to\bar\nu_e}(E_{\nu},L) \approx1-\sin^22\theta_{12}\sin^2\left(1.267\frac{\Delta m_{21}^2(\text{eV})L(\text{km})}{E_\nu(\text{GeV})}\right),
\end{equation}
where $E_{\nu}$ is the neutrino energy, $L$ is the distance from each reactor to the Jinping site,
and $\theta_{12}$ and $\Delta m_{21}^2$ are neutrino oscillation parameters.
$L$ is calculated using the longitude and latitude coordinates for each nuclear power plant and Jinping site, and
$\theta_{12}$ and $\Delta m_{21}^2$ are set to be 0.586 and $7.58\times10^{-5}$ eV$^2$, respectively.
The expected total differential spectrum at Jinping
is shown in Fig.~\ref{fig:Reactor}, which also includes the contribution from those under construction.
Table~\ref{tab:JPFlux} lists the numerical results for the fluxes.
\begin{table}[!hbp]
\centering
\begin{tabular}{cccccc} \hline\hline
Jinping                  & \multicolumn{2}{c}{In operation} & \multicolumn{2}{c}{Under Construction} & Total \\
                         & China & Others & China & Others  &       \\\hline
$\phi_\nu/(10^5$cm$^2$s$^{-1}$) & 3.71  & 2.73   & 6.20  & 0.35    & 12.99 \\\hline
\end{tabular}
\caption{Reactor neutrino flux at Jinping.}
\label{tab:JPFlux}
\end{table}

\subsubsection{Other non-$\bar \nu_e$ backgrounds}
Other possible backgrounds are the cosmic-ray muon induced $^9$Li and $^8$He,
($\alpha$, n) background, and the accidental coincidence background.
According to the recent publication by Borexino~\cite{BxNew}, the
signal-to-non-$\bar \nu_e$-background ratio is $\sim$100,
so these backgrounds were ignored in this study.


\subsection{Fiducial mass for geo-neutrinos}
The detection process for $\bar \nu_e$ is the IBD process in Eq.~\ref{IBDProcess},
where the delayed-coincidence technique can be applied to identify the prompt and delay signal pair
and it can improve significantly the ratio of signal to background.
Assuming a 2,000-ton fiducial mass for the solar $\nu_e$ detection, a much larger volume for
$\bar \nu_e$ can be applied, which will give us 3,000-ton fiducial mass.

\subsection{Sensitivity for geo-neutrinos}
\label{sec:geosen}
The event rates of geo-neutrino signal, reactor neutrino background, and the sensitivity of observing geo-neutrinos and
determining the U/Th ratio are discussed in this section.

\subsubsection{Signal and background rates and spectra}
\begin{figure}[h]
\centering
\includegraphics[height=6cm, clip]{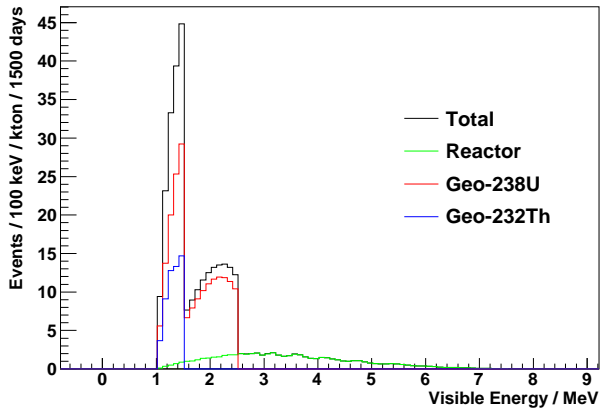}%
\caption{Geo-neutrino and reactor neutrino visible energy spectra at Jinping.}
\label{fig:CanRate}
\end{figure}

The detectable spectra from the IBD process can be calculated as
\begin{equation}
R_{Jinping}(E_{\nu}) = \phi_{Jinping}(E_{\nu}) \times \sigma(E_{\nu}).
\end{equation}
With a modest setup, \emph{i.e.} 1 kiloton fiducial volume and 1,500 days' data-taking, the differential spectra of geo-neutrino and reactor background are shown in Fig.~\ref{fig:CanRate}, and the event rates are summarized in Table~\ref{tab:CanRate}.
The signal to background ratio is rather promising.

\begin{table}[!h]
\centering
\begin{tabular}{ccccc}   \hline\hline
                         & \multicolumn{3}{c}{Geo-neutrinos} & Reactors \\
                         & $^{238}$U & $^{232}$Th  & Total   &         \\\hline
Rate/kiloton$/1,500$ day      & 184     & 45        & 229   &  64   \\\hline
\end{tabular}
\caption{Geo-neutrino and reactor neutrino event rates at Jinping.}
\label{tab:CanRate}
\end{table}

\subsubsection{Sensitivity of determining geo-neutrino signals}
\begin{figure}[!h]
\centering
\includegraphics[height=6cm, clip]{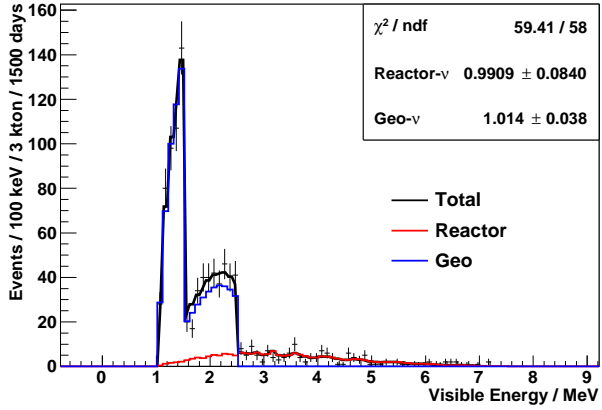}%
\caption{Likelihood fit result for both the geo-neutrino signals and background with the Th/U ratio fixed to 3.9.
We assume 3 kiloton of target mass and 1,500 days of data-taking in this study.
Numbers shown in the top-right corner are $\chi^2$/ndf and the ratios
of the fit result to the nominal value for the reactor background
and geo-neutrinos.
}
\label{fig:GeoFit1}
\end{figure}

\begin{figure}[!h]
\centering
\includegraphics[height=6cm, clip]{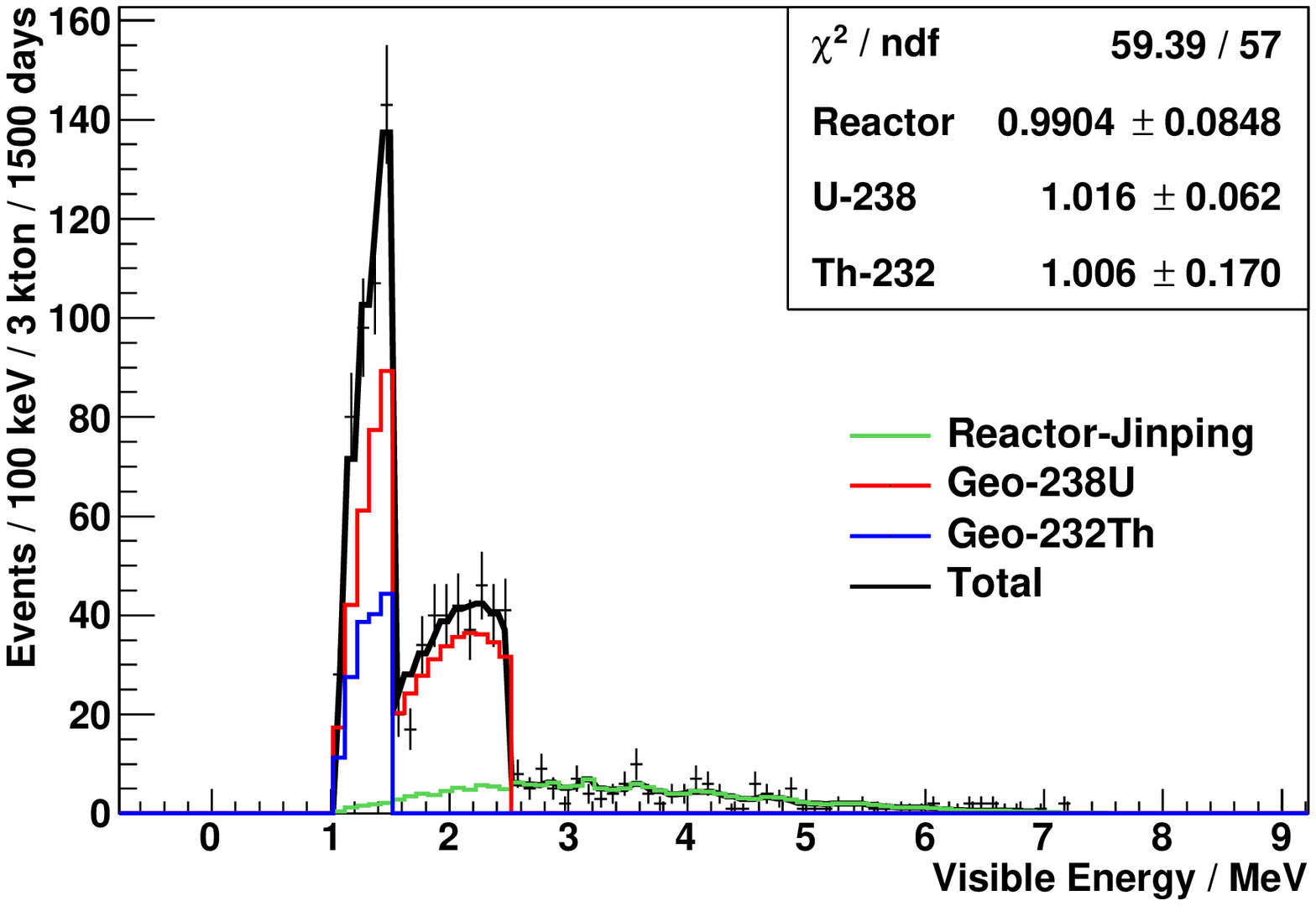}%
\caption{Likelihood fit result for both the geo-neutrino signals and background with the Th and U components free.
We assume 3 kiloton of target mass and 1,500 days of data-taking in this study.
Numbers shown in the top-right corner are $\chi^2$/ndf and the ratios
of the fit result to the nominal value for the reactor background,
$^{238}$U geo-neutrino, and $^{232}$Th geo-neutrino components.
}
\label{fig:GeoFit2}
\end{figure}

According to the true signal and background spectra in Fig.~\ref{fig:CanRate},
we randomly sampled the spectra and performed a likelihood fit with both signals and background
with 3 kiloton of target mass and 1,500 days of data-taking.
One example fit with Th/U ratio fixed to the known value is given in Fig.~\ref{fig:GeoFit1}.
The precision of the total geo-neutrino flux can be determined down to 4\%.
The other fit with the Th and U fractions free is shown in Fig.~\ref{fig:GeoFit2}.
The $^{238}$U fraction can be determined to 6\%, and
the $^{232}$Th fraction's precision can reach 17\%.

\subsubsection{U/Th ratio}
With 5,000 tests of random sampling and fitting,
we estimated the expected precision to determine the ratio of U to Th components, i.e. how well we can
determine the expected chondritic mass Th/U ratio of 3.9.
The result gives that the precision for the ratio of U/Th can be well decided to 10\%.

\subsubsection{Sensitivity to geo-neutrino models}
\begin{figure}[!h]
\centering
\includegraphics[height=8cm, clip]{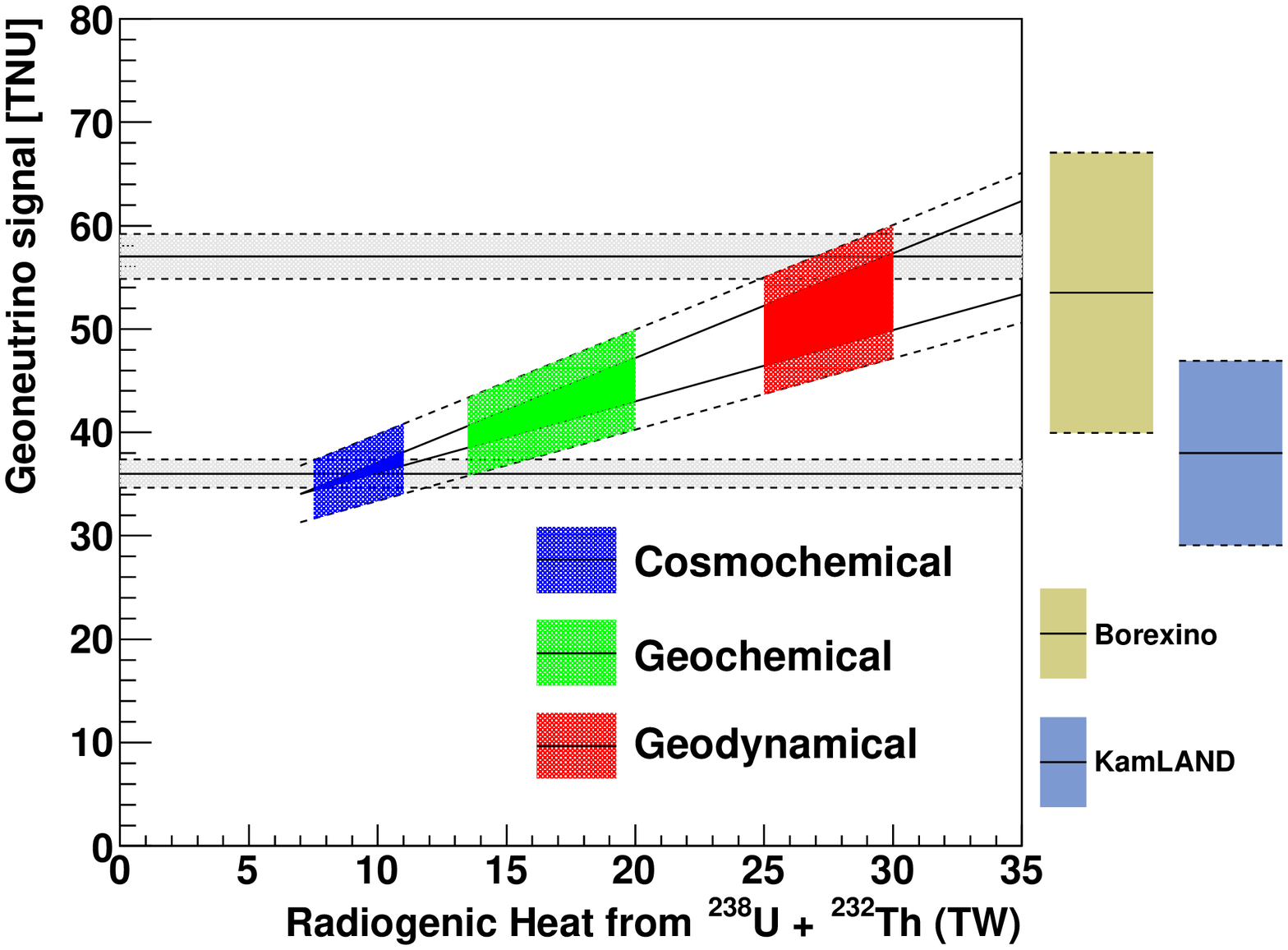}%
\caption{
Jinping's sensitivity in measuring the U and Th components of geo-neutrinos separatively (top).
The geo-neutrino sensitivity vs geo-neutrino model predictions at Jinping (bottom).
The three filled regions in the plot delimit, from the left to the right, the cosmochemical, geochemical, and geodynamical models, respectively.
On the right, the Borexino and KamLAND measurements are shifted up according to the differences of their crust geoneutrino fractions with Jinping.
The two horizontal bars are plotted at two possible geo-neutrino flux assumptions with Jinping geo-neutrino measurement sensitivity.
}
\label{fig:TNUcompare}
\end{figure}

With the 4\% precision in determining the total geo-neutrino flux,
the result can be compared with the model predictions.
Shown in Fig.~\ref{fig:TNUcompare} are two possible outputs with uncertainties
overlaid on the model predictions of geo-neutrino flux as a function of heat production.
The prediction has mainly two uncertainties, one is 10\% uncertainty of
the crust neutrino flux, and the other one is the distribution of the mantle neutrinos,
which could be uniform in the mantle or in the extreme case only concentrate around the boarder of
the mantle and the core.
With the expected improvement of the geo-neutrino flux, it is possible to
accept or reject some geo-neutrino predictions.
But, certainly, a more precise geology survey of the near-by crust,
and, if possible, a better understanding of the mantle neutrino distribution are necessary
to decrease the prediction uncertainty.

\subsection{Sensitivity for geo-reactor}
\label{sec:georct}
The geo-reactor neutrinos were supposed to have a similar
energy distribution as artificial reactors.
The number of events over 2.9 MeV was counted.
According to Poisson statistics, a 95\% upper limit
was deduced for the total power of the geo-reactor in the Earth core:
\begin{equation}
P_{reactor} < 2.3\ \rm{TW}~(95\%~\rm{CL}),
\end{equation}
which can conclusively confirm or exclude the 3-10 TW georeactor proposal.

\subsection{Discussion and summary}
With the ideal location of Jinping, the reactor neutrino background can be suppressed significantly.
As discussed in Section~\ref{sec:geosen},
the expected total number of geo-neutrino candidates can reach over 600 for a 3 kiloton target mass with 1,500-liveday exposure assumption,
while the number of background from nuclear power plants is evaluated to be less than 100 below 2.9 MeV.
The precision of the flux of the geo-neutrino will be determined down to 4\% and the ratio of Th/U will be determined better than 10\%.
As shown in Section~\ref{sec:georct},
a conclusive confirmation or rejection of geo-reactor with 3-10 TW can be made.

%% file: SRN.tex
\newpage
\section{Supernova relic neutrino}

Supernova relic neutrinos (SRN's), also known as Diffuse Supernova Neutrino Background (DSNB), is highly interesting for neutrino astronomy and neutrino physics.
When a massive star ($>8M_{\odot}$) is on its way to the end, it collapses into a high-density and -pressure core, and the core has a possibility to become a supernova.
An enormous amount of neutrinos are emitted when the supernova explodes and carry away almost 99\% of its gravitational energy.
The chance to detect supernova burst neutrinos is rather rare, probably once per  century~\cite{1p1century}, and the neutrinos from SN1987A so far are the only ones being recorded.
However neutrinos emitted from the past core-collapse supernovae accumulated and formed a continuum diffused background, and the chance to discover supernova relic neutrinos is
relatively higher.
An observation of supernova relic neutrinos will reveal the process of stellar evolution and the history of our universe,
and it is a unique tool for astronomy research. Experimental searches have been carried on by Super Kaminokande~\cite{SK0-srn, SK1-srn, SK2-srn},
KamLAND~\cite{KamLAND-srn}, Borexino~\cite{Borexino-srn}, and SNO~\cite{SNO-srn}. However, no SRN signal has been found yet. After the discovery of solar neutrinos,
supernova relic neutrinos as another extraterrestrial neutrinos are attracting more and more attention.
In this section, we will estimate the future prospect of Jinping underground experiment.

\subsection{SRN Models}
The SRN spectrum can be predicted according to the following formula~\cite{JBeacom}:
\begin{equation}
\DF{d\phi}{dE_{\nu}} (E_{\nu}) = \int_0^{\infty} [(1+z) \varphi [E_{\nu}(1+z)]] [R_{SN}(z)] \left[ \left| \DF{c\ dt}{dz} dz \right| \right ],
\end{equation}
where the first term within the integral is an average supernova neutrino spectrum $\varphi [E_{\nu}]$ and then appropriately redshifted,
the second term, $R_{SN}(z)$, is a core-collapse rate density as a function of redshift $z$, and
the last term is the known cosmological line-of-sight factor.

\begin{figure}[!h]
\centering
\includegraphics[width=8cm]{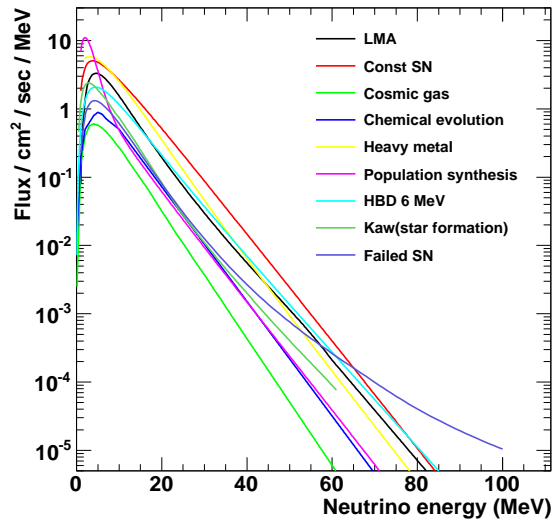}
\caption{(Color online) Model dependence of $\bar\nu_e$ energy spectra for supernova relic neutrinos}
\label{fig:SRNflux}
\end{figure}

Many models have been proposed to predict the SRN flux and spectrum.
In this analysis we will compare the following model predictions with the sensitivity of Jinping:
LMA\cite{LMASRN}, Constant SN\cite{ConstantSN}, Cosmic gas\cite{CosmicGas}, Chemical evolution\cite{ChemicalEvo}, Heavy metal\cite{HeavyMetal1}\cite{HeavyMetal2},
Population synthesis\cite{Population}, HBD 6 MeV\cite{HBD6}, Star formation rate\cite{StarForm}, and Failed SN\cite{FailedSN}. The most interesting $\bar\nu_e$
energy spectra from the above models are shown in Fig.~\ref{fig:SRNflux}. The average energy of supernova relic neutrinos is highly red-shifted from ~20 MeV
to below ~5 MeV for all models, and the total flux predictions among the models are within about an order of magnitude.

\subsection{Detection}
There are three flavors of neutrinos and anti-neutrinos in SRN, among which $\bar\nu_e$'s are mostly likely to be detected because of the large cross section of inverse
beta decay (IBD), $\bar\nu_e+p \rightarrow e^+ + n$,  in hydrogen (free proton) rich material within the energy region of several tens of MeV and the powerful rejection with delayed coincidence with neutron capture.

Liquid scintillator and water-based liquid scintillator (WbLS)~\cite{WaterLS} were studied as the detecting material,
since both have a high efficiency for tagging the delayed neutron
to reject accidental backgrounds~\cite{KamLAND-srn, Borexino-srn, SK2-srn}.
Gadolinium loaded water technique~\cite{WaterGd} was temporarily skipped in this study.

In liquid scintillator detectors, the high $dE/dx$ deposited by protons and alphas makes it possible to be differentiated from beta and gamma by a pulse shape analysis~\cite{LENApaper}.

In WbLS, low energy protons, alphas, and muons in the interesting energy region cannot produce Cherenkov light
in the search of SRN.
If WbLS can be used to distinguish Cherenkov and scintillation light, it will be a more powerful tool to suppress backgrounds.

\begin{figure}[h]
\centering
\includegraphics[width=8cm]{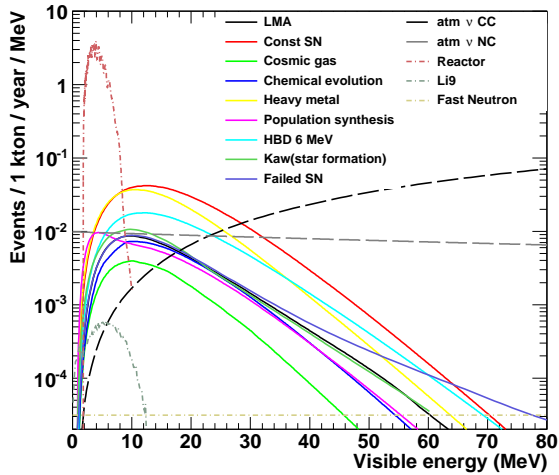}
\caption{(Color online) Visible energy spectra of supernova relic neutrinos for different models and visible energy spectra of all the possible backgrounds.}
\label{fig:SRNvis}
\end{figure}

The SRN event rate to be detected via the IBD process can be calculated by
\begin{equation}
\frac{dR}{dE_{\nu}}=\frac{d\phi}{dE_{\nu}}\times \sigma(E_{\nu})\times N_p\times T,
\end{equation}
where $\sigma(E_{\nu})$ represents the differential
IBD cross section~\cite{IBD-XSEC}, $N_p$ is the number of free protons in the target, and $T$ is the data-taking time.
The threshold of the reaction is 1.8 MeV, and the kinetic energy of the neutrino is almost  transferred to the positron.
The visible energy of the reaction is the kinetic energy of the positron plus two annihilation photons and can be related to the neutrino energy by
\begin{equation}
E_{vis} = E_\nu - 0.78\ \rm{MeV}.
\end{equation}
The differential visible energy spectra per kiloton detector per year is shown in Fig. \ref{fig:SRNvis}, and the expected event rates in the most interested $E_{vis}$ region of 10 - 30 MeV are listed in Table~\ref{tab:SRN}.

\subsection{Fiducial mass for SRN detection}
The target mass for the IBD signal will be about 3 kiloton, which is much larger than that of the solar signal search due to the background rejection ability of the delayed coincidence technique as discussed for the geo-neutrinos detection.

\subsection{Backgrounds}

There are mainly six types of backgrounds for the SRN detection: 1) accidental coincidence; 2) reactor $\bar\nu_e$;
3) fast neutrons induced by energetic cosmic-ray muons;
4) $^9$Li or $^8$He radioactive isotopes induced by energetic cosmic-ray muons; 5) atmospheric neutrino background through a charge current (CC) process; and 6) atmospheric neutrino background through a neutral current (NC) process.
The estimation of the background rates are explained below.

\subsubsection{Accidental coincidence}
Accidental coincidence background was considered to be negligible given the ideal location and cleanness of the neutrino detector.

\subsubsection{Reactor $\bar\nu_e$}
Reactor $\bar\nu_e$ events are identical to the SRN signal except for the energy spectrum, which
is below 10 MeV as shown in Fig.~\ref{fig:SRNvis}. A requirement on energy above 10 MeV together with the fact that there is no nuclear power plant in close proximity will significantly suppress the reactor $\bar\nu_e$ background down to a negligible level.

\subsubsection{Fast neutron}
Cosmic-ray muon or its secondary products may collide with the nuclei in target or surrounding materials, and knock out energetic fast neutrons. The recoiled proton by the fast neutron can minic a prompt positron signal, while the scattered neutron is quickly thermalized and captured by a nucleus, forming a delayed neutron signal.

The fast neutron background was scaled from KamLAND measurement $(3.2\pm3.2)$/(4.53 kton-year) = $(0.7\pm0.7)/$kton-year~\cite{KamLAND-srn} to the Jinping site. The muon rate at Jinping is about 1,000 times lower than KamLAND, as shown in Fig.~\ref{fig:MuFlux}. As a result, the fast neutron background is reduced to a rather low level, which is estimated to be $(0.7\pm0.7)\times10^{-3}$/kton-year for those with visible energy from 10 to 30 MeV.

Some minor factors may be considered in order to give  a more precise estimation.
Usually a neutrino detector is perfect for giving a muon self-veto. Muons passing surrounding insensitive materials, i.e.\ rock, are the main cause for fast neutrons to be problematic. A more careful consideration should be done to scale the background level according to the surface area ratio of two detectors. Since the KamLAND detector mass, 1 kiloton, may be close to the Jinping expectation, this is a minor correction.
Secondly, the fast neutron yield increases as an exponential function of muon energy, $yield=(E_{Jinping}/E_{KamLAND})^{0.77}$~\cite{SpaPowerlaw}. With the 351 GeV of average muon energy at Jinping and 260 GeV at KamLAND, the increase is insignificant.
The shape of the prompt signal for the fast neutron background has been understood~\cite{DYBnGd, DYBnH} and is basically a flat distribution below 100 MeV as shown in Fig.~\ref{fig:SRNvis}.

\subsubsection{Spallation product $^9$Li/$^8$He}
The spallation products $^9$Li/$^8$He induced by cosmic-ray muons can decay via a beta delayed neutron emission. The beta signal together with the delayed neutron signal have a similar signature as the IBD signal.
The half-life and Q-value of $^9$Li are 173 ms and 14 MeV, respectively, and 119 ms and 11 MeV for $^8$He, respectively.
The visible energy spectra of these two background sources are shown in Fig.~\ref{fig:SRNvis}.
It was noted that $^9$Li/$^8$He backgrounds were usually generated locally and could not fly very far,
so that the muons concerned were in the sensitive target region.
With an efficient muon detection and a long enough time veto, these backgrounds can be effectively removed.

With a suppression factor of 1,000 from the large overburden, the KamLAND measurement is scaled to the Jinping case, 4.0/4.53~(kton-year)/1,000 = $1\times10^{-3}$/kton-year, for the visible energy within the range from 10 to 30 MeV.

\subsubsection{Atmospheric neutrino background through the charged current}
In the energy region interested for the SRN search, another anti-neutrino background source is atmosphere neutrinos\cite{AtmNu1, AtmNu2}.
The atmospheric $\bar\nu_e$ background through the charged current interaction (CC) is  irreducible, which, however, is not dominant.

High energy atmospheric $\bar\nu_{\mu}$'s and $\nu_{\mu}$'s can produce low-energy muons and delayed neutrons,
and thus contaminate the IBD signal region.
The muons can decay to Michel electrons, with a triple coincidence tagging of the prompt event, muon decay, and neutron capture, This muon background can be rejected in theory.
However, for a water Cherenkov detector, the muon momentum can be below the Cherenkov light production threshold,
and become an invisible muon background.
This problem can be resolved in liquid scintillator and WbLS detectors.
Because of the inefficiency of the triple tagging and a small fraction of negative muon capture untagged, the CC background induced by $\bar\nu_{\mu}$ and $\nu_{\mu}$ is dominant.

The rate estimation in KamLAND, 0.9/(4.53~kton-year)~=~0.2/kton-year,
can be applied to Jinping.
With the WbLS technique, it was assumed to be 0.1/kton-year with the extra capability of particle identification for both prompt electron and delayed gamma signals.

\subsubsection{Atmospheric neutrino background through the neutral current}
The dominant background for SRN's is from the neutral current process (NC) of atmospheric neutrinos as studied in the KamLAND experiment.
Neutrinos at higher energies beyond the signal region may collide with $^{12}$C in the target and knock out a neutron.
The neutron scattering off protons or particles emitted in the de-excitation of the remaining
nuclei causes a prompt signal and the neutron capture will give a delayed signal.

Some new analysis techniques may help to further suppress this background~\cite{C11thesis, LENApaper}.
To have a delay neutron produced, 2/3 of the chance a $^{11}$C at ground state is produced. The half-life and Q-value of the $^{11}$C ground state are 20 min and 2 MeV, respectively. A triple tagging of the prompt signal, neutron capture, and $^{11}$C decay may help to veto this background, but 5\% of the background still survive. For the other 1/3 cases, $^{11}$C is at its excited states and decays through neutron, proton or alpha emissions. A pulse shape discrimination technique can also be applied to suppress the background probability down to 1\% level.

It was expected that the background level  16.4/(4.53~kton-year)~=~3.6/kton-year at KamLAND could be suppressed coarsely by a factor of 16=1/(5\%+1\%) at Jinping with a background level similar to the CC background 0.2/kton-year.
We expect that in a WbLS detector the above level can be further suppressed down to  0.1/kton-year with the extra capability of particle identification.

\subsection{SRN Detection Sensitivity and summary}
The events rate in $E_{vis}$ range around 10 - 30 MeV are summarized in Table~\ref{tab:SRN},
and the results with 3 kton $\times$ 3.3 years and 3 kton $\times$ 6.7 years are also shown.
With a WbLS detector, a neutrino experiment at Jinping is very promising
to make a discovery.

\begin{table}[h]
\centering
\small
\begin{tabular}{llll} \hline\hline
Expected event rate      & 1/kton-year & 1/(3 kton $\times$ 3.3 years) & 1/(3 kton $\times$ 6.7 years) \\ \hline
Signal                   & 0.05 - 0.66        & 0.5 - 6.6         & 1 - 13           \\ \hline
Accidental coincidence   & 0 & 0 & 0 \\
Reactor background       & 0 & 0 & 0 \\
Fast neutron             & $0.7\times10^{-3}$ & $7\times10^{-3}$  & $14\times10^{-3}$  \\
Spallation $^9$Li/$^8$He & $1\times10^{-3}$   & $10\times10^{-3}$ & $20\times10^{-3}$  \\
Atmospheric CC            & 0.2 (0.1)          & 2 (1)             & 4 (2)              \\
Atmospheric NC            & 0.2 (0.1)          & 2 (1)             & 4 (2)              \\
Total background         & 0.4 (0.2)          & 4 (2)             & 8 (4)              \\
\hline
\end{tabular}
\caption{Event rates for the supernova relic neutrinos and the corresponding backgrounds within $E_{vis}$ in range around 10 - 30 MeV.
For the signal, the range of several models' predictions is printed.
Background rates are calculated assuming a liquid scintillator target, and the
atmospheric CC and NC background rates in parentheses are the results with a WbLS target.}
\label{tab:SRN}
\end{table}

%% file: SNBurst.tex
\section{Supernova burst neutrino}

\subsection{Introduction}
On 1987 February 23, about two dozen supernova (SN) burst neutrinos were observed in the Kamiokande II, IMB, and Baksan experiments from the stellar collapse SN 1987A, resulting from the star Sanduleak -69202 exploded in the Large Magellanic Cloud, about 50 kpc away from the Earth~\cite{Hirata1,Hirata2,Bionta,Bratton,Alekseev1,Alekseev2}. This was the first observation of a supernova neutrino burst and SN 1987A remains the only known astrophysical neutrino source since then except for the Sun. SN burst neutrinos carry away almost all of the gravitational binding energy of a stellar collapse, which are important in studying the core-collapse supernova (ccSN) mechanism~\cite{Raffelt}. The SN neutrinos can also provide a large range of physical limits on neutrino properties~\cite{book,Lunardini,Masshierarchy}. Since ccSN explosions are likely strong galactic sources of gravitational waves, joint observations of both SN burst neutrinos and gravitational waves could provide deep insight into ccSN explosions as well as other fundamental physics~\cite{Gravwave}.

The detection of SN burst neutrinos is so important, however, galactic SN explosions occur with a rate of only a few per century~\cite{Rate}, which makes the detection a once-in-a-lifetime opportunity. SN neutrinos are expected to arrive at the Earth a few hours before the visual SN explosion, which enables a precious early warning for a SN observation~\cite{Raffelt}. The Supernova Early Warning System (SNEWS)~\cite{snews,SNEWSpaper} collaborates with experiments sensitive to ccSN neutrinos, to provide the astronomical community with a very high-confidence early warning of a SN occurrence, pointing more powerful telescopes or facilities to the event.

Here, the ccSN model is of the SN 1987A-type, of which all features are compatible with SN 1987A. The SN burst neutrinos has three main phases, which are prompt $\nu_e$ burst, accretion, and cooling, respectively~\cite{Raffelt}. The duration of 10 seconds covers 99\% of the luminosity carried off by all flavors of neutrinos and antineutrinos in a SN explosion. The energy spectrum of SN burst neutrinos follows a quasithermal distribution~\cite{SNnuspec},
\begin{equation}\label{spectrum}
  f_{\nu}(E)\propto E^{\alpha}e^{-(\alpha+1)E/E_{av}},
\end{equation}
where $E_{av}$ is the average energy and $\alpha$ a parameter describing the amount of spectral pinching. In this study, $E_{av}$ is set to be 12.28~MeV and $\alpha$ to be 2.61, which correspond to the cooling phase for SN burst $\bar{\nu}_e$'s and thus we choose a 10-second window searching for SN burst neutrinos.

The SN burst neutrinos are emitted in the few-tens-of-MeV range, and the detected neutrinos are dominated by IBD events~\cite{SNdetection} in a liquid scintillator detector with a fraction of about 90\%. The coincidence of IBD prompt signal from the positron (a 0.78-MeV downward shift of neutrino energy in general) with the delayed gamma emission ($\sim$2.2 MeV) of the IBD neutron capture on H provides a clear $\bar{\nu}_e$ signature against the uncorrelated backgrounds. Based on the chemical decomposition, the IBD cross section, and SN burst neutrino flux~\cite{Raffelt}, the expected number of SN burst neutrinos can be determined by
\begin{equation}
\label{SNevent}
  N = N_0\times\frac{L_{\bar{\nu}_e}}{5\times10^{52}\rm{erg}}\times(\frac{10\rm{kpc}}{D})^2\times(\frac{TM}{1\rm{kt}}),
\end{equation}
where $N_0$ corresponds to the expected number ($\sim$300) of SN burst neutrinos at a distance ($D$) of 10 kpc and a target mass ($TM$) of 1 kiloton. Generally, the luminosity ($L$) emitted is fixed for the study, which may vary with models.

Note that the total neutrino flux could be measured by using elastic neutrino - proton scattering as proposed in~\cite{nuProton}. For that the quenching of protons in LAB has to be known~\cite{protonQuenching}, which was measured recently. Hence a new opportunity could also be available.

\subsection{Supernova Trigger at Jinping}
The China Jinping Underground Laboratory is the deepest laboratory across the world with quite low cosmogenically backgrounds. Therefore, the signal-to-background ratio of SN burst neutrinos can be sufficiently high throughout a large range of distances from the Earth. A supernova trigger system can be designed and implemented with one or several detectors to be spatially built (due to the deep rock cover), online looking for any increase of IBD signals within a sliding 10-second window. The experience and techniques could be referred to from the supernova trigger system at Daya Bay~\cite{DYBSNtrigger}. The Jinping experiment is also aiming to be part of SNEWS in the future.

For the IBD selection in a liquid scintillator detector at Jinping, a prompt signal energy cut for SN burst neutrinos is assumed to be 10-50 MeV. Notice that the limit can be even lower due to the low background rates, increasing the selection efficiency of SN burst neutrinos and covering more SN models with soft neutrino energy spectra. Here the energy window 10-50~MeV is a conservative option to demonstrate the sensitivity (detection probability) of the proposed supernova trigger at Jinping. Based on the energy spectrum in Eq.~\ref{spectrum}, the selection efficiency of prompt energy cut is $\sim$88\%. The other selection criteria are the same as those in the nH analysis of Daya Bay $\theta_{13}$ measurement~\cite{DYBnH-SNB}, which included a 3$\sigma$ delayed energy cut for the 2.2-MeV gamma peak from neutron capture on H, a 1-400 $\mu$s prompt-delayed time coincidence cut, and a 500-mm prompt-delayed vertex distance cut. The product of the efficiencies of these selection criteria can be obtained from~\cite{DYBnH-SNB, DYBSNtrigger}. As a result, the final selection efficiency of SN burst neutrinos is $\sim$50\%. This selection efficiency will be considered in Eq.~\ref{SNevent} to correct the number of expected SN burst neutrinos when we calculate the detection probability of the supernova trigger (defined as the probability that a supernova neutrino burst will trigger or be detected).

Within a sliding 10-second window, the supernova trigger is determined from the IBD event distribution among detectors. As the background rate is estimated to be $<$1/yr based on our selection criteria, the expected number of backgrounds is rationally assumed to be ZERO within a 10-second window. Considering electronic noise or unexpected backgrounds, the supernova trigger is issued when $\geq$2 IBD signals are observed in one detector or two detectors within 10-second window. For two detectors, this trigger strategy means the case with no IBD signal or just 1 IBD signals in one of the two detectors ignored.
The detection probability of the supernova trigger will be demonstrated below in two scenarios of two 1.5-kiloton liquid scintillator detectors (total 3 kiloton as the default option) and one 1.5-kiloton liquid scintillator detector.

\begin{figure}[htb]
\centering
\includegraphics[width=12cm]{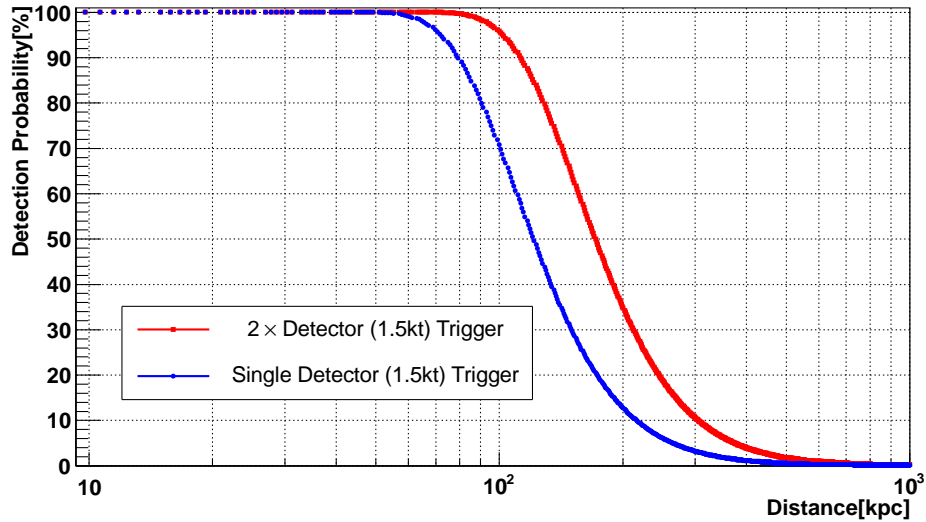}
\caption{The red curve corresponds to two 1.5-kt liquid scintillator detectors and the blue curve corresponds to a 1.5-kt liquid scintillator detector.}
\label{fig:detectionprobability}
\end{figure}

With the expected number of SN burst neutrinos corrected with the selection efficiency and the trigger strategy mentioned above, assuming the number of events follows a Poisson distribution in one detector and different detectors are mutually independent, the detection probability is shown in Fig.~\ref{fig:detectionprobability} as a function of distance to the Earth.
Notice that the most distance edge of the Milky Way is just 23.5 kpc from the Earth and SN 1987A exploded at a distance of 50 kpc. The larger distances will cover more SN explosions.

\subsection{Discussions}
From the description of the supernova trigger at Jinping above, the selection criteria of IBD events could be optimized for the prompt energy cut and even other selection criteria due to the quite low background rate. Besides, the trigger strategy may be kind of conservative at present. Thus, the realistic detection probability of the supernova trigger at Jinping may be 100\% throughout 100 kpc distance while Fig.~\ref{fig:detectionprobability} shows a greater than 95\% detection probability within 100 kpc in the scenario of two 1.5-kt detectors. Notice that the Super-K has 100\% detection probability of the supernova neutrino bursts out to 100 kpc which is the same level as our proposed supernova trigger at Jinping.

Due to the deep rock cover, small detectors may only be allowed to distribute in Jinping comparing to the size of the Super Kamiokande detector. However, the coincidence detection from multiple detectors is robust against spatially uncorrelated backgrounds, which enables a better detection probability than a single detector. The experience and techniques can be referred to the supernova trigger at Daya Bay~\cite{DYBSNtrigger}.

Offline analysis of the supernova neutrino bursts will also benefit from the low background rate and multiple detectors to be spatially distributed. And a smaller total target mass may provide a quite good limit of the rate of supernova neutrino bursts with sufficient live time.

A water-based liquid scintillator~\cite{NewLS} was in R\&D recently in which scintillating organic molecules and water are co-mixed using surfactants. This WbLS study has made a great progress and the extensive development of the WbLS chemical cocktail is ongoing at BNL. The pure water detectors are primarily sensitive to high energy interactions creating particles above the Cherenkov threshold. The pure scintillator detectors (isotropy without directionality) are sensitive to low-energy events. With the WbLS, the pointing of the SN neutrino bursts may be achieved by identifying the direction of Cherenkov light from the neutrino-electron scattering interactions like Super-K which is the only experiment with pointing capability at present.


%% file: DM.tex
\section{Dark Matter}
\newcommand{\be}{\begin{equation}}
\newcommand{\ee}{\end{equation}}
\newcommand{\bea}{\begin{eqnarray}}
\newcommand{\eea}{\end{eqnarray}}

\newcommand{\dq}{{\mathbf q} }

\subsection{Neutrinos from dark matter annihilation in the halo}

Neutrinos can be copiously produced via dark matter (DM) annihilation or decay in the
Galactic DM halo.
For annihilating DM,
the resulting neutrino energy spectrum can be a delta function if
neutrino is the direct final state of DM annihilation, i.e.\ $\chi\chi\to\nu\nu$;
it can also be a continuous energy spectrum, if DM annihilates into standard
model fermions which subsequently decay producing neutrinos.
In this section, we will focus on the case where the neutrino spectrum is a delta function,
since the mono-energetic neutrinos are readily distinguished from the background.
The differential flux of the anti-electron-neutrino in the $\chi\chi\to\nu\nu$ case is given by
(neglecting neutrino oscillations)
\be
{d \phi_{\bar\nu_e}(E_{\bar\nu_e}=m_\chi,\psi) \over d\Omega}
= {1 \over 2} {\langle \sigma_{\chi\chi\to\nu\nu} v  \rangle \over 4 \pi m_\chi^2} {1\over 3}
\int_\text{los} dx \rho_\chi^2(r(x,\psi))
\equiv  {\langle \sigma_{\chi\chi\to\nu\nu} v  \rangle \over 24 \pi m_\chi^2} J(\psi),
\ee
where the factor $(1/3)$ averages over the three flavors,
the factor $(1/2)$ pertains to identical DM particle,
 the integral is carried out along the line of sight (los),
$\rho_\chi(m_\chi)$ is DM density (mass),
$\psi$ is the angle away from the galactic center (GC),
$\Omega$ indicates the direction of DM annihilation,
$r(x,\psi)=(x^2+R_\odot^2-2xR_\odot \cos(\psi))^{1/2}$
is the distance to the GC,
$R_\odot=8.5$ kpc is the distance from the GC to the solar system,
$x$ is the distance between us and the location of DM annihilation.
For analysis without directional information, one can obtain the total anti-electron-neutrino flux
\be
\phi_{\bar\nu_e}(E_{\bar\nu_e}=m_\chi)
= {\langle \sigma_{\chi\chi\to\nu\nu} v  \rangle \over 24 \pi m_\chi^2} \int d\Omega J(\psi)
\equiv {\langle \sigma_{\chi\chi\to\nu\nu} v  \rangle \over 6 m_\chi^2}  J_\text{avg}
\ee
where $J_\text{avg}$ is the averaged $J$ factor over the whole sky, which
has a rather weak dependence
on the details of the dark matter density distribution in the halo, for some
commonly used dark matter profiles:
Navarro-Frenk-White (NFW) \cite{Navarro:1995iw},
Moore \cite{Moore:1999gc},
and Kravtsov \cite{Kravtsov:1997dp}.
We take the value $J_\text{avg}/(R_\odot \rho_\odot^2)=5$ \cite{Yuksel:2007ac},
assuming $\rho_\odot=0.4$ GeV/cm$^3$.
Thus the anti-electron-neutrino flux at $E_{\bar\nu_e}=m_\chi$,
is given by
\be
\phi_{\bar\nu_e} (E_{\bar\nu_e} = m_\chi) \simeq 1.1 \times 10^2\, \text{cm}^{-2}\, \text{s}^{-1}
\cdot {\text{MeV}^2 \over m_\chi^2}
\cdot {\langle \sigma_{\chi\chi\to\nu\nu} v \rangle \over 3 \times 10^{-26}\, \text{cm}^3\, \text{s}^{-1} }
\label{eq:neutrino_flux}
\ee

The monoenergetic feature of the neutrinos due to dark matter annihilation considered here, makes them
quite easy to be detected over the continuous backgrounds. The number of events due to the
anti-electron-neutrino are  given by \cite{Wurm:2011zn}
\be
{\cal N} \simeq\, \sigma_\text{det}\, \phi_{\bar\nu_e}\, N_\text{target}\, t\, \epsilon
\ee
where the detection cross section $\sigma_\text{det}$ needs to evaluated at
$E_{\bar\nu_e} = m_\chi$ for dark matter annihilation, the total neutrino flux $\phi_{\bar\nu_e}$
is given in Eq.\ (\ref{eq:neutrino_flux}), $N_\text{target}$ is the number of
target particles in the detector, $t$ is the total time-exposure,
and $\epsilon$ is the detector efficiency.

\begin{figure}[tbhp]
\vspace{0.5cm}
\centering
\includegraphics[width=0.5\textwidth]{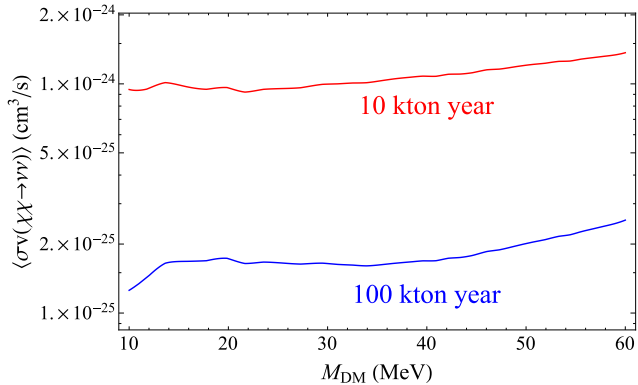}
\caption{The discovery reach of the DM annihilation cross section for MeV mass range.
The limits is derived with criteria $S=5\sqrt{B}$ or 10 events, whichever is larger.
Three backgrounds: supernova relic neutrino, atmospheric neutrino (both CC and NC)
are considered. We assume 100\% detection efficiency here.
}
\label{fig:neutrino_dm}
\end{figure}

The neutrino experiment at CJPL can search for neutrinos in the energy range
$E\sim(1-100)$ MeV. The dominant backgrounds come from reactor neutrino,
supernova relic neutrino, and atmospheric neutrinos in the energy range of interest.
For simplicity, we only consider DM mass above $\sim 10$ MeV,
to avoid the reactor neutrino background.
Electron antineutrinos can be detected via the inverse beta decay process $\bar{\nu}_e+p\to e^{+} + n$.
The energy resolution can be estimated as $\delta E / E = 8\%$.
The signal events due to DM in the energy bin around the DM mass with bin width
equal to twice energy resolution is computed, which is then compared to the background events
to derive the discovery limits.
As shown in Fig.\ (\ref{fig:neutrino_dm}), one can probe the DM annihilation
cross section to $\sim 10^{-24}$ ( $10^{-25}$) cm$^3$/s with 10 (100) kton-year
exposure.
Current exclusion limits on DM
annihilation into neutrinos are given by  KamLAND \cite{Collaboration:2011jza}
and Super-Kamiokande \cite{PalomaresRuiz:2007eu}.

\subsection{Neutrinos from dark matter annihilation in the Sun}

Another promising signal for indirect detection of dark matter is to look
for energetic neutrinos from annihilation of dark matter that have
accumulated in the Sun and/or Earth
(for early discussions, see e.g.\ \cite{Silk:1985ax} \cite{Krauss:1985ks}
\cite{Freese:1985qw} \cite{Krauss:1985aaa} \cite{Gaisser:1986ha}).
When the solar system moves through the dark matter halo, a dark matter
particle can scatter off a nucleus in the Sun or Earth and lose its velocity to
be lower than the escape velocity, and thus becomes gravitationally trapped.
The dark matter particle undergoes various scatterings in the Sun and eventually
settles to the core, after the capture. Over the lifetime of the Sun, a sufficient
amount of dark matter can accumulate in the core, so the equilibrium between
capture and annihilation (or evaporation) is expected. Unlike other standard
model particles, the neutrinos produced via dark matter annihilation can escape
easily from the Sun and can be detected in neutrino experiments on Earth.
The number of dark matter inside the Sun, $N_\chi$, is described by the differential equation
\be
{dN_\chi \over dt} = C_C - C_A N_\chi^2 -C_E N_\chi
\ee
where the three constants describe capture ($C_C$), annihilation ($C_A$), and
evaporation ($C_E$). For dark matter greater than the evaporation mass (which
is typically 3-4 GeV \cite{Griest:1986yu} \cite{Gould:1987ju}), the $C_E$ term can be
ignored. The dark matter annihilation rate is given by \cite{Wikstrom:2009kw}
\be
\Gamma_A \equiv {1\over 2 C_A N_\chi^2} = {1\over 2} C_C \tanh^2(t/\tau)
\ee
where $\tau\equiv 1/\sqrt{C_CC_A}$. The present dark matter annihilation rate is
found for $t=t^\odot\simeq 4.5\times 10^9$ years. When $t^\odot\gg \tau$,
annihilation and capture are in equilibrium, so one has $\Gamma_A = C_C/2$.
Thus, in equilibrium, the $\Gamma_A$ only depends on the capture rate.
Therefore, the resulting neutrino flux depends on the dark matter-nucleus cross section
in the capture process, not on the annihilation cross section. The dark matter
spin-dependent cross section, $\sigma^\text{SD}$, can be written as
\cite{Wikstrom:2009kw} \cite{Tanaka:2011uf}
\be
\sigma^\text{SD} = \kappa_f^\text{SD}(m_\chi) \phi_\mu^f
\ee
where $\kappa_f^\text{SD}(m_\chi)$ is the conversion factor between
the spin-dependent cross section and the muon flux.
The $\kappa_f^\text{SD}(m_\chi)$ can be obtained from Figure 3 of
ref.\ \cite{Wikstrom:2009kw} for standard model final states,
$f=W^+W^-, \tau^+\tau^-, t\bar{t}, b\bar{b}$, from which
the neutrinos come from.
These curves are based on calculations using DarkSUSY \cite{Gondolo:2004sc}.
Super-Kamiokande \cite{Choi:2015ara} experiment has the best constraint on dark matter spin-dependent cross section
from neutrino telescope experiments.

\subsection{Discussion}
Search for neutrinos from dark matter annihilations in the Sun and halo is totally possible with the Jinping neutrino experiment proposal, however the target mass is still a limiting factor for the final sensitivity.
Liquid scintillator or water-based liquid scintillator detectors will have a higher efficiency than water detectors at low energy, tens of MeV. If the direction information is available with water-based liquid scintillator, the sensitivity may be further enhanced. More studies are in progress on this thread.

%% file: Acknowledgement.tex
\section{Acknowledgement}
This work is supported in part by, the National Natural Science Foundation of China (No. 11235006 and No. 11475093), the Tsinghua University Initiative Scientific Research Program (20121088035, 20131089288, and 20151080432), the Key Laboratory of Particle \& Radiation Imaging (Tsinghua University), the CAS Center for Excellence in Particle Physics (CCEPP). We acknowledge Yalong River Hydropower Development Company, Ltd. (Yalong Hydro) for building the underground laboratory and continuing efforts on the future development and maintenance of the CJPL.   

%% file: jinping.bbl
\begin{thebibliography}{99}
\bibitem{CJPL1}      K. J. Kang \emph{et al.}, Journal of Physics: Conference Series \textbf{203}, 012028 (2010).
\bibitem{CDEX1}      Q. Yue \emph{et al.}, Phys. Rev. D (R) \textbf{90}, 091701 (2014).

\bibitem{PdX1}       M. J. Xiao \emph{et al.}, Sci China-Phys Mech Astron, \textbf{57}, 2024 (2014).

\bibitem{Jianmin}  Jianmin Li, Xiangdong Ji, Wick Haxton, and Joseph S.Y. Wang, Physics Procedia \textbf{61}, 576 (2015).

\bibitem{zh4} C. Zhang \emph{et al.}, Int. J. Rock Mech. Min. Sci. \textbf{52}, 139 (2012).
\bibitem{zh5} X. T. Feng \emph{et al.}, 8th Asian Rock Mech. Sym. (2014).
\bibitem{zh6} N. Liu \emph{et al.}, Chin. J. Rock Mech. Eng. \textbf{32}, 2235 (2013).

\bibitem{JPRock} Ke-Jun Kang \emph{et al.}, Front. Phys., \textbf{8(4)}, 412 (2013).
\bibitem{SNORock} F. Duncan, A.J. Noble, and D. Sinclair, Annu. rev. Nucl. Part. Sci. \textbf{60}, 163 (2010).
\bibitem{BXRock} C. Arpesella, Appl. Radiat. Isot. \textbf{47}, 991-996, (1996).
\bibitem{KMRock} Precision Measurement of Neutrino Oscillation Parameters and Investigation of Nuclear Georeactor Hypothesis with KamLAND, Chao Zhang, PhD. Thesis, California Institute of Technology (2011)

\bibitem{Mountain}   Y. C. WU, Chinese Phys. C \textbf{37}, 086001 (2013).
\bibitem{HomestakeS}  Corey Adams \emph{et al.} (LBNE Collaboration), arXiv:1307.7335 (2013). 

\bibitem{IAEA}       International Atomic Energy Agency, http://www.iaea.org/ (2015).
\end{thebibliography}

\begin{thebibliography}{99}

\bibitem{Minfang} M.~Yeh {\it et al}., Nucl. Instr. Meth. A \textbf{51} (2011) 660.
\bibitem{20L}     M.~H.~Li {\it et al}., arXiv:1511.09339 (2015).
\end{thebibliography}

\begin{thebibliography}{99}



\bibitem{jnb1}       J.~Bahcall, SLAC Beam Line 31N1, 2-12 (2001) [arXiv:astro-ph/0009259 (2000)].
\bibitem{HaxtonNt}   W.C.~Haxton, Nature \textbf{512}, 378 (2014).
\bibitem{McDonald}   A.B.~McDonald, New Journal of Physics \textbf{6}, 121 (2004).
\bibitem{thirty}     Solar neutrinos the first thirty years, edited by J.~Bahcall, \emph{et al.}, Westview Press (1994).

\bibitem{JBHomepage} J.~Bahcall home page: http://www.sns.ias.edu/\~{}jnb/.

\bibitem{UniMetal}   Neutrino Astrophysics, J.~Bahcall, Cambridge University Press (1989).
\bibitem{HaxtonAb}   W.C.~Haxton, R.G.~and A.~Serenelli, Astrophys. J. \textbf{687}, 678 (2008).
\bibitem{CompS}      A.~Serenelli, C.~Pe\~na-Garay, and W.C.~Haxton, Phys. Rev. D \textbf{87}, 043001 (2013).

\bibitem{fusionI}    E.G.~Adelberger \emph{et al.}, Rev. Mod. Phys. \textbf{70}, 1265 (1998).
\bibitem{fusionII}   E.G.~Adelberger \emph{et al.}, Rev. Mod. Phys. \textbf{83}, 195 (2011).

\bibitem{LUNA}       Home page of LUNA: http://luna.lngs.infn.it/.
\bibitem{JUNA}       Home page of JUNA: http://www.juna.ac.cn/.
\bibitem{WPLiu}      W.~Liu \emph{et al.}, Nuclear Physics A \textbf{616}, 131 (1997).

\bibitem{pontecorvo} B. Pontecorvo, Sov. Phys. JETP \textbf{6}, 429 (1957) and \textbf{26}, 984 (1968).
\bibitem{mns}        Z. Maki, M. Nakagawa, and S. Sakata, Prog. Theor. Phys. \textbf{28}, 870 (1962).
\bibitem{MSW-W}      L. Wolfenstein, Phys. Rev. D \textbf{17}, 2369 (1978).
\bibitem{MSW-MS}     S.P. Mikheev and A.Y. Smirnov, Sov. J. Nucl. Phys. \textbf{42}, 913 (1985); Nuovo Cimento \textbf{9C}, 17 (1986).

\bibitem{Homestake}  B.T. Cleveland \emph{et al.}, Astrophys. J. \textbf{496}, 505 (1998).
\bibitem{SAGE}       J.N. Abdurashitov \emph{et al.}, (SAGE Collaboration), Phys. Rev. C \textbf{80}, 015807 (2009).
\bibitem{GALLEX}     P. Anselmann \emph{et al.}, (GALLEX Collaboration), Phys. Lett. B \textbf{285}, 376 (1992).
\bibitem{GNO}        M. Altmann \emph{et al.}, (GNO Collaboration), Phys. Lett. B \textbf{616}, 174 (2005).
\bibitem{Kamiokande} Y. Fukuda \emph{et al.}, (Kamiokande Collaboration), Phys. Rev. Lett. \textbf{77}, 1683 (1996).
\bibitem{SuperK}     S. Fukuda \emph{et al.}, (Super-Kamiokande Collaboration), Phys. Lett. B \textbf{539}, 179 (2002).
\bibitem{SNO}        Q.R. Ahmad \emph{et al.}, (SNO Collaboration), Phys. Rev. Lett. \textbf{89}, 011301 (2001).


\bibitem{BxPhaseI}   G. Bellini, \emph{et. al.} (Borexino Collaboration), Phys. Rev. D \textbf{89}, 112007 (2014).

\bibitem{Helio}      R.B.~Leighton, R.W.~Noyes, and G.W.~Simon, Astrophys. J. \textbf{135}, 474 (1962).

\bibitem{HaxtonProg} W.C. Haxton, R.G. Hamish Robertson, and A. Serenelli, Annu. Rev. Astron. Astrophys. \textbf{51}, 21 (2013).
\bibitem{OsciForm}   J. Bahcall and C. Pe\~na-Garay, New J. of Phys. \textbf{6}, 63 (2004).
\bibitem{jnb2}       J. Bahcall and R. K. Ulrich, Rev. Mod. Phys. \textbf{60}, 297 (1988).


\bibitem{Roadmap}    J. Bahcall and C. Pe\~na-Garay, JHEP \textbf{11}, 004 (2003).

\bibitem{NoStandard} P. C. de Holanda and A. Yu. Smirnov, Phys. Rev. D \textbf{69}, 113002 (2004).
\bibitem{Bonventre}  R. Bonventre, \emph{et. al.}, Phys. Rev. D \textbf{88}, 053010 (2013).
\bibitem{Friedland}  A. Friedland, C. Lunardini, and C. Pe\~na-Garay, Phys. Lett. B \textbf{594}, 347 (2004).
\bibitem{newRev}     M. Maltoni and A. Yu. Smirnov, arXiv:1507.05287 (2015).

\bibitem{Bright}    J. N. Bahcall and P. I. Krastev, Phys. Rev. C \textbf{56}, 2839 (1997).
\bibitem{SurfaceDen}J. N. Bahcall, P. I. Krastev, and A. Yu. Smirnov, Phys. Rev. D \textbf{60}, 093001 (1999).
\bibitem{SK}        A. Renshaw, \emph{et. al.}, Phys. Rev. Lett. \textbf{112}, 091805 (2014).

\bibitem{MetalProb}  A. Serenelli, \emph{et. al.}, Astrophys. J. \textbf{705}, L123 (2009).
\bibitem{MetalProb2} A. Serenelli, W. C. Haxton, and C. Pe\~na-Garay, Astrophys. J. \textbf{743}, 24 (2011).

\bibitem{LowMetal}   A. Serenelli, S. Basu, J.W. Ferguson, and M. Asplund, Astrophys. J. \textbf{705}, L123 (2009).

\bibitem{10000model} J. Bahcall, A. Serenelli, and S. Basu, Astrophys. J. Suppl. Ser., \textbf{165}, 400 (2006).

\bibitem{Liao}      P.C. de Holanda, Wei Liao, and A.Yu. Smirnov, Nucl. Phys. B \textbf{702}, 307 (2004).

\bibitem{n_e}        J. Bahcall, M. H. Pinsonneault, and S. Basu, Astrophys. J. \textbf{555}, 990 (2001).
\bibitem{decoherent} G. L. Fogli, E. Lisi, D. Montanino, and A. Palazzo1, Phys. Rev. D \textbf{62}, 113004 (2000).
\bibitem{dc-Be7}     E. Lisi, A. Marrone, D. Montanino, A. Palazzo, and S. T. Petcov, Phys. Rev. D \textbf{63}, 093002 (2000).
\bibitem{solar-3f}   S. Goswami and A. Y. Smirnov, Phys. Rev. D \textbf{72}, 053011 (2005).
\bibitem{Earth1}     C. Giunti, \emph{et. al.}, Nucl. Phys. B \textbf{521}, 3 (1998).
\bibitem{Earth2}     Theory of the earth, D. L. Anderson, Blackwell Scintific Publications .
\bibitem{SKSolarI}   J. Hosaka, \emph{et. al.} (Super-Kamiokande Collaboration), Phys. Rev. D \textbf{73}, 112001 (2006).
\bibitem{NueXSec}    C. Giunti and C. W. Kim, Fundamentals of Neutrino Physics and Astrophysics, Oxford university press (2007).
\bibitem{SnoCoverage}B. Aharmim, \emph{et. al.} (SNO Collaboration) Phys. Rev. C \textbf{72}, 055502 (2005).
\bibitem{BxCoverage} L. Oberauera, C. Grieba, F. Feilitzscha, and I. Mannoc, Nucl. Instr. and Meth. A \textbf{530}, 453 (2004).
\bibitem{SuperK2}    K. Abe, \emph{et. al.} (Super-Kamiokande Collaboration), Phys. Rev. D \textbf{83}, 052010 (2011).

\bibitem{Yeh}        M. Yeh, \emph{et al.}, Nucl. Inst. \& Meth. A \textbf{660}, 51 (2011).

\bibitem{BxBe7I}     C. Arpesella, \emph{et. al.} (Borexino Collaboration), Phys. Rev. Lett. \textbf{101}, 091302 (2008).
\bibitem{BxBe7II}    G. Bellini, \emph{et. al.} (Borexino Collaboration), Phys. Rev. Lett. \textbf{107}, 141302 (2011).
\bibitem{BxPep}      G. Bellini, \emph{et. al.} (Borexino Collaboration), Phys. Rev. Lett. \textbf{108}, 051302 (2012).
\bibitem{BxB8}       G. Bellini, \emph{et. al.} (Borexino Collaboration), Phys. Rev. D \textbf{82}, 033006 (2010).
\bibitem{BxPp}       G. Bellini, \emph{et. al.} (Borexino Collaboration), Nature \textbf{512}, 383 (2014).
\bibitem{SnoSumm}    B. Aharmim, \emph{et. al.} (SNO Collaboration), Phys. Rev. C \textbf{88}, 025501 (2013).

\bibitem{nonlinear}  F. P. An, \emph{et. al.} (Daya Bay Collaboration), Phys. Rev. Lett. \textbf{112}, 061801 (2014).

\end{thebibliography}

\begin{thebibliography}{99}

\bibitem{Geology} William F. McDonough and Ond$\breve{r}$ej $\breve{S}$r$\acute{a}$mek, Environ. Earth Sci. \textbf{71}, 3787 (2014).
\bibitem{HeatFlow}Jaupart C, Labrosse S, Mareschal J-C, Treatise on geophysics, \textbf{7}, 253 (2007).
\bibitem{GeoReview} G. Bellini, \emph{et. al.}, Prog. Part. Nucl. Phys. \textbf{73}, 1 (2013).
\bibitem{YuHuang}   Yu Huang, \emph{et. al.}, Geochem. Geophys. Geosyst. \textbf{14}, 2003 (2013).
\bibitem{CrustErr}  Bassin, C.,  G. Laske, and  T. G. Masters (2000), The current limits of resolution for surface wave tomography in North America, \emph{EOS Trans}.
\bibitem{CrustErr2} Shapiro, N. M., and M. H. Ritzwoller, Geophysical Journal International, \textbf{151}, 88 (2002).
\bibitem{CrustErr3} Negretti, M., M. Reguzzoni, and D. Sampietro, A web processing service for GOCE data
    exploitation, in First International GOCE Solid Earth workshop, edited, Enschede, The
    Netherlands (2012).

\bibitem{McNew} Huang, Y., V. Strati, F. Mantovani, S. B. Shirey, and W. F. McDonough, Geochem. Geophys. Geosyst., \textbf{15}, 3925 (2014).

\bibitem{KLGeo1} T. Araki, \emph{et. al.} (KamLAND Collaboration), Nature \textbf{436}, 499 (2005).
\bibitem{KLGeo2} A. Gando, \emph{et. al.} (KamLAND Collaboration), Nature Geoscience \textbf{4}, 647 (2011).
\bibitem{KLGeo3} A. Gando, \emph{et. al.} (KamLAND Collaboration), Phys. Rev. D \textbf{88}, 033001 (2013).
\bibitem{BXGeo1} G. Bellini, \emph{et. al.} (Borexino Collaboration), Phys. Lett. B \textbf{687}, 299 (2010).
\bibitem{BXGeo2} G. Bellini, \emph{et. al.} (Borexino Collaboration), Phys. Lett. B \textbf{722}, 295 (2013).
\bibitem{BxNew}  M. Agostini, \emph{et. al.} (Borexino Collaboration), Phys. Rev. D \textbf{92}, 031101(R) (2015).

\bibitem{GeoReactor} D. F. Hollenbach and J. M. Herndon, Proc. Nat. Acad. Sci, \textbf{98}, 11085 (2001).

\bibitem {GeoSpec} S. Enomoto, Neutrino geophysics and observation of geo-neutrinos at KamLAND, Ph.D. Thesis, Tohoku University (2005).

\bibitem{XSEC}   P. Vogel and J. F. Beacom, Phys. Rev. D \textbf{60}, 053003 (1999).

\bibitem {industry} International Atomic Energy Agency, http://www.iaea.org/ (2015).
\bibitem {RctEst} M. Baldoncini, \emph{et. al.}, Phys. Rev. D \textbf{91}, 065002 (2015).


\end{thebibliography}

\begin{thebibliography}{}

\bibitem{1p1century}  S. Ando, J.F. Beacom, H. Y$\ddot{u}$ksel, Phys. Rev. Lett. \textbf{95}, 171101 (2005).
\bibitem{SK0-srn}     M. Malek \emph{et al.} (Super-Kamiokande Collaboration), Phys. Rev. Lett. \textbf{90}, 061101 (2003).
\bibitem{SK1-srn}     K. Bays \emph{et al.} (Super-Kamiokande Collaboration), Phys. Rev. D \textbf{85}, 052007 (2012).
\bibitem{SK2-srn}     H. Zhang \emph{et al.} (Super-Kamiokande Collaboration), Astropart. Phys. \textbf{60}, 41 (2015).
\bibitem{KamLAND-srn} A. Gando \emph{et al.} (KamLAND Collaboration), Astrophys. J. \textbf{745}, 193 (2012).
\bibitem{Borexino-srn}G. Bellini \emph{et al.} (Borexino Collaboration), Phys. Lett. B \textbf{696}, 191 (2011).
\bibitem{SNO-srn}     B. Aharmim \emph{et al.} (SNO Collaboration), Astrophys. J. \textbf{653}, 1545 (2006).
\bibitem{JBeacom} J. F. Beacom, Annu. Rev. Nucl. Part. Sci. \textbf{60}, 439 (2010).

\bibitem{LMASRN} S. Ando, K. Sato and T. Totani, Astropart. Phys. \textbf{18}, 307 (2003); J. Phys. Soc. Jpn. (Suppl. B) \textbf{77}, 9 (2008).
\bibitem{ConstantSN} T. Totani and K. Sato, Astropart. Phys. \textbf{3}, 367 (1995).
\bibitem{CosmicGas} R. A Malaney, Astropart. Phys. \textbf{7}, 125 (1997).
\bibitem{ChemicalEvo} D. H. Hartmann and S. E. Woosley, Astropart. Phys. \textbf{7}, 137 (1997).
\bibitem{HeavyMetal1} M. Kaplinghat, G. Steigman and T. P. Walker, Phys. Rev. D \textbf{62}, 043001 (2000).
\bibitem{HeavyMetal2} L. Strigari, M. Kaplinghat, G. Steigman and T. Walker, JCAP \textbf{0403}, 007 (2004).
\bibitem{Population} T. Totani, K. Sato and Y. Yoshii, Astrophys. J. \textbf{460}, 303 (1996).
\bibitem{HBD6} S. Horiuchi, J. F. Beacom and E. Dwek, Phys. Rev. D \textbf{79}, 083013 (2009).
\bibitem{StarForm} M. Fukugita and M. Kawasaki, Mon. Not. Roy. Astron. Soc. \textbf{340}, L7 (2003).
\bibitem{FailedSN} C. Lunardini, Phys. Rev. Lett. \textbf{102}, 231101 (2009).

\bibitem{WaterLS}    M. Yeh \emph{et al.}, Nucl. Inst. \& Meth. A \textbf{660}, 51 (2011).
\bibitem{WaterGd}    J. F. Beacom and M. R. Vagins, Phys. Rev. Lett. \textbf{93}, 171101 (2004).
\bibitem{IBD-XSEC}   P. Vogel and J. F. Beacom, Phys. Rev. D \textbf{60} 053003 (1999).

\bibitem{SpaPowerlaw} S. Abe, \emph{et al.} (KamLAND Collaboration), Phys. Rev. C \textbf{81}, 025807 (2010).
\bibitem{DYBnGd}     F. P. An, \emph{et al.} (Daya Bay Collaboration), Chin. Phys. C \textbf{37}, 011001 (2013).
\bibitem{DYBnH}      F. P. An \emph{et al.} (Daya Bay Collaboration), Phys. Rev. D (R) \textbf{90}, 071101 (2014).

\bibitem{AtmNu1} O. Peres and A. Smirnov, Phys. Rev. D \textbf{79}, 113002 (2009).
\bibitem{AtmNu2} G. Battistoni, A. Ferrari, T. Montaruli and P.R. Sala, Astropart. Phys. \textbf{23}, 526 (2005).

\bibitem{C11thesis} R. Moellenberg, Ph. D. thesis, (2009).
\bibitem{LENApaper} M. Wurm, Astropart. Phys. \textbf{35}, 685 (2012).


\end{thebibliography}

\begin{thebibliography}{9}

\bibitem{Hirata1}
K.S. Hirata, \emph{et al.}, Phys.\ Rev.\ Lett. \textbf{58}, 1490 (1987).

\bibitem{Hirata2}
K.S. Hirata, \emph{et al.}, Phys.\ Rev.\ D \textbf{38}, 448 (1988).

\bibitem{Bionta}
R.M. Bionta, \emph{et al.}, Phys.\ Rev.\ Lett. \textbf{58}, 1494 (1987).

\bibitem{Bratton}
C.B. Bratton, \emph{et al.}, Phys.\ Rev.\ D \textbf{37}, 3361 (1988).

\bibitem{Alekseev1}
E.N. Alekseev, \emph{et al.}, JETP Lett. \textbf{45}, 589 (1987).

\bibitem{Alekseev2}
E.N. Alekseev, \emph{et al.}, Phys.\ Lett.\ B \textbf{205}, 209 (1988).

\bibitem{Raffelt}
G.G. Raffelt, arXiv:1201.1637 (2012).

\bibitem{book}
R.N. Mohapatra and P.B. Pal, \emph{Massive Neutrinos in Physics and Astrophysics},
World Scientific Publishing Co. Pte. Ltd. 2004.

\bibitem{Lunardini}
C.~Lunardini and A.Y.~Smirnov, JCAP \textbf{0306}, 009 (2003).

\bibitem{Masshierarchy}
P.D.~Serpico, S.~Chakraborty, T. Fischer, \emph{et al.}, Phys.\ Rev.\ D \textbf{85}, 085031 (2012).

\bibitem{Gravwave}
C.D.~Ott, E.P.~O'Connor, S.~Gossan, \emph{et al.}, Nucl.\ Phys.\ B (Proc.\ Suppl.) \textbf{235-236}, 381 (2013).

\bibitem{Rate}
S.~Ando, J.F.~Beacom, H.~Y{\"{u}}ksel, Phys.\ Rev.\ Lett. \textbf{95}, 171101 (2005).

\bibitem{snews}
SNEWS,
\emph{http://snews.bnl.gov (2015)},
{\emph{Supernova Early Warning System}}

\bibitem{SNEWSpaper}
K.~Scholberg, \emph{et al.}, New J.\ Phys. \textbf{6}, 114 (2004).

\bibitem{SNnuspec}
G.~Raffelt, I. Tamborra, B. M{\"{u}}ller, L. H{\"{u}}depohl, H.T. Janka, Phys.\ Rev.\ D \textbf{86}, 125031 (2012).

\bibitem{SNdetection}
K.~Scholberg, Annu.\ Rev.\ Nucl.\ Part.\ Sci. \textbf{62}, 81 (2012).

\bibitem{nuProton}
John F. Beacom, Will M. Farr, Petr Vogel, Phys. Rev. D \textbf{66}, 033001 (2002).

\bibitem{protonQuenching} B. von Krosigk et al., Eur. Phys. J. C \textbf{73}, 2390 (2013).

\bibitem{DYBSNtrigger}
Hanyu Wei, \emph{et al.},
Astropart.Phys. 75 (2016) 38.

\bibitem{DYBnH-SNB}
F.P. An, \emph{et al.} (Daya Bay Collaboration), Phys.\ Rev.\ D  (R) \textbf{90}, 071101 (2014).

\bibitem{NewLS}
J. R. Alonso {\it et al.}, arXiv:1409.5864v3 (2014).


\end{thebibliography}

\begin{thebibliography}{100}

\bibitem{Collaboration:2011jza}
  A.~Gando {\it et al.}  [The KamLAND Collaboration],
  Astrophys.\ J.\  {\bf 745}, 193 (2012).

\bibitem{PalomaresRuiz:2007eu}
  S.~Palomares-Ruiz and S.~Pascoli,
  Phys.\ Rev.\ D {\bf 77}, 025025 (2008).

\bibitem{Yuksel:2007ac}
  H.~Yuksel, S.~Horiuchi, J.~F.~Beacom and S.~Ando,
  Phys.\ Rev.\ D {\bf 76}, 123506 (2007).

\bibitem{Navarro:1995iw}
  J.~F.~Navarro, C.~S.~Frenk and S.~D.~M.~White,
  Astrophys.\ J.\  {\bf 462}, 563 (1996).

\bibitem{Moore:1999gc}
  B.~Moore, T.~R.~Quinn, F.~Governato, J.~Stadel and G.~Lake,
  Mon.\ Not.\ Roy.\ Astron.\ Soc.\  {\bf 310}, 1147 (1999).

\bibitem{Kravtsov:1997dp}
  A.~V.~Kravtsov, A.~A.~Klypin, J.~S.~Bullock and J.~R.~Primack,
  Astrophys.\ J.\  {\bf 502}, 48 (1998).

\bibitem{Wurm:2011zn}
  M.~Wurm {\it et al.}  [LENA Collaboration],
  Astropart.\ Phys.\  {\bf 35}, 685 (2012).

\bibitem{Silk:1985ax}
  J.~Silk, K.~A.~Olive and M.~Srednicki,
  Phys.\ Rev.\ Lett.\  {\bf 55}, 257 (1985).


\bibitem{Krauss:1985ks}
  L.~M.~Krauss, K.~Freese, W.~Press and D.~Spergel,
  Astrophys.\ J.\  {\bf 299}, 1001 (1985).

\bibitem{Freese:1985qw}
  K.~Freese,
  Phys.\ Lett.\ B {\bf 167}, 295 (1986).

\bibitem{Krauss:1985aaa}
  L.~M.~Krauss, M.~Srednicki and F.~Wilczek,
  Phys.\ Rev.\ D {\bf 33}, 2079 (1986).

\bibitem{Gaisser:1986ha}
  T.~K.~Gaisser, G.~Steigman and S.~Tilav,
  Phys.\ Rev.\ D {\bf 34}, 2206 (1986).

\bibitem{Griest:1986yu}
  K.~Griest and D.~Seckel,
  Nucl.\ Phys.\ B {\bf 283}, 681 (1987)
  [Erratum-ibid.\ B {\bf 296}, 1034 (1988)].

\bibitem{Gould:1987ju}
  A.~Gould,
  Astrophys.\ J.\  {\bf 321}, 560 (1987).

\bibitem{Wikstrom:2009kw}
  G.~Wikstrom and J.~Edsjo,
  JCAP {\bf 0904}, 009 (2009).


\bibitem{Choi:2015ara}
  K.~Choi {\it et al.}  [Super-Kamiokande Collaboration],
  Phys.\ Rev.\ Lett.\  {\bf 114}, no. \textbf{14}, 141301 (2015).

\bibitem{Aartsen:2012kia}
  M.~G.~Aartsen {\it et al.}  [IceCube Collaboration],
  Phys.\ Rev.\ Lett.\  {\bf 110}, no. \textbf{13}, 131302 (2013).

\bibitem{Boliev:2013ai}
  M.~M.~Boliev, S.~V.~Demidov, S.~P.~Mikheyev and O.~V.~Suvorova,
  JCAP {\bf 1309}, 019 (2013).

\bibitem{Tanaka:2011uf}
  T.~Tanaka {\it et al.}  [Super-Kamiokande Collaboration],
  Astrophys.\ J.\  {\bf 742}, 78 (2011).


\bibitem{Gondolo:2004sc}
  P.~Gondolo, J.~Edsjo, P.~Ullio, L.~Bergstrom, M.~Schelke and E.~A.~Baltz,
  JCAP {\bf 0407}, 008 (2004).




\end{thebibliography}
